\documentclass{aa}  
\usepackage{lscape}
\usepackage{graphicx}
\usepackage{dcolumn}

\usepackage{txfonts}
\usepackage[colorlinks,bookmarks]{hyperref}
\usepackage{color}
\definecolor{linkblue}{rgb}{0,0,0.8}
\definecolor{linkgreen}{rgb}{0,0.5,0}
\definecolor{gold}{rgb}{1,0.85,0}
\definecolor{silver}{rgb}{0.7,0.7,0.7}
\hypersetup{pdfpagemode=UseNone, pdfstartview=FitH, linkcolor=linkblue, %
  citecolor=linkblue, urlcolor=linkblue}

\newcommand{\m}{\mathrm}
\newcommand{\angstrom}{\mbox{\normalfont\AA}} 

\usepackage{booktabs,tabularx}
\newcolumntype{Y}{>{\raggedright\arraybackslash}X}
\newcommand\ph{$\phantom{1}$} 

\usepackage[normalem]{ulem}

\begin{document} 

   \title{Lyman Continuum Leaker Candidates at $z\sim3-4$ in the HDUV Based on a Spectroscopic Sample of MUSE LAEs}
   \titlerunning{LyC Leaker Candidates in the HDUV Based on a Sample of MUSE LAEs}
   
   \author{J. Kerutt \thanks{\email{kerutt@astro.rug.nl}}
          \inst{1}\fnmsep\inst{2}
          \and
          P. A. Oesch
          \inst{2}\fnmsep\inst{3}
          \and
          L. Wisotzki
          \inst{4}
          \and
          A. Verhamme
          \inst{2}
          \and 
          H. Atek
          \inst{5}
          \and
          E. C. Herenz
          \inst{6}
          \and
          G. D. Illingworth
          \inst{7}
          \and 
          H. Kusakabe
          \inst{2}
          \and 
          J. Matthee
          \inst{8}
          \and
          V. Mauerhofer
          \inst{1}
          \and
          M. Montes
          \inst{9}
          \and 
          R. P. Naidu
          \inst{10}
          \and 
          E. Nelson
          \inst{11}
          \and
          N. Reddy
          \inst{12}
          \and
          J. Schaye
          \inst{6}
          \and
          C. Simmonds
          \inst{13}\fnmsep\inst{14}
          \and
          T. Urrutia
          \inst{4}
          \and
          E. Vitte
          \inst{2}\fnmsep\inst{15}
          }

   \institute{Kapteyn Astronomical Institute, University of Groningen, P.O. Box 800, 9700 AV Groningen, The Netherlands
        \and  
            Department of Astronomy, Université de Genève, 51 Ch. Pegasi, 1290 Versoix, Switzerland
        \and 
            Cosmic Dawn Center (DAWN), Niels Bohr Institute, University of Copenhagen, Jagtvej 128, K\o benhavn N, DK-2200, Denmark
        \and 
            Leibniz-Institut f\"ur Astrophysik Potsdam (AIP), An der Sternwarte 16, 14482 Potsdam, Germany 
        \and  
            d'Astrophysique de Paris, CNRS, Sorbonne Universit\'e, 98bis Boulevard Arago, 75014, Paris, France
        \and  
            Inter-University Centre for Astronomy and Astrophysics, Pune 411 007, India
        \and  
            Department of Astronomy and Astrophysics, University of California, Santa Cruz, CA 95064, USA
        \and  
            Department of Physics, ETH Z{\"u}rich, Wolfgang-Pauli-Strasse 27, Z{\"u}rich, 8093, Switzerland
        \and 
            Instituto de Astrof\'{\i}sica de Canarias, c/ V\'{\i}a L\'actea s/n, E-38205 - La Laguna, Tenerife, Spain \\
            Departamento de Astrof\'isica, Universidad de La Laguna, E-38205 - La Laguna, Tenerife, Spain
        \and 
            MIT Kavli Institute for Astrophysics and Space Research, 77 Massachusetts Ave., Cambridge, MA 02139, USA
        \and 
            Department for Astrophysical and Planetary Science, University of Colorado, Boulder, CO 80309, USA
        \and 
            Department of Physics and Astronomy, University of California, Riverside, 900 University Avenue, Riverside, CA 92521, USA
        \and 
            Kavli Institute for Cosmology, University of Cambridge, Madingley Road, Cambridge, CB3 0HA, UK.
        \and  
            Cavendish Laboratory, University of Cambridge, 19 JJ Thomson Avenue, Cambridge, CB3 0HE, UK
        \and 
            European Southern Observatory, Av. Alonso de Córdova 3107, 763 0355 Vitacura, Santiago, Chile
    }   

   \date{Received ; accepted }

  \abstract
   {In recent years, a number of Lyman continuum (LyC) leaker candidates at intermediate redshifts have been found, providing insight into how the Universe was reionised at early cosmic times.
   Around a hundred LyC leakers at all redshifts are known by now, which enables us to analyse their properties statistically.}
   {Here we identify new LyC leaker candidates at $z\approx 3-4.5$ and compare them to objects from the literature to get an overview of the different observed escape fractions and their relation to the properties of the Lyman $\alpha$ (Ly$\alpha$) emission line. 
   The aim of this work is to test indicators (or proxies) for LyC leakage suggested in the literature and to improve our understanding of the kind of galaxies from which LyC radiation can escape.}
   {We use data from the Hubble Deep Ultraviolet (HDUV) legacy survey to search for LyC emission based on a sample of $\approx 2000$ Ly$\alpha$ emitters (LAEs) detected previously in two surveys with the Multi-Unit Spectroscopic Explorer (MUSE), MUSE-Deep and MUSE-Wide. Based on their known redshifts and positions, we look for potential LyC leakage in the WFC3/UVIS F336W band of the HDUV. The escape fractions are measured and compared in different ways, including based on spectral energy distribution (SED) fitting performed using the CIGALE software.}
   {We add twelve objects to the sample of known LyC leaker candidates (five highly likely leakers and seven potential ones), one of which was previously known, and compare their Ly$\alpha$ properties to their escape fractions. We find escape fractions between $\sim 20\%$ and $\sim 90\%$, assuming a high transmission in the intergalactic medium (IGM). We show a method to use the number of LyC leaker candidates we find to infer the underlying average escape fraction of galaxies, which is $\approx 12\%$.}
  {Based on their Ly$\alpha$ properties, we conclude that LyC leakers are not very different from other high-z LAEs and suggest that most LAEs could be leaking LyC even if this can not always be detected due to the direction of emission and the transmission properties of the IGM.}

   \maketitle
%


\section{Introduction} \label{Introduction} 

The Epoch of Reionisation (EoR) is the last phase transition of the universe, where the intergalactic medium (IGM) went from a neutral to a mostly ionised state (see review \citealp{Wise2019} and e.g. \citealp{Fan2006,Bouwens2007,Ouchi2009b,Ouchi2009a,Bouwens2015,Haiman2016}). This mostly coincides with the formation of the first galaxies. While recent years have brought new insights, there are still aspects we need to understand about this crucial phase in the development of the universe. 

What we can constrain relatively well is the time of the EoR ($z \approx6 \-- 8$, \citealp{Fan2006}), for example using the drop in the fraction of Ly$\alpha$ emitters (LAEs, e.g. \citealp{,Ouchi2008,Kashikawa2011,Caruana2012,Kusakabe2020}), which is also seen in simulations (e.g. \citealp{Garel2021}). This could either be due to the intrinsic evolution of the LAE population or an effect of the IGM. While Lyman continuum (LyC) emission is absorbed by the neutral hydrogen in the IGM at the EoR, Ly$\alpha$ (Ly$\alpha$) is scattered out of the line of sight and is, therefore, an indicator of the neutral fraction of the IGM.

In recent years, observational results pointed toward star-forming galaxies such as LAEs being the best candidates for providing the ionising emission at and after the EoR (\citealp{Fontanot2014,Onoue2017,Matsuoka2018,Japelj2017,Naidu2020,Matthee2022,Naidu2022,Begley2022,Matsuoka2022}), with only a minimal contribution by active galactic nuclei (AGN, \citealp{Cowie2009,MadauHaardt2015,Smith2020,Trebitsch2021}). Since the number density of AGN decreases rapidly at $z > 3$ (e.g.\ \citealp{Masters2012}) and the escape fraction of ionising photons from AGN is not as high as needed (e.g.\ \citealp{Micheva2017}), the only possibility for AGN to contribute significantly to the EoR would be through a large number of low luminosity AGN, as has been claimed for example by \citet{Giallongo2015} and \citet{Grazian2018}. Recent James Webb Space Telescope (JWST) observations (\citealp{Kocevski2023,Ubler2023,Harikane2023}) seem to confirm this possibility, while \citet{Parsa2018} found significantly fewer faint AGN at redshifts $z > 4$ (see also recent results using Subaru/Suprime-Cam and Hyper Suprime-Cam in \citealp{Onoue2017,Matsuoka2018,Matsuoka2022} at redshifts $z \approx 6$). It is also possible, that AGN contribute significant amounts of LyC emission starting at $z\approx2-3$, after the EoR (e.g. \citealp{Becker2015,MadauHaardt2015,Faucher2020,Trebitsch2021}), which means the question of the AGN contribution to the (re)ionisation of the universe is not fully answered yet.

Having thus mostly ruled out AGN as the sources of reionisation, the focus has shifted towards star-forming galaxies. An escape fraction of ionising photons of $\approx10\%$ is needed to explain the EoR (\citealp{Pawlik2009,Vanzella2012,Mitra2015,Giallongo2015,Robertson2015,MadauHaardt2015,Feng2016} but also \citealp{Faucher2008,Matthee2017a}, who find lower necessary escape fractions under certain assumptions).

Since the neutral fraction of hydrogen in the IGM is rising towards the EoR (\citealp{Madau1995,Inoue2014}), it is not possible to directly observe the LyC radiation responsible for the ionisation of the IGM. However, assuming the properties of the sources of the EoR do not evolve much with redshift, we can study them at lower redshifts, where the neutral fraction of the IGM allows the LyC emission to get through (\citealp{Hu1998,Steidel2001,Shapley2006,Ouchi2008,Blanc2011,Vanzella2012}). 

There have been several discoveries of LyC leaker candidates at $z\approx3-4$, for example the well-studied Ion 1 \citet{Vanzella2010a,Ji2020}, Ion 2 \citet{Vanzella2015,Vanzella2020} and Ion 3 \citet{Vanzella2018} (see table \ref{tab:LyC_leakers_ind_high} in the appendix). A few LyC leakers at those redshifts have very high escape fractions (e.g. $f_{\m{esc}}^{\m{LyC}} > 50\%$ in \citealp{Vanzella2015, deBarros2016}, $f_{\m{esc}}^{\m{LyC}} = 52\%$ in \citealp{saxena2022}, $f_{\m{esc}}^{\m{LyC}} = 90\%$ in \citealp{Marques-Chaves2022}, $f_{\m{esc}}^{\m{LyC}} = 100\%$ in \citealp{Rivera-Thorsen2022}). These high escape fractions could be the result of orientation effects, as we might see them at the right angle where their LyC escapes, since in general, escape fractions from individual objects (and stacks) tend to be low. For example \citet{Grimes2009} find no LyC from a sample of 16 local starburst galaxies, \citet{Rutkowski2016} find non-detections with upper limits of $f_{\m{esc}}^{\m{LyC}} < 2.1\%$ for objects at $z\approx1$, and \citet{Grazian2016} only find two LyC leakers in 37 galaxies at $z\approx3.3$. There also seems to be a trend with redshift (\citealp{Mitra2013,Fontanot2014,Faisst2016,Khaire2016,Japelj2017}).
However, even at $z\approx4$, the possible absorption by the IGM has to be taken into account when measuring escape fractions (e.g. \citealp{Bassett2021}), which is not easy considering the stochastic nature of the different lines of sight (\citealp{Madau1995,Inoue2014}) and makes measurements of LyC escape fractions rather uncertain. 

At slightly lower redshifts, $z\approx 2-3$, where the IGM transmission is reasonably high, there have been many successful detections of LyC leakers (e.g. \citealp{Vanzella2010b,Mostardi2015,Grazian2016,Shapley2016,Steidel2018,Fletcher2019,Rivera-Thorsen2019,Saha2020}). However, it is important to exclude lower-redshift interlopers, as they can be responsible for several assumed LyC leaker candidates (\citealp{Shapley2006,Vanzella2010a,Siana2015,Mostardi2015}), which is why reliable techniques to exclude such interlopers are important (see e.g. \citealp{Pahl2021}, who suggest a colour selection technique on resolved photometry).

Going to even lower redshifts can be a solution to avoid interlopers, and many low-redshift analogues of high-redshift star-forming galaxies (in particular LAEs and Lyman Break Galaxies, LBGs) have been studied in recent years. Examples of such analogues are Green Peas (\citealp{Cardamone2009}), which sometimes exhibit a high relative escape fraction of ionising emission, as high as $\m{f}_{\m{esc}}^{\m{LyC}} = 73\%$ (\citealp{Izotov2016a}, \citealp{Izotov2016b}, \citealp{Izotov2018a}, \citealp{Izotov2018b}), while \cite{Izotov2021} find 15 LyC leakers among a sample of 20 Green Peas. Other low-z galaxies that are leaking LyC emission are for example Haro 11 (\citealp{Bergvall2006,Hayes2007,Leitet2011,Leitet2013,Keenan2017,Rivera-Thorsen2017})  with $f_{\m{esc}}^{\m{LyC,rel}} = 16.6^{+7.4}_{-6.5}\%$ (\citealp{Leitet2013}), Tol 1247 with $f_{\m{esc}}^{\m{LyC}} = 2.4\%$ (\citealp{Leitet2013}), Mrk54 with $f_{\m{esc}}^{\m{LyC}} = 6.2\%$ (\citealp{Deharveng2001}), and J0921 with $f_{\m{esc}}^{\m{LyC}} \sim 1\%$ (\citealp{Borthakur2014}, but see table \ref{tab:LyC_leakers_low} in the appendix). 
Recently, the number of known LyC emitters at low-redshifts has been significantly increased with the Low-Redshift Lyman Continuum Survey (LzLCs, \citealp{Wang2021,Flury2022a,Flury2022b,Saldana-Lopez2022}), which includes 35 LyC emitters at $z=0.2-0.4$ with escape fractions of up to $50\%$ and notably 12 objects with $f_{\m{esc}}^{\m{LyC}} > 5\%$. A caveat to many of the studies of low-redshift LyC leakers is that they are often selected to have specific properties that make LyC leakage likely. Therefore, it is not certain whether they are representative of the general galaxy population.

Individual detections of LyC leakers at high redshifts could introduce a selection bias in the interpretation of the escape fraction since they could be extreme and rare objects. However, since the IGM is stochastic and highly variable along different lines of sight, it is also possible that they have particularly low foreground opacities and are not different from the general population. Several studies, therefore, used stacking to obtain the mean escape fraction, with the first direct detection of LyC emission from star-forming galaxies at high redshift from stacked spectra of 29 LBGs at $z=3.4$ (\citealp{Steidel2001}) and an escape fraction of $f_{\m{esc}} \gtrsim 50 \%$. Other studies followed, usually finding much lower escape fractions (e.g. \citealp{Siana2010,Grazian2017,Fletcher2019,Steidel2018}), often using LBGs and LAEs to find LyC emission (e.g.\ \citealp{Grazian2016,Japelj2017,Grazian2017,Naidu2018,Marchi2018} and \citealp{Rutkowski2016,Rutkowski2017} at lower redshifts). Results from these stacking analyses are not conclusive yet, as for example \citet{Steidel2018} find $f_{\m{esc}}^{\m{LyC}} = 9\%$ using LBGs without prior selection for LyC leakage, barely enough to explain reionisation. However, they find that fainter galaxies could have higher escape fractions, which highlights the importance of the sample selection.
In a sample targeting preferentially [\ion{O}{III}] emitters (which are thought to be good candidates for LyC leakage), \citet{Fletcher2019} use 61 LAEs and find $20\%$ individual LyC leakers, but when excluding the individually-detected objects from the stack, they find an upper limit in the LyC escape fraction of $0.3\%$. This could mean that only some objects have a high escape fraction, but most have close to 0, or that most IGM transmission lines are not transparent for LyC. Indeed, in a similar study, \citet{Bian2020} do not find any individual LyC leaker candidates among a sample of 54 LAEs at $z\approx3.1$ in the GOODS-S field and no detection in a stack of those objects either. In contrast, \citet{Begley2022} find an escape fraction of $f_{\m{esc}}=0.07\pm0.02$ in a stack of 148 star-forming galaxies at $z\simeq3.5$.
Individual detections are of course dependent on the depth of the data and thus somewhat arbitrary unless the same data and detection criteria are compared.
Another aspect is that from simulations we expect the escape fraction of ionising photons to vary over time (by more than six orders of magnitude, \citealp{Trebitsch2017}) and between different sight-lines and especially with inclination (e.g. \citealp{Trebitsch2017,Smith2022}).

The varying results on escape fractions from the literature can also in part be attributed to the different methods and assumptions made to derive the escape fractions. Not only is the absorption in the IGM uncertain for individual objects, but also the stellar population properties such as metallicity, age, star formation history, and initial mass function as well as the dust absorption. In addition, only some studies take the effects of the circumgalactic medium (CGM) into account (e.g. \citealp{Reddy2016b,Steidel2018,Pahl2021}), while most group them with those of the IGM, although significant $\ion{H}{I}$ absorption has been found up to $\approx500\,\m{km/s}$ for LAEs at $z=2.9-3.8$ (\citealp{Muzahid2021}).

It would, therefore, be ideal to infer the LyC emission from other observables, which are less affected by the IGM and model assumptions. Several such proxies for LyC have been suggested (e.g. \citealp{Dijkstra2014,Verhamme2015,Verhamme2017,Marchi2018,Izotov2018b}), including a connection with Ly$\alpha$ emission, e.g. through the equivalent widths (\citealp{Micheva2017,Marchi2018,Steidel2018,Fletcher2019,Pahl2021,Reddy2022,Flury2022b}) or double peaks (\citealp{Verhamme2017,Vanzella2020,Izotov2021}). Contrary to such intuitive trends, one of the highest redshift LyC leakers, Ion1 at $z=3.8$, has Ly$\alpha$ in absorption (\citealp{Ji2020}). This can happen for moderate neutral hydrogen column densities (between log[N$_{\ion{H}{I}}\,\m{cm}^{-2}$] $\approx13- 17$), where the gas would be optically thin for LyC, but Ly$\alpha$ is already scattered. Similarly, the strong LyC leaker from \citet{Marques-Chaves2022} has a surprisingly wide peak separation of $\Delta v(\m{Ly}\alpha) = 680 \pm 70\, \m{km}\, \m{s}^{-1}$, indicating a lot of scattering for Ly$\alpha$ and thus a rather large neutral hydrogen column density. 

Nevertheless, some degree of connection between Ly$\alpha$ and LyC emission is expected through their linked production mechanisms and similar liability to interact with neutral hydrogen. A prerequisite for Ly$\alpha$ emission is the production of LyC, which ionises neutral hydrogen, and then recombines to emit a Ly$\alpha$ photon (in $\approx$ 2/3 of the cases for a temperature of $10^4\,\m{K}$, e.g. \citealp{Dijkstra2014}). Simplified, more LyC photons could therefore mean more Ly$\alpha$ photons. While Ly$\alpha$ then scatters in neutral hydrogen, LyC will be absorbed, which means that if there is LyC emission, they both benefit from a relatively free path (meaning a low neutral hydrogen column density) through the interstellar medium (ISM) to easily escape the galaxy. Not only the creation of Ly$\alpha$ thus depends on the presence of LyC, but due to its resonant nature, the shape of the Ly$\alpha$ line traces the neutral hydrogen column density in the ISM and CGM. A high neutral hydrogen column density will prevent LyC from escaping and imprint itself on the Ly$\alpha$ line shape through the FWHM and peak separation. Therefore, both properties have been proposed as tracers of LyC emission. Thus, LAEs could be ideal candidates to look for LyC emission (see e.g. \citealp{Rivera-Thorsen2017,Vanzella2018,Gazagnes2020,Izotov2021,Matthee2022}), which is what we test in this study.

Based on a sample of LAEs (see \citealp{Kerutt2022}) found with the Multi-Unit Spectroscopic Explorer (MUSE, \citealp{Bacon2010}), we look for individual LyC emitters in the Hubble Deep Ultraviolet (UV) Legacy Survey (HDUV, \citealp{Oesch2018}) in the GOODS-South and North (\citealp{Giavalisco2004}).
\citet{Naidu2017} already found six candidates in the HDUV (three of which might contain an AGN) using the two HST filter bands WFC3/UVIS F336W and F275W at $z\approx2$, but since the Lyman break is in the middle of F275W, computing the escape fraction is challenging. \citet{Jones2018} use F275W in GOODS-North for objects at $z\approx2.4$, which guarantees that only LyC will fall in the filter band, and find six candidates, four of which are interlopers. A more recent study with the same filter band WFC3/UVIS F275W focused on the GOODS-South and resulted in five LyC leaker candidates, two of which might be interlopers (\citealp{Jones2021}). This demonstrates the need for carefully analysing the potential leakers to estimate the contamination, which is what we strive for in this paper. This allows us to better understand if the proposed connection between the LyC and Ly$\alpha$ holds up at higher redshifts. If not, this might be an indication that the Ly$\alpha$ and LyC emission is not necessarily coming from the same region in the galaxy.

The paper is structured as follows: In Sect. \ref{Data} we discuss the data we use for this study, from the HDUV survey and the MUSE-Wide and -Deep surveys. In Sect. \ref{Sample Selection} we search for LyC emission from our selected sample of LAEs and discuss potential contamination. Having defined a sample of LyC leaker candidates, we then measure their spectral energy distributions (SEDs) and derive LyC escape fractions, which we discuss in Sect. \ref{Measuring Escape fractions}. We compare the escape fractions to Ly$\alpha$ properties in Sect. \ref{Lyman alpha properties of LyC leaker candidates} in order to find correlations. We discuss our results in Sect. \ref{Discussion} and give a summary and conclusion in Sect. \ref{Summary and Conclusions}. 
Throughout this paper, we use AB magnitudes and physical distances and assume flat $\Lambda\m{CDM}$ cosmology with $\m{H}_0 = 70\,\m{km}\,\m{s}^{-1}\,\m{Mpc}^{-1}$, $\Omega_{\m{m}} = 0.3$ and $\Omega_{\Lambda}=0.7$.


\section{Data} \label{Data} 

\begin{figure}
  \centering
  \includegraphics[width=0.49\textwidth]{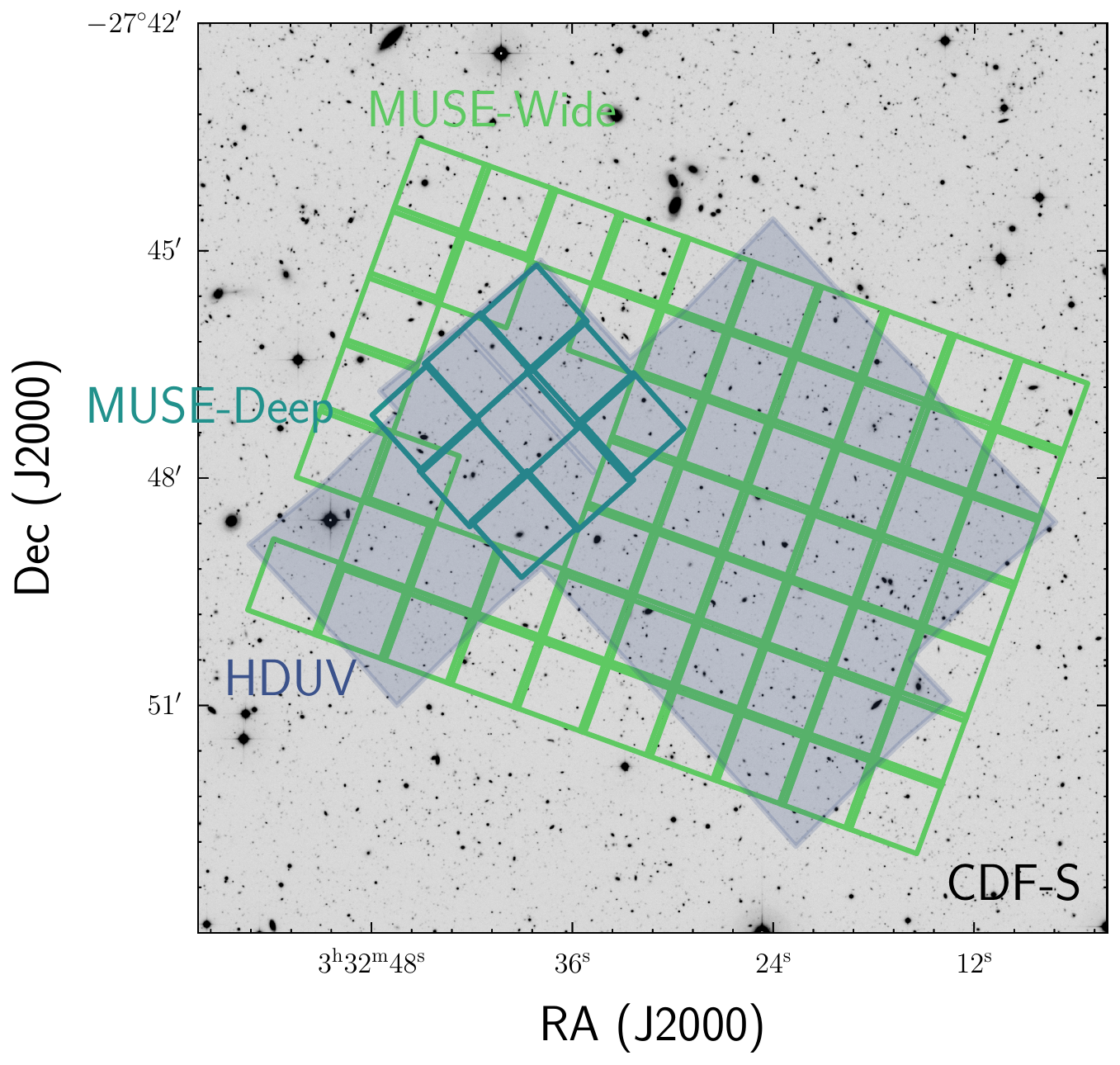}
  \caption{Footprint of MUSE and HDUV pointings in the CDFS region. The background is a V-band image taken from the Garching-Bonn Deep Survey (GaBoDS, \citealp{Hildebrandt2006}). The 60 individual MUSE-Wide pointings relevant here are shown in bright green, and the nine MUSE-Deep fields (\citealp{Bacon2017, Inami2017}, also including the UDF 10) are marked blue. The HDUV footprint is shown as the blue shaded area and covers most of the MUSE-Wide fields as well as MUSE-Deep.}
  \label{fig:fields_cdfs}
\end{figure}

We base our search for LyC emission on known LAEs from the MUSE-Wide survey (\citealp{Urrutia2019,Kerutt2022}), using their positions and redshifts as selection criteria to look for possible LyC leakage in the HST filter bands WFC3/UVIS F275W and F336W, which we take from the HDUV survey (\citealp{Oesch2015,Oesch2018}).
Both surveys are located in the Chandra Deep Field South (CDFS, \citealp{Giacconi2001}) region, which is complemented by the Great Observatories Origins Deep Survey (GOODS; \citealp{Giavalisco2004}) consisting of multiple deep observations of the Hubble Space Telescope (HST) and Spitzer. In Fig.~\ref{fig:fields_cdfs} we show the footprints of the HDUV, the MUSE-Deep and -Wide surveys, overlaid on an inverted HST image. As can be seen, the HDUV survey and the MUSE-Wide and -Deep surveys overlap significantly, with most of the HDUV being covered by MUSE observations.

\subsection{Spectroscopy from MUSE} \label{Photometry}

There are two main observational techniques for detecting star-forming galaxies at redshifts $z=3-6$: Using photometry to detect the Lyman break, resulting in the detection of LBGs and using spectroscopy to detect emission lines, at those redshifts mostly the Ly$\alpha$ line, resulting in LAEs. The former method is potentially biased against finding strong LyC leaker candidates, which would make the Lyman break less pronounced. A high LyC escape fraction could result in an unusual SED and a less steep Lyman break, making it hard to identify LyC leakers (e.g. \citealp{Cooke2014,Vanzella2016}, but see also \citealp{Steidel2018} who do find LyC emission in LBGs). Furthermore, the search for LyC emission is often complicated by the lack of precise redshift information (e.g. when using the Lyman break in photometry). Therefore, in order to make use of the redshift information from spectroscopy gained from the Ly$\alpha$ line, and because this line is expected to correlate with LyC emission, we use a pre-selection of LAEs from MUSE to start our search. 

Integral field spectroscopy is ideal for finding emission line objects such as LAEs without proxies (such as HST detections in the rest-UV), allowing for a relatively unbiased sample selection (when it comes to galaxy properties such as star formation rate, other emission lines or UV properties, which are often used at low redshifts to select LyC leaker candidates). Thanks to the Ly$\alpha$ line, we also have a direct estimate of the redshift, which can be slightly shifted with respect to the systemic redshift (e.g. \citealp{Verhamme2018}) but is good enough to search for LyC emission in broadband data. 

Therefore, we use as the basis for our search a sample of high-redshift LAEs that were found with MUSE (\citealp{Bacon2010}) installed at the ESO Very Large Telescope (VLT) in Chile. MUSE has a spectral range covering $4750\, \angstrom$ to $9350\, \angstrom$, allowing the detection of Ly$\alpha$ at $1215.67\, \angstrom$ in the redshift range of $2.9<z<6.7$. 

Several studies within the MUSE consortium are concerned with finding emission line galaxies, in particular LAEs, in MUSE data (e.g. \citealp{Herenz2017,Claeyssens2019}). Here we use data from the MUSE-Wide survey\footnote{For the first 44 fields, the data and data products such as cut-outs, mini-cubes and extracted spectra as well as emission line catalogues are publicly available and can be found at \url{https://musewide.aip.de/project/}.} (\citealp{Urrutia2019, Kerutt2022}), which consists of 100 MUSE pointings of one hour exposure time, including 60 pointings in CDFS region (see Fig.~\ref{fig:fields_cdfs}), with a field of view of one arcmin$^2$ each, overlapping by $\approx4\arcsec$. Among the 100 pointings, there are also observations from the MUSE-Deep\footnote{The catalogue of objects in the MUSE-Deep fields used here (and presented by \citealp{Inami2017}) is available at \url{http://cdsarc.u-strasbg.fr/viz-bin/qcat?J/A+A/vol/page}, the second data release can be found in \citet{Bacon2023} and at \url{https://amused.univ-lyon1.fr/}.} survey (\citealp{Bacon2017, Inami2017, Bacon2023}), namely 9 mosaic fields with 10 hours exposure time each and one field of 31 hours (the MUSE UDF-10, see \citealp{Bacon2017,Bacon2023}).

The emission line catalogue for MUSE-Wide (which here also includes the MUSE-Deep observations) is constructed in a consistent way, using the two spatial and one spectral dimension of integral field spectroscopy, assuring the inclusion of potentially extended Ly$\alpha$ halos (which are ubiquitous, see \citealp{Wisotzki2016, Leclercq2017,Wisotzki2018}). The emission lines are detected using a matched filtering approach (\citealp{HerenzWisotzki2017}) and each object is classified by at least three people using a graphical user interface (\citealp{Kerutt2017}) that allows accessing all information from the MUSE datacube and additional HST broadband images. Criteria for classifying an object as an LAE were a typical asymmetric or double-peaked line (e.g. \citealp{Verhamme2006}), no other emission lines visible in the MUSE spectrum (except if they match the redshift), and no strong emission in a band bluewards of the Lyman break. Especially the last criterion might potentially bias the sample against strong LyC leakers, which is why we only used it as an indication in case there are multiple potential UV continuum counterparts visible in the HST data. In those cases, we chose the one with the more pronounced Lyman break. This assures the exclusion of low-redshift interlopers, which is especially important for our search for LyC emission (see Sect.~\ref{Sample Selection} where we describe further measures to reduce interlopers, such as confidence flags). However, there is a remaining possibility that a few, particularly strong LyC leakers could potentially have been falsely assigned lower redshifts. 
The $15\%$ completeness limit for the detection of LAEs in MUSE-Wide is $\m{log_{10}(L}_{\m{Ly}\alpha}\, \m{[erg/s]}) \approx 42.2$ ($\m{log_{10}(L}_{\m{Ly}\alpha}\,\m{[erg/s]}) \approx 42.7$) at a redshift of $z \approx 3$ ($z\approx6.5$), as discussed in \citet{Herenz2019}, who find a characteristic Ly$\alpha$ luminosity for their luminosity function of $\m{log}L^*\m{[erg/s]}=42.2^{+0.22}_{-0.16}$.

This catalogue of emission line objects was the basis for the selection of LAEs from \citet{Kerutt2022} used in this paper, which consists of 1920 LAEs identified in the MUSE-Wide and -Deep data (excluding known AGN). We also use the properties of the Ly$\alpha$ emission lines of those LAEs, which were measured from spectra extracted from the MUSE data, constructed by summing each spectral layer, and weighted by the wavelength-dependent \cite{Moffat1969} profile that best describes the MUSE point spread function (PSF, see \citealp{Urrutia2019}). The positions for the spectral extractions were the highest SN peak of the Ly$\alpha$ emission in MUSE. From these spectra we can gain information on the full width at half maximum (FWHM), the peak separation (in case of a double peak) and, in combination with broadband HST data, the rest-frame equivalent width ($\m{EW}_0$) of Ly$\alpha$, all of which we take from \citet{Kerutt2022}. 

\begin{figure*}
\centering
\includegraphics[width=0.8\textwidth]{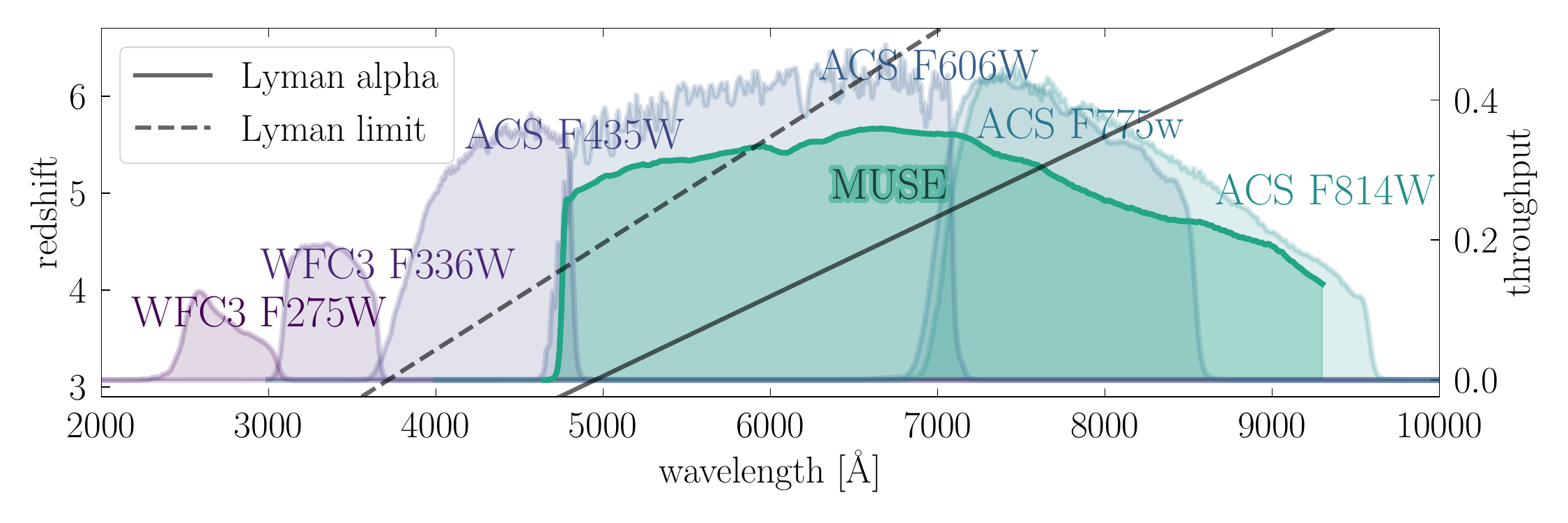}
\caption{HST bands and MUSE throughput used in this study. The $x$-axis shows the observed wavelength in \AA, the left $y$-axis shows the redshift and corresponds to the black dashed and solid lines, and the right $y$-axis shows the filter throughput and corresponds to the coloured filter curves. The solid line shows where the Ly$\alpha$ line would fall depending on redshift, and the dashed line shows the Lyman limit at $912\,\angstrom$, delimiting the LyC, which is bluewards of this line. Only the redshift range where Ly$\alpha$ is visible for MUSE is shown. The two bands with the lowest wavelengths (WFC3/UVIS F275W and F336W) are from the HDUV legacy survey and are used in this study to identify LyC emission.} 
\label{fig:band_width_plot}
\end{figure*}


\subsection{The HDUV survey}

To identify possible LyC emission in our sample of MUSE LAEs, we use data from the HDUV legacy survey, (\citealp{Oesch2015,Oesch2018}), which includes the UVUDF (\citealp{Teplitz2013,Rafelski2015}) and covers an area of $\approx100\,\m{arcmin}^2$ in the two GOODS/CANDELS-Deep (\citealp{Giavalisco2004, Koekemoer2011}) fields and an area of $43.4\,\m{arcmin}^2$ in GOODS-S (\citealp{Oesch2018}). The HDUV survey provides the two HST filters WFC3/UVIS F275W and F336W (see Fig.~\ref{fig:band_width_plot}), reaching a limiting magnitude of $27.5$ to $28\, \m{mag}_{\m{AB}}$ at $5\, \sigma$ in $0\farcs4$ apertures (\citealp{Oesch2018}) and covering a wavelength range from $2435 - 3032\, \angstrom$ and $3096 - 3639\, \angstrom$, respectively. 
Thus, the Lyman break at $912\,\angstrom$ falls in the HST filter band WFC3/UVIS F275W at a redshift range of $z=1.67-2.3$ and in the band WFC3/UVIS F336W at a redshift range of $z=2.4-3.0$, in other words, at redshifts $z>2.3$ the band F275W ($z>3.0$ for F336W) is uncontaminated by non-LyC emission (see Fig.~\ref{fig:band_width_plot}). We therefore use the HST filter band F336W to search for LyC emission from the MUSE LAEs, which have redshifts $z>2.9$.


\subsection{Additional HST data}

To measure the LyC escape fraction, we need measurements of the LyC emission and the UV continuum (see Sect. \ref{Measuring Escape fractions}). Since we will base our $f_\mathrm{esc}$ measurements on SED fitting (see Sect. \ref{SED Fitting}), we need flux measurements in various wavelength bands, which we get through aperture photometry. Even though most of our objects already have measurements in e.g. the 3DHST catalogue (\citealp{Skelton2014}), we want to keep the flux measurement method consistent in all bands, including the LyC band WFC3/UVIS F336W. Therefore, we perform aperture photometry (using an aperture of $0\farcs35$ radius, see Sect.~\ref{Measuring the LyC emission and SN}) not only for LyC but also for other HST bands, including WFC3/UVIS F275W, ACS F435W, F606W, F775W, F814W, F850LP, WFC3/IR F105W, F125W, F140W and F160W. Except for WFC3/UVIS F275W and F336W, the data was taken from the Hubble Legacy Fields (HLF\footnote{The HST images for all bands except the two HDUV ones WFC3/UVIS F275W and F336W were taken from \url{https://archive.stsci.edu/prepds/hlf/}.}, \citealp{Illingworth2016}) GOODS-S data release 2.0. 
For the aperture photometry, we need to account for the different PSFs in the bands, which increase in size with increasing wavelength. Thus, the filter band with the largest PSF is the WFC3/IR F160W band. For measuring the photometry, we use images of each filter band that are matched to the PSF in WFC3/IR F160W band\footnote{The PSF matched images can be found here: \url{https://archive.stsci.edu/hlsps/hlf/v2.0/60mas_conv/}}. To create these images, \citet{Whitaker2019} use a linear combination of Gaussian-weighted Hermite polynomials based on a stack of isolated stars to determine the PSF in WFC3/IR F160W. The images of the other bands were then convolved with the matching kernel.
We take the positions of the UV continuum counterparts from \citet{Kerutt2022}, which were determined by considering all detections in the band ACS F814W within a radius of $0\farcs5$ from the maximum SN of the Ly$\alpha$ emission in MUSE. Then at least two people examined the potential counterparts, taking other bands into account, and decided on the correct counterparts (see \citealp{Kerutt2022} for details). 


\section{Sample Selection} \label{Sample Selection}

In this section, we describe how we select candidates to detect LyC emission. We then divide the found candidates into two groups, a gold ($\mathrm{SN}_{\m{F336W}}>3$) and a silver ($\mathrm{SN}_{\m{F336W}}>2$) sample, indicating their quality as potential LyC leakers. 

\subsection{Redshift and Quality Cuts}

To construct the sample, we use the redshift information from the MUSE catalogue (\citealp{Kerutt2022}) in the MUSE-Wide fields as a basis to look for LyC emission in the HDUV data. For the LAEs we find in MUSE, which are in a redshift range $2.9<z<6.7$, the Lyman break at $912\,\angstrom$ (shortwards of which is the LyC) lies in the observed wavelength range $3563 - 7014\,\angstrom$, which means the LyC falls into the wavelength range of the filter WFC3/UVIS F336W (see the dashed line in Fig.~\ref{fig:band_width_plot}). For redshifts $z<3$, there is an overlap of non-LyC emission in the WFC3/UVIS F336W band, which is why we apply a redshift cut of $z>3$ for our sample selection.
Another restriction to keep in mind is the IGM transmission, which decreases drastically towards higher redshifts (e.g. \citealp{Inoue2014}), from a mean IGM transmission at $900\,\AA$ of $\tau_{\mathrm{IGM}}=0.56$ at $z=3$ to  $\tau_{\mathrm{IGM}}=0.17$ at $z=4.5$, which is why we do not consider any objects beyond $z>4.5$ for our search for LyC leaker candidates. In this redshift range ($3<z<4.5$), there are $743$ LAEs in the MUSE-Wide survey that overlap with the GOODS-S part of the HDUV survey. 
The redshifts for most of those objects are based on a single line, the Ly$\alpha$ line. Due to radiative transfer processes in the neutral hydrogen in the ISM, it can be asymmetric or double-peaked, sometimes mimicking the \ion{O}{II} doublet. Therefore, low SN Ly$\alpha$ lines are not easy to identify, which is why we add a cut in the confidence parameter given in the MUSE-Wide survey. This subjective parameter represents the certainty of the redshift classification, ranging from zero (not certain at all) to three (very sure). We exclude objects with a confidence below two, which leaves us with a final sample of $621$ LAEs at $z=3.0-4.5$. 


\subsection{Measuring the LyC emission and signal-to-noise}  \label{Measuring the LyC emission and SN}

To determine possible LyC leakers, we measure the flux in the WFC3/UVIS F336W band at the same position as the UV continuum of each object. Since we perform SED fitting later (see Sect. \ref{SED Fitting}), we apply the same procedure for all other bands as well.

We use an aperture of $0\farcs35$ radius to perform aperture photometry on the images convolved to the WFC3/IR F160W PSF, after subtracting the local background (using the median of the 3 sigma clipped values in a cutout of $2^{\prime\prime} \times 2^{\prime\prime}$ arcseconds, excluding the aperture itself). Following \citet{Skelton2014}, we apply an aperture correction $21\%$ to total fluxes to account for an aperture with a radius of $0\farcs35$.

For the position of the apertures, we use the same location as the UV continuum emission, which is where the LyC emission is expected as well. Unlike e.g. Ly$\alpha$, which is resonantly scattered, it will be detectable in the line of sight where it was first emitted. Therefore, we do not use the Ly$\alpha$ positions from the MUSE detections, but rather the UV continuum counterpart positions, which were determined using the \texttt{Galfit} software (\citealp{Galfit,Galfit2}) in the ACS F814W band (\citealp{Giavalisco2004, Koekemoer2011, Grogin2011}), which goes down to a magnitude of 28.92 $\mathrm{mag}_{\mathrm{AB}}$ in MUSE-Wide and 29.16 $\mathrm{mag}_{\mathrm{AB}}$ in MUSE-Deep (for a $2\,\sigma$ detection, see \citealp{Kerutt2022} for more details). This implies that we only consider objects that have a detection in the UV continuum, but since the WFC3/UVIS F336W filter band is a bit shallower and only reaches a magnitude of 28 $\mathrm{mag}_{\mathrm{AB}}$, this criterion will not exclude any potential LyC leakers.
In case the LyC emission originates from a single star-forming region in the galaxy, its position can still be slightly offset with respect to the maximum of the UV continuum, but still be captured by our $0\farcs35$ aperture photometry. 

To measure the signal-to-noise (SN) of the LyC flux, we need to take into account that the noise in the HDUV data is correlated. Therefore, we determine the local noise properties by applying 100 random, non-overlapping apertures of the same size as for the flux measurement to estimate the noise in an area of $8^{\prime\prime} \times 8^{\prime\prime}$ around the object, which we do in the original, unconvolved data.
To not be influenced by a real signal, we mask objects and bright/hot pixels. For the latter, we use a SN cut of three for individual pixels. To mask objects, we use the segmentation map created in the WFC3/IR F160W band in the 3DHST survey (\citealp{Skelton2014}). This band has a wider PSF than WFC3/UVIS F336W and lower redshift interlopers tend to be brighter and more extended in this wavelength range. To be sure not to include any emission that might be more extended in WFC3/UVIS F336W than in the WFC3/IR F160W band the segmentation map is based on, we grow the mask by one pixel in each direction. The standard deviation of the 100 apertures provides us with an estimate of the noise and the SN is the flux measurement in the original, unconvolved data over this standard deviation. 

We set the SN cut for our pre-selection to the relatively low value of SN=2, since we have the prior knowledge from the MUSE catalogues that there is an object at the correct redshift at the position we are examining in the LyC band WFC3/UVIS F336W (see also below in Sect.~\ref{Identification of LyC Counterparts} for a detailed discussion). Including the redshift cut $3<z<4.5$ mentioned above, this leaves us with a total number of 42 objects to take into consideration). 


\subsection{Identification of LyC Counterparts and Contamination} \label{Identification of LyC Counterparts}

Having measured the SN ratio in the LyC, we use this as an indicator of the possible presence of a LyC leaker. However, even with a SN cut of two, the signal could still be created by various other influences such as random noise, close neighbours, or low-redshift interlopers. Therefore, we use several criteria to narrow down our sample of 42 possible LyC leakers and also to separate our remaining candidates into likely leakers, which we call the gold sample, and possible leakers, the silver sample, as done for example in \citet{Fletcher2019}. The difference between the gold and silver samples is that for a silver object, an SN cut of two in the WFC3/UVIS F336W band is applied, while for a gold object, we require $\mathrm{SN}>3$.

We use several criteria to make sure our sample of LyC leakers is not contaminated by low-redshift interlopers or noise, as discussed in many pioneering studies searching for individual LyC leakers (see e.g. \citealp{Shapley2006,Iwata2009,Vanzella2010a,Nestor2011,Nestor2013,Mostardi2015}). For our study, we have the ideal situation to be able to use integral field spectroscopy to obtain reliable redshifts, in addition to the high resolution of the space-based HST data. 

\begin{figure}
\centering
\includegraphics[width=\linewidth]{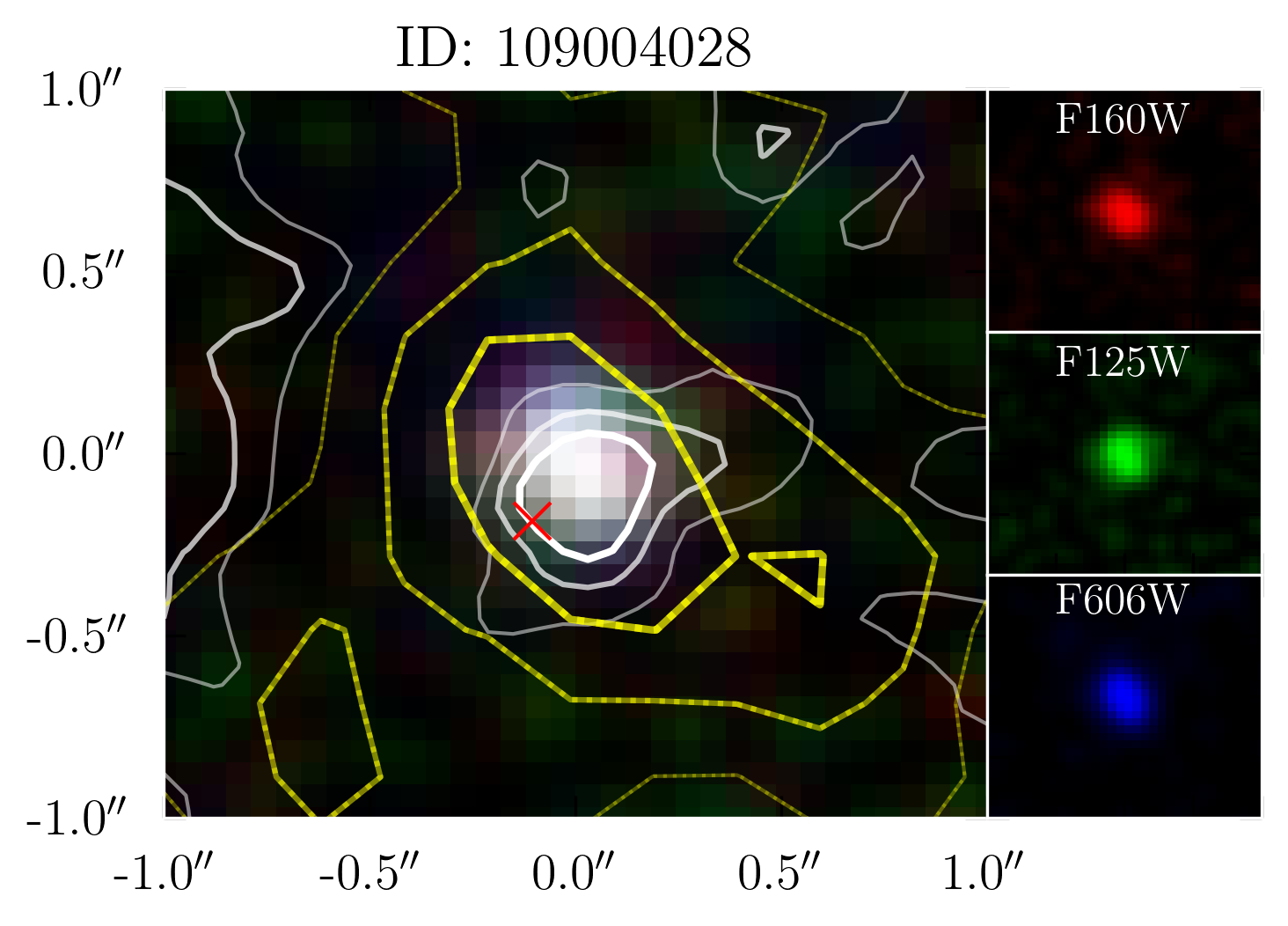}
\caption{RGB image for one of the LyC leaker candidates. The three images that were combined for the main image are shown to the right of each pane and are the HST bands WFC3/IR F160W (red), F125W (green) and ACS F606W (blue). The white contours in the main panels show the signal in HST band WFC3/UVIS F336W, where the LyC emission can be found. The contours show SN of 3, 2 and 1 (strongest to lightest) for an image smoothed with a Gaussian kernel of $\sigma=1$ pixel. The yellow dotted contours show the extent of the Ly$\alpha$ emission in the MUSE data, also smoothed with a Gaussian kernel of $\sigma=1$ pixel. It has to be kept in mind, that the PSF of MUSE is larger than that of HST, however, it has been shown that Ly$\alpha$ emission is usually ten times more extended than the UV continuum (e.g. \citealp{Wisotzki2016, Leclercq2017,Wisotzki2018}). The red cross in the main panel shows the position of the pixel with the highest S/N in the Ly$\alpha$ emission found with MUSE.}
\label{fig:rgbs1}
\end{figure}

\begin{figure*}
\centering
\begin{minipage}{.49\textwidth}
  \centering
  \includegraphics[width=\linewidth]{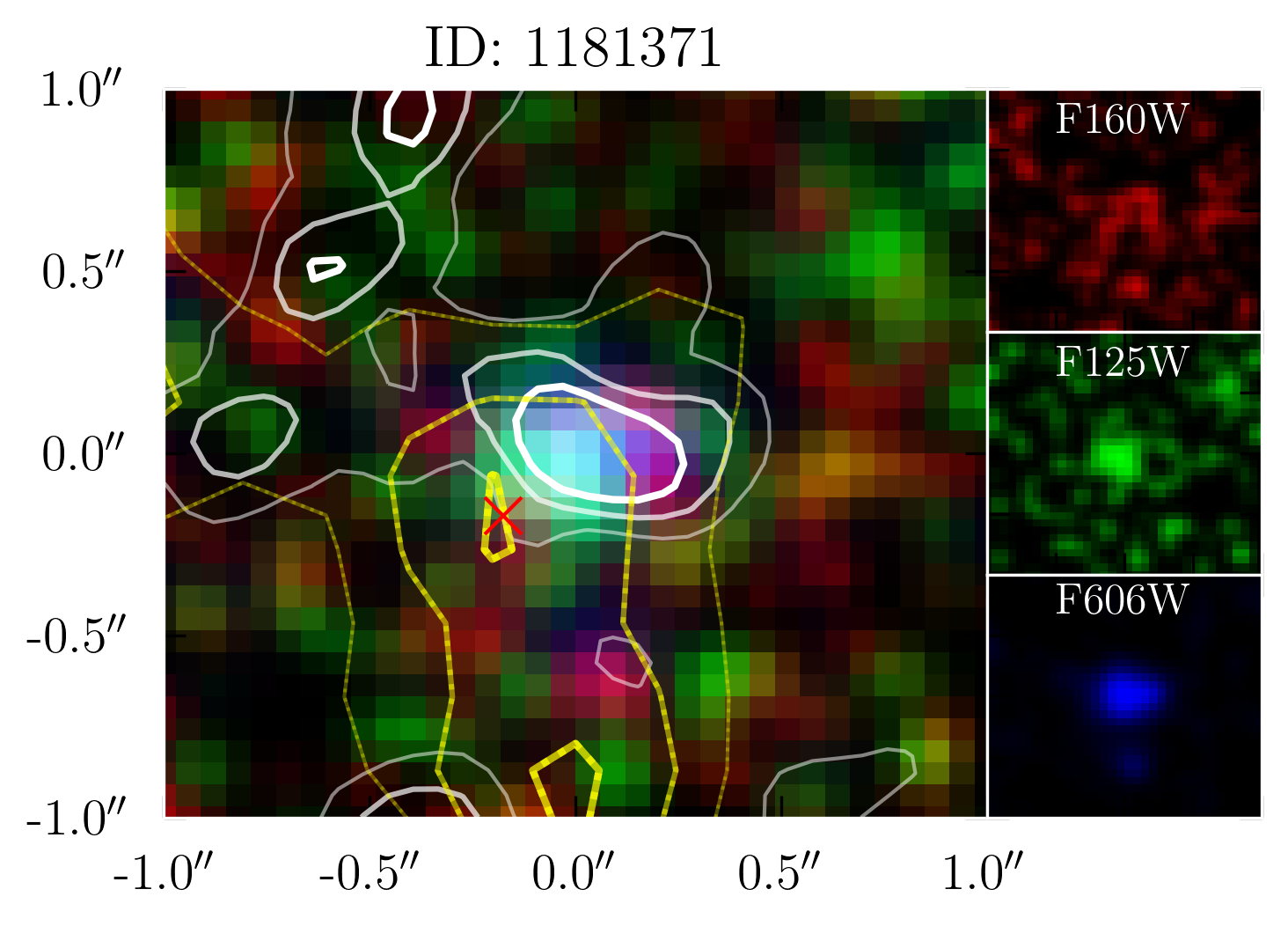}
  \includegraphics[width=\linewidth]{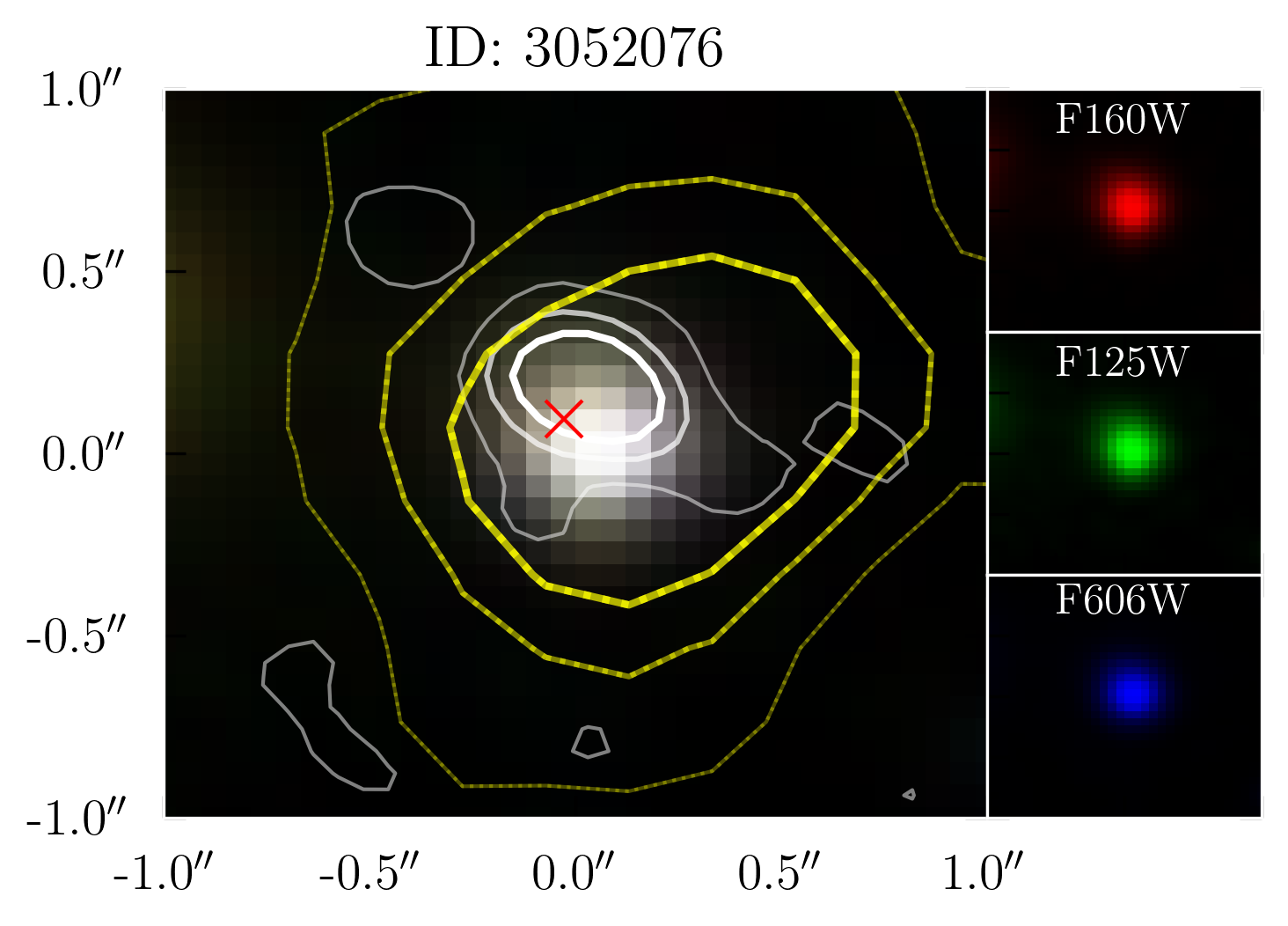}
\end{minipage}%
\begin{minipage}{.49\textwidth}
  \centering
  \includegraphics[width=\linewidth]{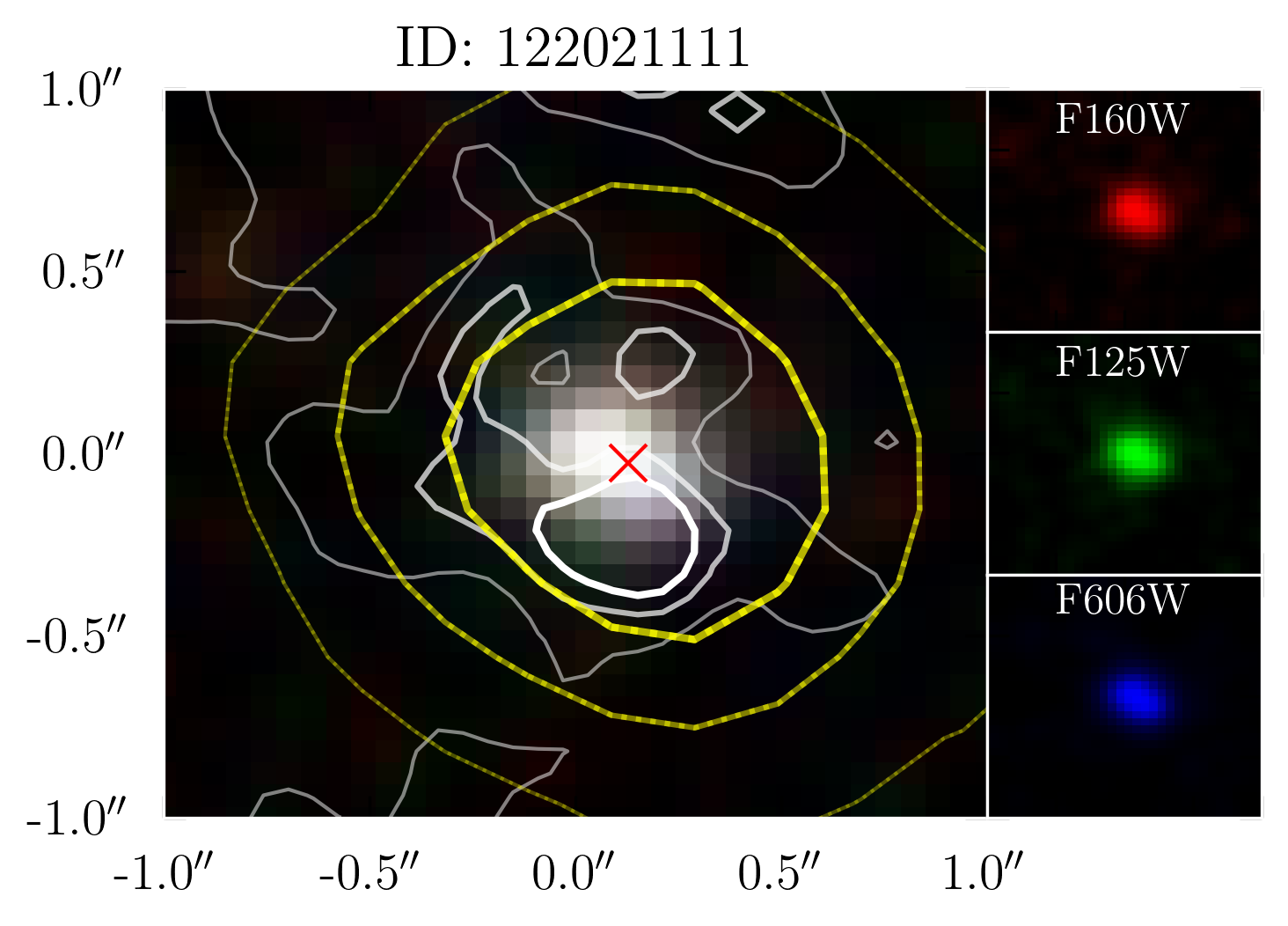}
  \includegraphics[width=\linewidth]{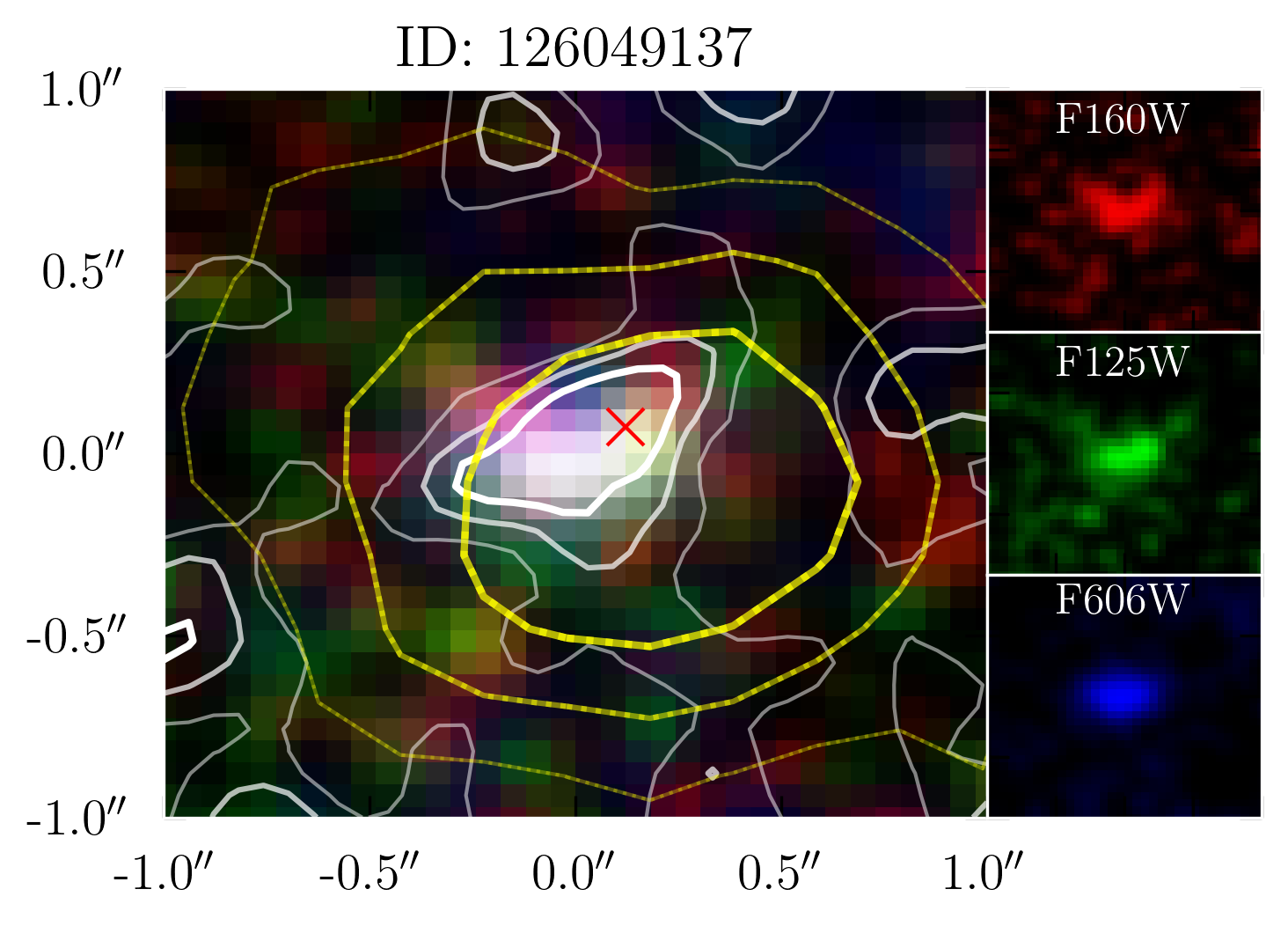}
\end{minipage}
\caption{Same as Fig.~\ref{fig:rgbs1} for the rest of the gold sample LyC leaker candidates.}
\label{fig:rgbs2}
\end{figure*}

We visually vet all 42 potential LyC leakers using red-green-blue (rgb) images (see Figs. \ref{fig:rgbs1} and \ref{fig:rgbs2}) of three HST filter bands (ACS F606W, WFC3/IR F125W and F160W), overlaid by SN contours in the WFC3/UVIS F336W band. In these rgb images, we can see if the object has a similar morphology in the infrared as in the UV and if it matches the detection in the LyC band. If there are several clumps with different colours, this is an indication of a potential chance alignment of different objects (which was the case for six objects). Different areas of the same object can have different colours depending on its properties. However, we try to distinguish between this case and clearly different clumps. If they do have similar colours, this could indicate an ongoing merger, which could trigger star formation and thus also the production of LyC photons. 

In addition to using rgb images, we look at cutouts of the objects in the WFC3/UVIS F336W band and use the MUSE catalogue from \citet{Kerutt2022} as well as the catalogue from \citet{Skelton2014} to highlight all other known objects in the same cutout (see Figs.~\ref{fig:overview_goldsample} and \ref{fig:overview_silversample}). This allows us to exclude contamination by close, projected neighbours, that might only be visible in the MUSE catalogue through their emission lines (which was the case for 2 objects). Another indication of low-redshift contamination is a strong signal in the band WFC3/UVIS F275W with SN>3. Even though LyC can be stronger at lower wavelengths, here we expect the LyC to be significantly reduced by the IGM (which is true for all but seven objects, three of which were obvious contaminants). 

Another criterion is the overlap between the signal found in LyC in WFC3/UVIS F336W and the UV continuum, in this case, ACS F775W. For this, we use segmentation maps made with {\textsc SExtractor} (\citealp{SExtractor}) and determine the overlap between the two bands by the percentage of pixels belonging to the LyC emission which are also above the segmentation map threshold of $1\sigma$ in the UV continuum. A non-zero overlap is required for us to consider an object a LyC leaker candidate. 

This leaves us with a final sample of seven silver candidates and five gold candidates. An overview of the final sample with IDs from different existing catalogues and redshifts can be found in table \ref{tab:goodss}.

\subsection{Comparison to previous Studies}

There have been previous studies of LyC leakage in the same region, one by \citet{saxena2022}, who found 11 LyC leaker candidates in the CDFS using narrow-band imaging data from \citet{Guaita2016} and the U-band from \citet{Nonino2009}. Of their 11 candidates, five are in the field of view of the HDUV with MUSE-Deep IDs 1087, 8035, 6666, and 7820 (one is not in the MUSE catalogue, see table \ref{tab:goodss} for the IDs). Of those, only two have potential detections in the HDUV WFC3/UVIS F336W band, IDs 1087 and 8035. The former is among our gold sample, the latter (ID 13385 in \citealp{saxena2022} at $z=3.431$ with a SN in WFC3/UVIS F336W of $\approx1.8$) was assigned a lower redshift of $z=0.523$ in the MUSE-Deep catalogue (and was not detected in MUSE-Wide), although with a low confidence of 1, indicating that the redshift determination from MUSE might not be reliable. However, the MUSE redshift of the latter agrees quite well with the photometric redshift of \citet{Skelton2014} at $z=0.586$ (ID 25631). 

Another recent study in the same field was done by \citet{Rivera-Thorsen2022}, also using data from HST and MUSE. Unlike previous studies, they first run a source detection software on three HST UV filters (F225W, F275W and F336W) and then determine whether the detected signal could be LyC emission. They find six new LyC leaker candidates and confirm the object from \citet{saxena2022} that is also among our candidates (ID 3052076 with a SN$_{336}=4.07$). Of the other six, three are below our redshift cut of $z>3$ imposed by the wavelength range of the WFC3/UVIS F336W filter band that we use for detecting LyC emission. Another of their candidates is above our redshift upper limit of $z<4.5$ in our catalogue but might be a different, close-by object. The last candidate does not have a close enough counterpart in our sample based on \citet{Kerutt2022}.

\begin{table*} 
\begin{center} 
\caption{Overview of candidates} 
\begin{tabular}{ l l l l l l l l l l l l l} 
\hline\hline 
ID$_{\m{MW}}$ & ID$_{\m{MD}}$ & ID$_{\m{Skel}}$ & ID$_{\m{Raf}}$ & ID$_{\m{Guo}}$ & RA & Dec & z$_{\mathrm{MW}}$ & z$_{\mathrm{MD}}$ &z$_{\mathrm{Skel}}$ & c$_{\mathrm{MW}}$ & c$_{\mathrm{MD}}$ & SN$_{336}$ \\ \hline 
gold candidates &   &   &   &   &   &   &   &   &   &   &   &   \\
\bf{\textcolor{gold}{1181371}} & 7193 & 24480 & 6470 & - & 53.1358 & -27.7955 & 3.084 & 3.085 & 3.441 & 2 & 2 & 5.01 \\ 
\bf{\textcolor{gold}{3052076}} & 1087 & 24193 & 3506 & 12448$^{*}$ & 53.1679 & -27.798 & 3.457 & 3.462 & 1.711 & 3 & 3 & 4.07 \\ 
\bf{\textcolor{gold}{109004028}} & - & 15601 & - & 7570 & 53.0994 & -27.8392 & 3.267 & - & 1.492 & 2 & - & 3.03 \\ 
\bf{\textcolor{gold}{122021111}} & - & 16523 & - & 8005 & 53.1389 & -27.8354 & 3.794 & - & 0.724 & 3 & - & 3.57 \\ 
\bf{\textcolor{gold}{126049137}} & - & 20189 & - & 10131 & 53.2042 & -27.8172 & 4.426 & - & 1.222 & 2 & - & 4.39 \\ \hline
silver candidates &   &   &   &   &   &   &   &   &   &   &   &   \\
\bf{\textcolor{silver}{1521589}} & 7169 & 26130 & 4873 & 13707 & 53.1283 & -27.7887 & 3.152 & 3.155 & 1.537 & 3 & 2 & 2.69 \\ 
\bf{\textcolor{silver}{3452147}} & 2134 & 23769 & 3431 & 12256 & 53.1541 & -27.7988 & 3.521 & 3.524 & 4.099 & 3 & 2 & 2.6 \\ 
\bf{\textcolor{silver}{4062373}} & 1360 & 27696 & 37765 & 14691 & 53.1792 & -27.7829 & 3.663 & 3.666 & 1.014 & 3 & 2 & 2.21 \\ 
\bf{\textcolor{silver}{4172404}} & 7121 & 27408 & 6974 & 14469 & 53.1851 & -27.7839 & 3.672 & 3.675 & 3.759 & 3 & 2 & 2.88 \\ 
\bf{\textcolor{silver}{5622786}} & 7137 & (20161?) & (781?) & - & 53.1604 & -27.8174 & 4.005 & 4.007 & - & 3 & 2 & 2.05 \\ 
\bf{\textcolor{silver}{119004004}} & - & 16269 & - & 7896 & 53.1891 & -27.8363 & 3.314 & - & 0.798 & 3 & - & 2.59 \\ 
\bf{\textcolor{silver}{122032127}} & - & 16000 & - & 7760 & 53.1326 & -27.8374 & 4.348 & - & 0.892 & 2 & - & 2.33 \\ 
\end{tabular}\label{tab:goodss} 
\end{center} 
\tablefoot{ID$_{\m{MW}}$: Identifier in the MUSE-Wide catalogue (\citealp{Urrutia2019,Kerutt2022} and Urrutia et al. in prep.). The colours indicate if the object is placed in the gold or silver category, as explained in Sect. \ref{Identification of LyC Counterparts}. ID$_{\m{MD}}$: Identifier in the MUSE-Deep catalogue (\citealp{Inami2017}). ID$_{\m{Skel}}$: Identifier from the 3D-HST CANDELS catalogue in \citet{Skelton2014}. ID$_{\m{Raf}}$: Identifier from the UDF catalogue in \citet{Rafelski2015}. ID$_{\m{Guo}}$: Identifier from the CANDELS GOODS-S catalogue in \citet{Guo2013}. RA, DEC: Right ascension and declination (\citealp{Kerutt2022}). z$_{\m{MW}}$: redshift from MUSE-Wide, based on the Ly$\alpha$ line, corrected to systemic using the full-width at half maximum (FWHM) or peak separation (if the line has a double peak), based on \citet{Verhamme2018}. z$_{\m{MD}}$: redshift from \citet{Inami2017}. z$_{\m{Skel}}$: redshift from \citet{Skelton2014}. c$_{\m{MW}}$: Classification confidence (0 lowest and 3 highest) of the redshift of the LAE based on three different investigators and a consolidation (see \citealp{Urrutia2019,Schmidt2021,Kerutt2022}). c$_{\m{MD}}$: confidence from \citet{Inami2017}. SN$_{336}$: Signal-to-noise in the LyC band WFC3/UVIS F336W. The object marked with $^*$ is already found in \citet{saxena2022} and further discussed in \citet{Rivera-Thorsen2022}.}  
\end{table*} 


\section{Measuring Escape Fractions} \label{Measuring Escape fractions}

In this section we will measure escape fractions in three different ways: assuming a fixed ratio between the UV continuum and the LyC, using a Binary Population and Spectral Synthesis (BPASS, \citealp{Eldridge2017,Stanway2018,Byrne2022}, version 2.3), and fitting the SEDs using the Code Investigating GALaxy Emission (\texttt{CIGALE}, \citealp{Burgarella2005,Noll2009,Boquien2019}). 

\subsection{Definition and ingredients of the escape fraction}

A common definition for the relative escape fraction $f_{\m{esc,rel}}$ was suggested by \citet{Steidel2001} (but see also e.g.\ \citealp{Siana2007,Bian2017,Grazian2018,Steidel2018}):

\begin{equation}
 f_{\m{esc, rel}} =  \left(\frac{f_{\m{LyC}}}{f_{\m{UVC}}}\right)_{\m{obs}} \times \left(\frac{L_{\m{UVC}}}{L_{\m{LyC}}}\right)_{\m{int}} \times \exp(\tau^{\m{IGM}}_{\m{LyC}})
\end{equation}

This relative escape fraction is a comparison between the observed flux ratio of the LyC and UV continuum (UVC) fluxes $(f_{\m{LyC}}/f_{\m{UVC}})_{\m{obs}}$ and the intrinsic luminosity ratio $(L_{\m{UVC}}/L_{\m{LyC}})_{\m{int}}$, corrected for the absorption of LyC photons by neutral hydrogen in the IGM, $\exp(\tau^{\m{IGM}}_{\m{LyC}})$. This definition is independent of dust and can be related to the absolute escape fraction by taking into account the dust attenuation of the non-ionising UV continuum, $A_{\m{UVC}}$:

\begin{equation} \label{eq:fesc_abs}
 f_{\m{esc, abs}} =  f_{\m{esc,rel}}  \times 10^{-0.4 \, A_{\m{UVC}}} 
\end{equation} 

This definition of the escape fraction depends on knowledge of the dust attenuation, for example, obtained through SED fitting or several photometric bands. 
If there is no significant dust attenuation, the relative and absolute escape fractions are the same (see for example the galaxy in \citealp{Shapley2016}).
Typically, the escape fraction is defined as the fraction between the flux values at $900\,\angstrom$ for the LyC and $1500\,\angstrom$ for the UV continuum.

For our escape fraction estimates, we use three different approaches, two of which use fixed intrinsic flux ratios $(L_{\m{UVC}}/L_{\m{LyC}})_{\m{int}}$ (see Sect. \ref{Fixed intrinsic ratios}) and the last one based on SED fitting (see Sect. \ref{SED Fitting}).


\textbf{\subsection{Observed flux ratio}}

The observed flux ratio $(f_{\m{LyC}}/f_{\m{UVC}})_{\m{obs}}$ can be obtained from the measurements in the LyC band WFC3/UVIS F336W, as described in Sect. \ref{Measuring the LyC emission and SN}.  Depending on the redshift of the individual candidates, the two bands used for the LyC and UV continuum respectively cover different wavelength ranges and their effective wavelengths are not necessarily at $900\,\angstrom$ and $1500\,\angstrom$. In the third column in Table \ref{tab:various_fesc}, we show the fraction $(f_{\m{LyC}}/f_{\m{UV}})_{\m{obs}}$ based directly on the measurements in the respective bands used for detecting the LyC and UV continuum. This fraction is free of any model assumptions and only contains measured values. It is used to compute the escape fraction for the fixed LyC to UV continuum ratio, while the escape fractions based on the BPASS model are corrected to $900\,\angstrom$ and $1500\,\angstrom$ (see Sect. \ref{Fixed intrinsic ratios} below).  
It has to be noted that except for the first object, all fractions are below one, meaning the LyC flux is lower than the UV continuum, as expected. The first object presents an interesting case and needs a special explanation. It could be that the nebular continuum at the Lyman break is responsible for creating a Lyman bump (see \citealp{Inoue2010}), resulting in an excess of LyC emission.  
\newline

\begin{table*} 
\begin{center} 
\caption{LyC leaker candidates, various measurements of f$_{\m{esc}}$} 
\begin{tabular}{  l l l l l l l l l l } 
\hline 
 ID & z & $\left( \frac{f_{\m{LyC}}}{f_{\m{UV}}} \right)_{\m{obs}}$ &  T$_{\m{IGM}}$ &  A$_{\m{V}}$ & method & $\left( \frac{L_{1500}}{L_{900}} \right)_{\m{int}}$ & f$_{\m{esc,rel}} [\%]$ & f$_{\m{esc,rel}}^{\m{IGM}} [\%]$ & f$_{\m{esc,abs}}^{\m{IGM}} [\%]$  \\ \hline 
  gold candidates & & & & & & & & & \\ \hline 
\bf{\textcolor{gold}{1181371}} & 3.08 & 1.17$\pm$0.22& 0.66 & 0.42$\pm$0.11 &  fixed & 3 & 350$\pm$66 & 530$\pm$100 & 360$\pm$77 \\ 
 &  &  &  &  & BPASS & 1.72 & 102$\pm$19 & 155$\pm$28 & 105$\pm$22 \\ 
 &  &  &  &  & CIGALE & 1.36 &  &  & \textbf{\ph88$\pm$7} (90$\pm$6) \\ \hline  
\bf{\textcolor{gold}{3052076}} & 3.46 & 0.10$\pm$0.02& 0.46 & 0.78$\pm$0.01 &  fixed & 3 & \ph30$\pm$6 & \ph66$\pm$13 & \ph32$\pm$6 \\ 
 &  &  &  &  & BPASS & 1.72 & \ph17$\pm$3 & \ph37$\pm$7 & \ph18$\pm$3 \\ 
 &  &  &  &  & CIGALE & 4.32 &  &  & \textbf{\ph23$\pm$5} (40$\pm$0) \\ \hline  
\bf{\textcolor{gold}{109004028}} & 3.27 & 0.14$\pm$0.04& 0.57 & 0.58$\pm$0.13 &  fixed & 3 & \ph43$\pm$12 & \ph76$\pm$21 & \ph45$\pm$14 \\ 
 &  &  &  &  & BPASS & 1.72 & \ph13$\pm$4 & \ph23$\pm$7 & \ph13$\pm$4 \\ 
 &  &  &  &  & CIGALE & 4.64 &  &  &  \textbf{\ph34$\pm$10} (47$\pm$6) \\ \hline  
\bf{\textcolor{gold}{122021111}} & 3.79 & 0.06$\pm$0.01& 0.30 & 0.03$\pm$0.04 &  fixed & 3 & \ph19$\pm$4 & \ph64$\pm$13 & \ph62$\pm$13 \\ 
 &  &  &  &  & BPASS & 1.72 & \ph12$\pm$3 & \ph40$\pm$10 & \ph39$\pm$10 \\ 
 &  &  &  &  & CIGALE & 3.98 &  &  & \textbf{\ph66$\pm$11} (77$\pm$12) \\ \hline  
\bf{\textcolor{gold}{126049137}} & 4.43 & 0.27$\pm$0.12& 0.05 & 0.89$\pm$0.24 &  fixed & 3 & \ph81$\pm$37 & 1473$\pm$673 & 649$\pm$329 \\ 
 &  &  &  &  & BPASS & 1.72 & \ph49$\pm$22 & 891$\pm$400 & 393$\pm$196 \\ 
 &  &  &  &  & CIGALE & 1.39 &  &  & \textbf{\ph69$\pm$10} (76$\pm$9) \\ \hline
 silver candidates & & & & & & & & & \\ \hline 
\bf{\textcolor{silver}{1521589}} & 3.15 & 0.37$\pm$0.11& 0.61 & 0.10$\pm$0.07 &  fixed & 3 & 110$\pm$34 & 180$\pm$56 & 164$\pm$52 \\ 
 &  &  &  &  & BPASS & 1.72 & \ph26$\pm$8 & \ph42$\pm$13 & \ph39$\pm$12 \\ 
 &  &  &  &  & CIGALE & 1.55 &  &  &   \textbf{\ph79$\pm$15} (86$\pm$10) \\ \hline  
\bf{\textcolor{silver}{3452147}} & 3.52 & 0.48$\pm$0.15& 0.46 & 0.70$\pm$0.18 &  fixed & 3 & 143$\pm$46 & 314$\pm$101 & 165$\pm$60 \\ 
 &  &  &  &  & BPASS & 1.72 & \ph69$\pm$22 & 152$\pm$48 & \ph80$\pm$28 \\ 
 &  &  &  &  & CIGALE & 2.70 &  &  &  \textbf{\ph47$\pm$14} (56$\pm$14) \\ \hline  
\bf{\textcolor{silver}{4062373}} & 3.66 & 0.35$\pm$0.13& 0.35 & 0.92$\pm$0.11 &  fixed & 3 & 106$\pm$40 & 301$\pm$113 & 129$\pm$50 \\ 
 &  &  &  &  & BPASS & 1.72 & \ph61$\pm$23 & 173$\pm$65 & \ph74$\pm$28 \\ 
 &  &  &  &  & CIGALE & 4.93 &  &  & \textbf{\ph74$\pm$13} (82$\pm$13) \\ \hline  
\bf{\textcolor{silver}{4172404}} & 3.67 & 0.52$\pm$0.15& 0.35 & 0.49$\pm$0.22 &  fixed & 3 & 156$\pm$45 & 443$\pm$128 & 282$\pm$100 \\ 
 &  &  &  &  & BPASS & 1.72 & \ph94$\pm$27 & 267$\pm$77 & 170$\pm$60 \\ 
 &  &  &  &  & CIGALE & 1.75 &  &  & \textbf{\ph31$\pm$15} (47$\pm$15) \\ \hline  
\bf{\textcolor{silver}{5622786}} & 4.00 & 0.36$\pm$0.15& 0.19 & 0.00$\pm$0.00 &  fixed & 3 & 108$\pm$44 & 555$\pm$225 & 555$\pm$225 \\ 
 &  &  &  &  & BPASS & 1.72 & \ph62$\pm$25 & 319$\pm$128 & 319$\pm$128 \\ 
 &  &  &  &  & CIGALE & 3.86 &  &  &  \textbf{\ph53$\pm$11} (71$\pm$10) \\ \hline  
\bf{\textcolor{silver}{119004004}} & 3.31 & 0.18$\pm$0.06& 0.57 & 0.81$\pm$0.26 &  fixed & 3 & \ph54$\pm$19 & \ph96$\pm$34 & \ph45$\pm$19 \\ 
 &  &  &  &  & BPASS & 1.72 & \ph28$\pm$10 & \ph51$\pm$18 & \ph24$\pm$10 \\ 
 &  &  &  &  & CIGALE & 2.81 &  &  &  \textbf{\ph26$\pm$14} (50$\pm$14) \\ \hline  
\bf{\textcolor{silver}{122032127}} & 4.35 & 0.17$\pm$0.06& 0.08 & 0.45$\pm$0.28 &  fixed & 3 & \ph52$\pm$19 & 674$\pm$246 & 445$\pm$199 \\ 
 &  &  &  &  & BPASS & 1.72 & \ph23$\pm$9 & 298$\pm$117 & 197$\pm$92 \\ 
 &  &  &  &  & CIGALE & 1.25 &  &  & \textbf{\ph77$\pm$14} (84$\pm$12) \\ \hline  
\end{tabular}\label{tab:various_fesc} 
\end{center} 
\tablefoot{Overview of different escape fraction values based on three different methods. (f$_{\m{LyC}}$/f$_{\m{UV}})_{\m{obs}}$: observed flux ratio between the LyC band WFC3/UVIS F336W and the UV continuum in ACS F775W, given in frequency space. T$_{\m{IGM}}$: LyC transmission in the IGM at the redshift of the object, using the mean of the $5\%$ lines of sight with the highest transmission. A$_{\m{V}}$: dust extinction from the SED fit of CIGALE. (L$_{1500}$/L$_{900})_{\m{int}}$: intrinsic luminosity ratio between the UV continuum at $1500\,\angstrom$ and the LyC at $900\,\angstrom$. The three different methods (in the method column) correspond to a fixed value of 3, which is often used in the literature, a value of 1.72 based on a BPASS model with a young age of 3 Myr and a metallicity of $\m{Z} = 0.002$, and the third value is the ratio measured from the SED model of CIGALE, using the stellar unattenuated model without IGM absorption. f$_{\m{esc,rel}}$: relative escape fraction for two different methods. The first corresponds to the intrinsic flux ratio of 3 and uses the fluxes directly measured from the respective bands (the third column in this table). The second escape fraction is based on the BPASS model and uses flux measurements which are corrected to the wavelengths $1500\,\angstrom$ and $900\,\angstrom$ using the model. There is no relative escape fraction without IGM absorption from the SED fits using CIGALE. f$_{\m{esc,rel}}^{\m{IGM}}$: relative escape fractions taking into account the IGM absorption shown in the fourth column of this table. The three escape fractions correspond again to the different methods, with the third one being the sightline-dependent escape fraction measured by CIGALE. f$_{\m{esc,abs}}^{\m{IGM}}$: absolute escape fractions taking the dust extinction A$_{\m{V}}$ (fifth column of this table) into account, again for the three different methods. 
Note that since CIGALE takes the dust attenuation in the UVC already into account, we directly get the absolute escape fraction. The value in brackets denotes the absolute escape fraction from CIGALE when the dust description is extended below $912\AA$ but only at a $10\%$ level.}
\end{table*}


\subsection{Intrinsic ratio of ionising to non-ionising UV luminosities} \label{Fixed intrinsic ratios}

The intrinsic ratio $(L_{\m{UVC}}/L_{\m{LyC}})_{\m{int}}$ between the LyC and the UV continuum at $900\,\angstrom$ and $1500\,\angstrom$ is not known as it cannot be measured directly but must be inferred from stellar population models. Common values in the literature range from $(L_{\m{1500}}/L_{\m{900}})_{\m{int}} = 3$ (frequency space $\nu$, with luminosity given in erg/s/Hz, see e.g.\ \citealp{Steidel2001, Grazian2016, Grazian2017, Japelj2017, Marchi2018}) to $5$ (see e.g.\ \citealp{Naidu2018}) and even as high as $6\--8$ (see e.g.\ \citealp{Siana2007}), which depends on the underlying assumptions on the star-formation history, metallicity, initial mass function (IMF) and age.

We use two fixed values for the intrinsic ratio. The first is chosen to compare to results from the literature as $(L_{\m{1500}}/L_{\m{900}})_{\m{int}} = 3$, which is what the first relative escape fraction f$_{\m{esc,rel}}$ for each object in table \ref{tab:various_fesc} is based on. For the second fixed value, we use an extreme BPASS model with a young age of 3 Myr, binary evolution, an IMF upper mass cutoff of $\m{M}_{\m{max}}=300\,\m{M}_{\odot}$ and a metallicity of $\m{Z} = 0.002$, which gives $(L_{\m{1500}}/L_{\m{900}})_{\m{int}} = 1.72$. The young age of the model is chosen such that we expect a large Lyman continuum luminosity (see e.g. \citealp{Rosdahl2022}), but not too young in order to have enough time to form escape channels for LyC. This way, the resulting escape fraction measurements (the second f$_{\m{esc,rel}}$ in table \ref{tab:various_fesc}) can be seen as a lower limit for our objects. The subsolar metallicity we choose here is in agreement with observations of high redshift, star-forming galaxies, see e.g. \citealp{Steidel2016,Cullen2020,Reddy2022}, and matches reasonably well with the results of the SED fitting for most objects (see table \ref{tab:CIGALE}).

Using this model, we can correct the measured flux values in the LyC and UV continuum bands to the values at $900\,\angstrom$ and $1500\,\angstrom$. For this, we convolve the model spectrum with the filter band throughput (de-redshifted to rest-frame wavelength) and get the flux ratio between the effective wavelength corrected to rest-frame and the desired wavelength. This ratio is used to correct the measured flux values, which are then used for the escape fraction. The resulting escape fraction $f_{\m{esc,rel}}$ is the second value in the eighth column in the table \ref{tab:various_fesc}.


\subsection{SED Fitting} \label{SED Fitting}

To better understand the properties of our objects and to obtain our third measurement of the escape fraction, we model their SEDs using CIGALE (\citealp{Burgarella2005,Noll2009,Boquien2019}). In this way, we can take the additional information from other photometric bands into account when modelling the underlying stellar population of the galaxy, instead of assuming the same model for all objects as we did in Sect. \ref{Fixed intrinsic ratios}. For the SED fitting, we use the HST bands F275W, F336W, F435W, F606W, F775W, F814W, F850LP, F105W, F125W, F140W, F160W as well as the HAWKI K-band and the IRAC channels 1, 2, 3 and 4.

\begin{figure*}
\centering
 \includegraphics[width=0.8\linewidth]{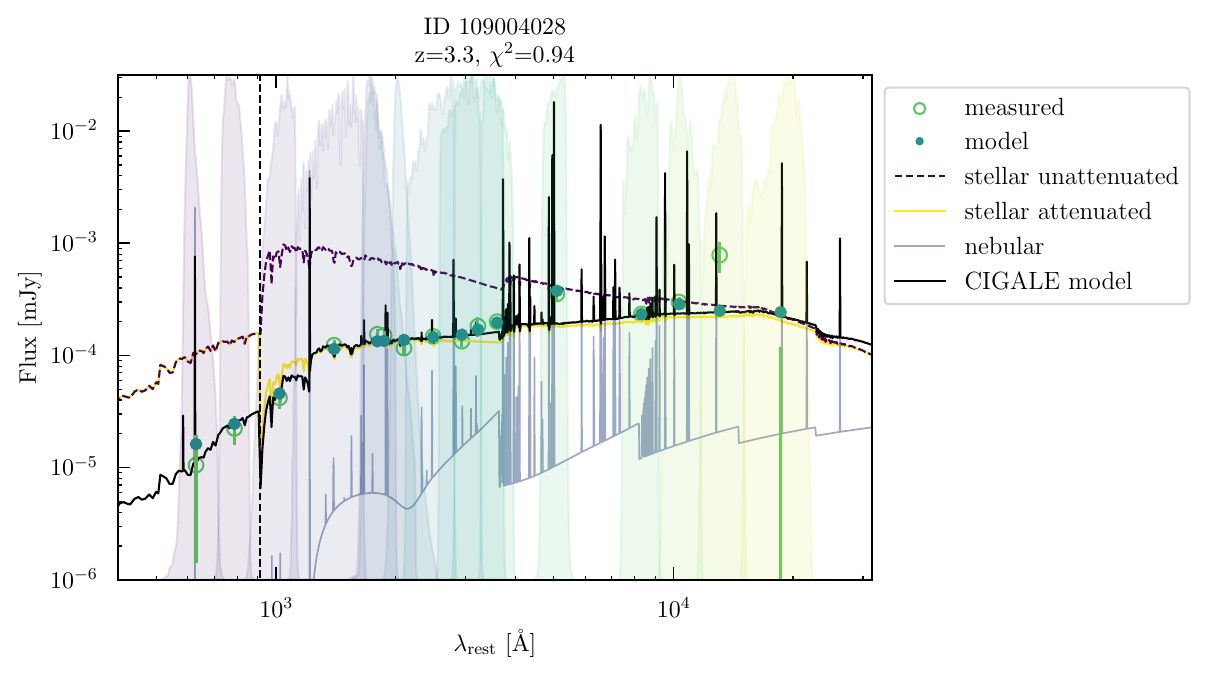}
\caption{Example of an SED fit using CIGALE (modified CIGALE plot using the output files). The x-axis shows the logarithmic rest-frame wavelength in Angstrom, the y-axis shows the logarithmic flux in micro Jansky. The purple dashed line shows the stellar unattenuated emission and the yellow line shows the stellar emission taking dust into account. The light blue line shows nebular emission and the solid black line shows the composite CIGALE model fit. The empty green dots are the measured values in the individual filter bands (shown as transparent areas in different colours from purple to yellow) and the dark green dots show the model values in the same filter bands. The vertical dashed line indicates the LyC break at $912\angstrom$. The object ID, redshift, and reduced $\chi^2$ are written on the top.}
\label{fig:SEDs1}
\end{figure*}

\begin{figure*}
\centering
\begin{minipage}{.5\textwidth}
  \centering
  \includegraphics[width=\linewidth]{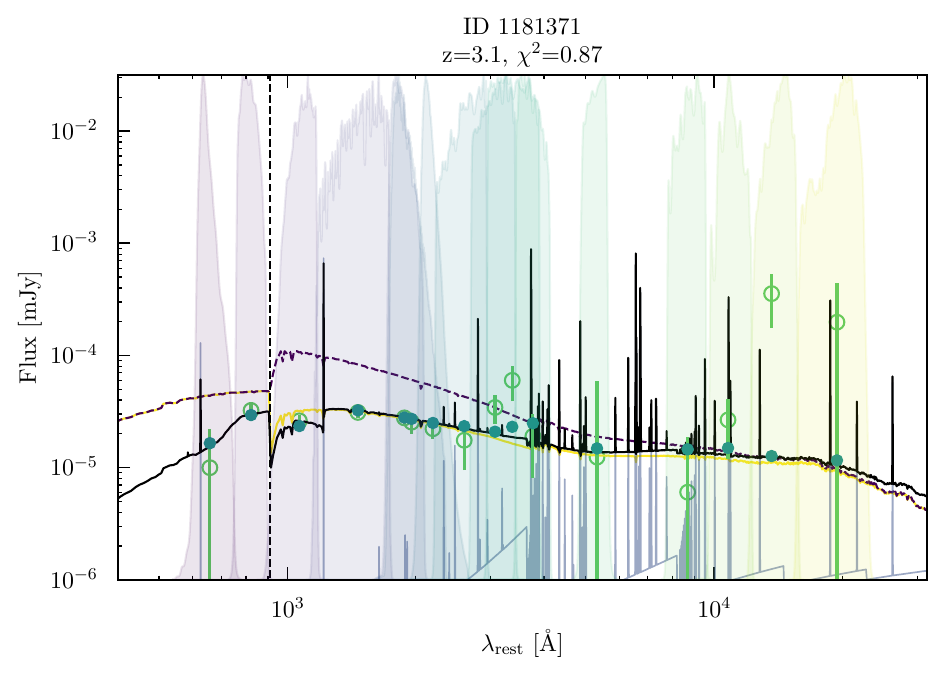}
  \includegraphics[width=\linewidth]{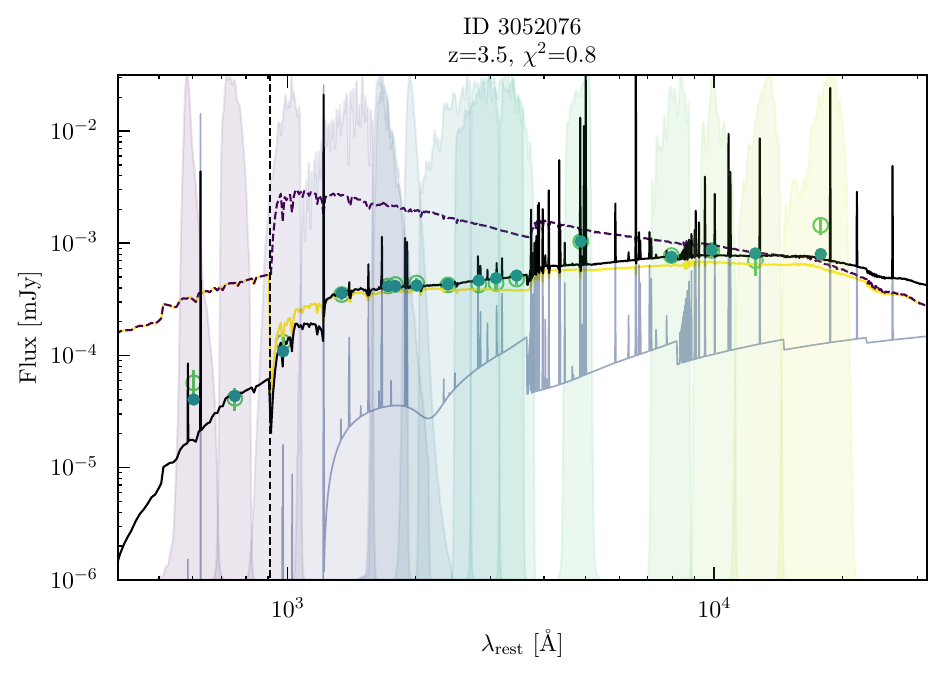}
\end{minipage}%
\begin{minipage}{.5\textwidth}
  \centering
  \includegraphics[width=\linewidth]{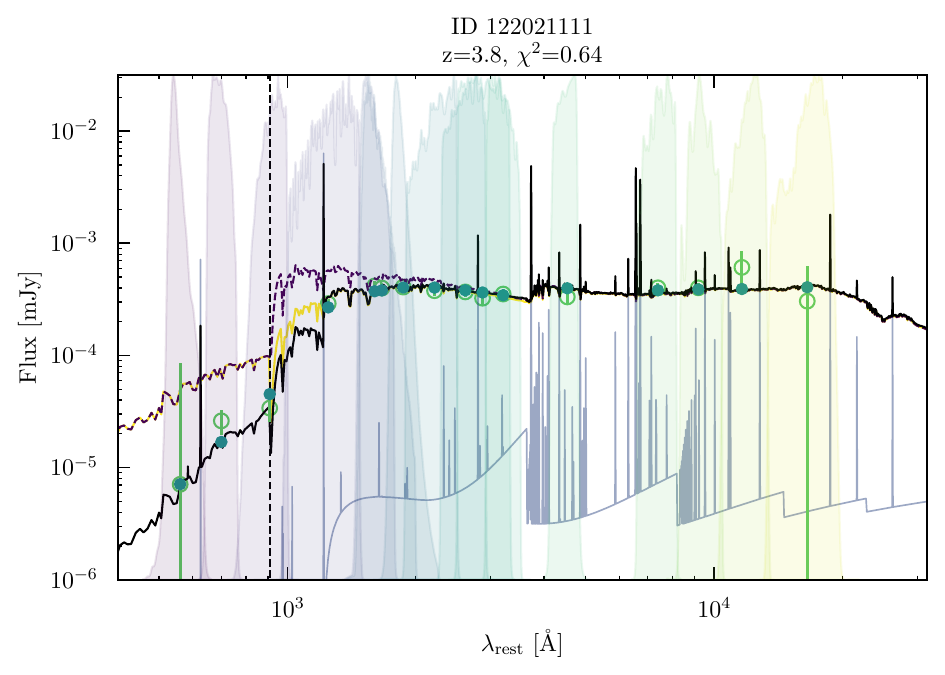}
  \includegraphics[width=\linewidth]{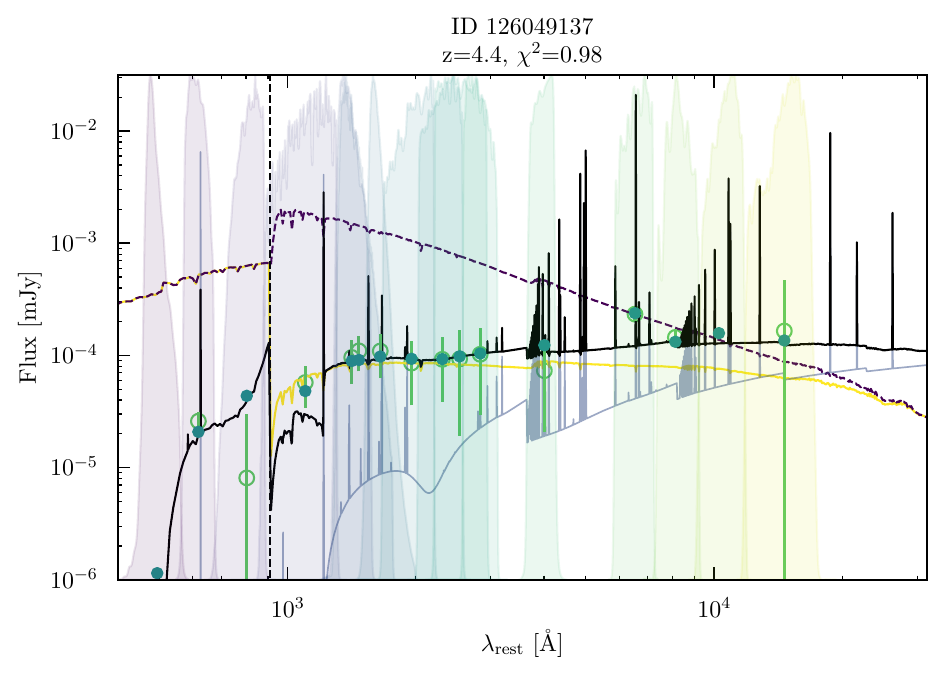}
\end{minipage}
\caption{Same as Fig.~\ref{fig:SEDs1} for the rest of the gold sample LyC leaker candidates.}
\label{fig:SEDs2}
\end{figure*}

The results for the five best LyC leaker candidates (gold sample) can be seen in Figs. \ref{fig:SEDs1} and \ref{fig:SEDs2}. For the dust attenuation law, we use the \texttt{CIGALE} module \url{dustatt_modified_starburst}, based on the \citet{Calzetti2000} starburst attenuation curve. The model allows for a more flexible approach to the dust attenuation (\citealp{Boquien2019,Boquien2022}), especially by changing the slope using the $\delta$ parameter and including options for the extinction of emission lines using an extinction curve based on the Milky Way (MW, \citealp{Cardelli1989,ODonnell1994}), and Small and Large Magellanic Cloud (SMC, LMC, based on \citealp{Pei1992}). 
It has been shown for example in \citealp{Reddy2018}, that for high-redshift LAEs, an SMC-like dust curve would be more realistic for the stellar continuum. To get as close as possible to the SMC curve, we use initial values for the $\delta$ parameter of $\delta\approx-0.5$ (as shown in \citealp{Boquien2022}), but leave the parameter-free for CIGALE to fit. When it comes to the extinction of emission lines, a MW-like dust curve is adapted here. However, we have verified that the choice of absorption curve for emission lines does not change the escape fraction results significantly.
For the star formation history, we use the module \url{sfh2exp}, which consists of two exponentials, one for the long-term star formation and one for the more recent starburst. We use a Chabrier initial mass function with the module \url{bc03} (based on \citealp{Bruzual2003}), leaving the metallicity as a free parameter (see table \ref{tab:CIGALE}).

The results of the SED fitting can be found in the appendix in table \ref{tab:CIGALE}. We note that our candidates have quite high specific star formation rates, which could contribute to their high escape fractions. 


\textbf{\subsection{IGM absorption}} \label{IGM absorption}

The last factor in the escape fraction calculation is the correction for the IGM absorption of LyC photons, using the IGM transmissivity at $900\,\angstrom$, $\tau_{\m{900,\m{IGM}}}$. Here we use the modelled line of sight transmissions from \citet{Inoue2011} and \citet{Inoue2014} for the different redshifts of our candidates. However, at redshifts above 3, a significant fraction of lines of sight through the IGM are already completely opaque for LyC photons. Using the median of the distribution of possible IGM transmission lines would result in a value of zero transmission. We argue that since we do detect LyC emission, this introduces a selection effect in the lines of sight, favouring ones with high transmission of LyC emission. For all escape fraction measurements we, therefore, choose an arbitrary cut in the transmission values, using the mean of only the highest $5\%$ of in the $10,000$ lines of sight from \citet{Inoue2014} at the specific redshift of the objects. The IGM transmission values $T_{\m{IGM}}=\exp(-\tau^{\m{IGM}}_{\m{900}})$ are given in the fourth column in table \ref{tab:various_fesc}. However, CIGALE assumes that there is no IGM transmission below the Lyman break, which is why we adjust the IGM transmission as explained above, using only the highest $5\%$.

Thus, we obtain IGM corrected escape fractions  $f_{\m{esc,rel}}^{\m{IGM}}$, see the second to last column in table \ref{tab:various_fesc}, for the assumption that the IGM transmission is high for all of our LyC leaker candidates. Following a similar thought, we assume that our LyC leaker candidates are not affected by absorption in the CGM. In the literature, the effects of IGM and CGM are often not treated separately, however, it has been shown by \citet{Muzahid2021} that LAEs at similar redshifts to ours, live in extended neutral hydrogen environments, up to a distance of $\approx500\,\m{km/s}$ or $\approx 7$ virial radii. This would make it harder for LyC emission to escape into the IGM. However, their results show that LAEs in groups are more likely to have extended neutral hydrogen around them, hinting at large-scale structures. While we cannot exclude that our candidates could be affected by CGM absorption, we apply the same argument as for the IGM absorption: The absorption is stochastic and depends on the line of sight (see Fig. 5 in \citealp{Muzahid2021}. The fact that we do see LyC emission means that we are looking at a relatively free line of sight or possibly a hole or low column density in the CGM. Just as with the IGM, the CGM coverage is difficult to treat for individual objects. We therefore assume a similarly unusual CGM configuration for our LyC leaker candidates as for the IGM and interpret the correction for the IGM absorption as including the CGM absorption as well.


\textbf{\subsection{Dust discussion}}

As can be seen in Figs. \ref{fig:SEDs1} and \ref{fig:SEDs2}, the LyC part of the spectrum is not affected by dust in our models, since there are no descriptions for that part. Although it would be possible to extrapolate the dust extinction law to smaller wavelengths, we decided against such a procedure for our main escape fraction measurements and keep the LyC unaffected by dust for several reasons. It is not guaranteed, that the stars that emit most of the UV continuum live in the same environment as the ones that produce the bulk of the LyC. It has been speculated that even a single star cluster containing a handful of massive, hot stars such as Wolf-Rayet stars or O-type stars, could be responsible for most of the ionising emission that we see (e.g. \citealp{Vanzella2020,Vanzella2022,Mestric2023}).
While it is true that our spatial resolution does not allow us to study this assumption, the fact that we detect LyC emission shows that there must be a clear path with a low neutral hydrogen column density ($\m{N}_{\ion{H}{I}}\lesssim10^{17}\,\m{cm}^{-2}$) through which LyC could escape (see e.g. \citealp{Reddy2016b,Steidel2018,Saldana-Lopez2022,Reddy2022}). 
Such a path could be blown in the ISM for example via supernovae (SN), however for young starbursts $\sim3$ Myr, there might not have been enough time yet for SN to explode. Instead, paths could be cleared by turbulence or radiative/stellar feedback (e.g. \citealp{Kakiichi2021}). In addition, a highly energetic ionising background would also destroy dust, leaving the dust density low in such an escape channel. Since dust and neutral hydrogen often coexist, we only need to take dust into account in more neutral regions. However, the cross-section of dust is four orders of magnitude smaller than that for photoionisation (e.g. \citealp{Kakiichi2021}), meaning the optical depth of dust is always smaller than that of neutral hydrogen for LyC photons (see \citealp{Reddy2016a,Ma2020,Rosdahl2022}), which is why the influence of dust on the LyC, even in neutral regions, is negligible. An exception could be photoionised channels in which the extinction by dust can become comparable to the absorption by photoionisation for higher metallicities (significantly above Z=$0.2\,Z_{\odot}$), however since LyC leakers at high redshifts are expected to be low metallicity systems and indeed most of our candidates have metallicities Z$\sim0.2\,Z_{\odot}$, we argue that this effect is not significant (\citealp{Kakiichi2021}).
It should also be kept in mind that using a unity dust covering fraction might not be ideal, since the covering fraction of neutral gas is not unity and the same processes that clear channels in the ISM would also clear out the dust, given that the neutral hydrogen and dust are dynamically coupled. 

Several theoretical studies have also shown that indeed the LyC escape fraction is not strongly influenced by the presence of dust. \citet{Kimm2019} show on the scale of individual star-forming clouds of $10^6\m{M}_{\odot}$ that the absorption by neutral hydrogen is the dominant factor, while \citet{Mauerhofer2021} confirm this for a simulated galaxy of $2.3 \times 10^9\,\m{M}_{\odot}$. \citet{Yoo2020} study disc galaxies with a stellar mass of $\sim10^9\m{M}_{\odot}$, which is the same mass range as our galaxies (see table \ref{tab:CIGALE} in the appendix). They find that dust changes the LyC escape fraction by only $\sim 10^{-3}-10^{-4}\%$, as they assume that only $1\%$ of dust can survive in ionised regions. Without this assumption, the escape fraction is reduced by $17\%$ ($37\%$ for solar metallicities). However, \citet{Ma2020} find that the decrease in escape fraction at higher masses of $\sim10^9\,\m{M}_{\odot}$ could be due to dust absorption. For those studies, the presence of dust at high temperatures plays an important role, as \citet{Ma2020} assume dust up to a temperature of $10^6$ K, while for other studies like \citet{Mauerhofer2021} and \citet{Kimm2019}, dust can only survive up to several $10^4$ K.

This is why, for our main escape fraction measurements, we do not include any dust extinction of the LyC in the SED fitting by CIGALE, which is why the stellar unattenuated and attenuated models in Figs. \ref{fig:SEDs1} and \ref{fig:SEDs2} are the same for wavelengths below $912\, \angstrom$. We are therefore treating the LyC and the rest of the spectrum differently because we know that for wavelengths $\lambda > 912\, \angstrom$ we see the whole stellar population of the galaxy, where dust has an influence, whereas, for $\lambda < 912\, \angstrom$, we only see emission that was unaffected by dust for the reasons explained above. This results in a discontinuity in the fitted spectra in Figs. \ref{fig:SEDs1} and \ref{fig:SEDs2} at $912\, \angstrom$, which however does not affect the measured properties.
The escape fraction from CIGALE is, therefore, a highly sightline-dependent measure, since we assume that the ionising photons escape through ionised channels which are low in dust. This is in contrast to the other measures of escape fraction, assuming a fixed ionising to non-ionising ratio for the whole galaxy.
We can use the measured UV dust extinction from the CIGALE models to correct the relative escape fractions measured from the two other methods to absolute escape fractions (shown in table \ref{tab:various_fesc} and described above in Sect. \ref{Fixed intrinsic ratios}), using fixed values for the intrinsic ratio, shown in the last column of table \ref{tab:various_fesc}. 
In order to estimate the potential influence of dust on our escape fraction measurements from CIGALE, we provide a second CIGALE measurement in table \ref{tab:various_fesc}. Here we extend the dust treatment of CIGALE below $912\AA$, but reduce it by $90\%$, thus accounting for the fact that since we do detect LyC, we can assume a relatively low dust extinction in the direction of escape, in addition to the arguments above. As expected, the LyC escape fraction increases when dust is taken into account, by factors between $2\%$ and $92\%$.
Keeping in mind the uncertainties concerning the influence of dust on the LyC, we can assume that the true escape fraction lies between the relative and the absolute ones given in table \ref{tab:various_fesc}. In this sense, the absolute escape fraction without dust absorption in the LyC can be seen as a lower limit. 


\textbf{\subsection{Escape fraction results}}

Having now measured the different LyC escape fractions with different methods, we see that, depending on the assumptions made, escape fractions for the same object can vary widely. For example, the first gold candidate has a relative and even absolute escape fraction above $100\%$ if we assume an intrinsic UV continuum to LyC ratio of three. Using the less steep intrinsic ratio of 1.72 from the BPASS model (see Sect.~\ref{Fixed intrinsic ratios}), we find more reasonable escape fractions (below $100\%$), while the CIGALE value is again high at $\approx90\%$. Indeed, the escape fractions based on the BPASS model usually result in the smallest derived escape fractions. 
 

\section{Lyman \texorpdfstring{$\alpha$}{Lg} properties of LyC leaker candidates} \label{Lyman alpha properties of LyC leaker candidates}

We can now look at the Ly$\alpha$ properties of the LyC leaker candidates and compare them to the general population of LAEs. We consider Ly$\alpha$ properties such as FWHM, peak separation and equivalent width, which we take from \citet{Kerutt2022}.

It has been suggested in the literature that Ly$\alpha$ emission could be a good indicator of LyC leakage (see e.g. \citealp{Micheva2017,Verhamme2017,Marchi2018,Steidel2018,Fletcher2019,Vanzella2020,Pahl2021,Reddy2022}). Since LyC emission is needed to ionise the neutral hydrogen in the ISM of a galaxy, which recombines to emit a Ly$\alpha$ photon in $\approx$ 2/3 of the cases (e.g. \citealp{Osterbrock1989,Dijkstra2014}), the two types of emission are closely linked. The more ionising radiation is absorbed by the neutral hydrogen in the ISM, the stronger the intrinsic Ly$\alpha$ emission. In the same vein, the escape channels of Ly$\alpha$ and LyC could be similar, as Ly$\alpha$ will be scattered and potentially absorbed by dust in a high neutral hydrogen column density environment, while the LyC would be absorbed by the neutral hydrogen as well. A low neutral hydrogen column density would thus permit an easier escape for both Ly$\alpha$ and the LyC. Therefore, a high Ly$\alpha$ equivalent width, as well as a narrow line and a narrow peak separation, have been proposed as good indicators of potential LyC leakage (e.g. \citealp{Verhamme2017,Vanzella2020,Izotov2021}, see Sect.~\ref{Discussion} for a discussion).


\subsection{Lyman \texorpdfstring{$\alpha$}{Lg} Properties of the LyC leaker candidates} \label{Properties of the LyC leaker candidates} 

\begin{figure*}[ht]
\begin{minipage}[t]{0.5\textwidth}
\includegraphics[width=\linewidth]{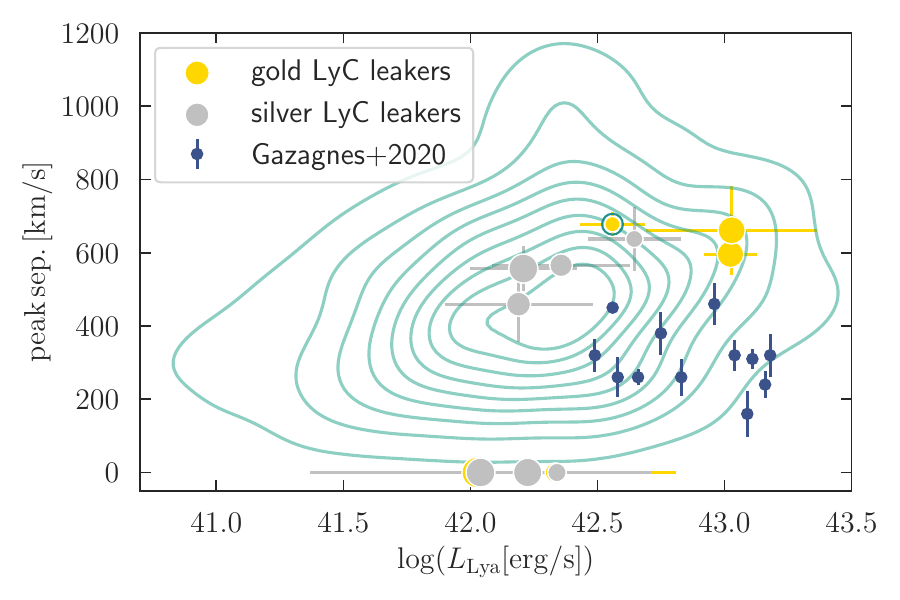}
\end{minipage}
\hfill
\begin{minipage}[t]{0.5\textwidth}
\includegraphics[width=\linewidth]{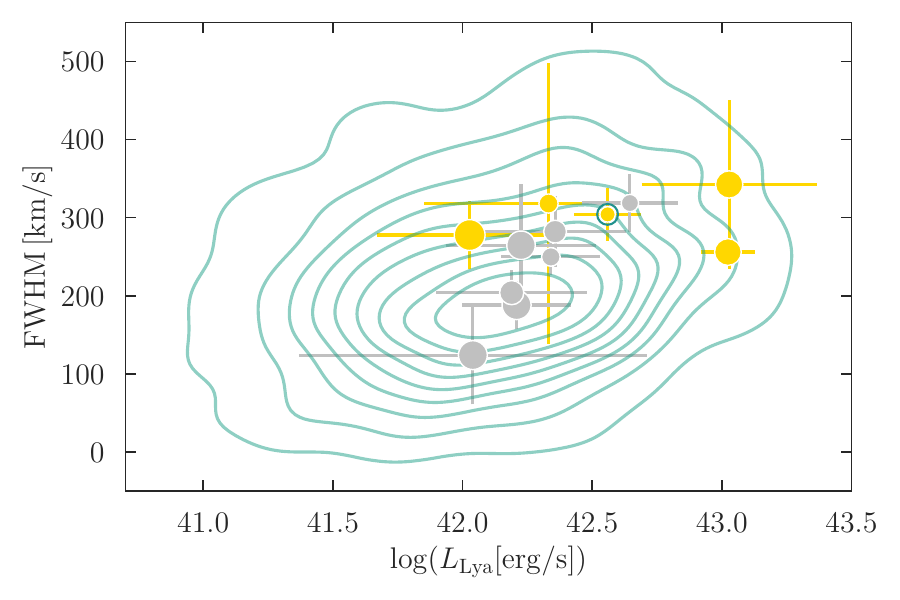}
\end{minipage}
\caption{Left: Peak separation of the Ly$\alpha$ line as a function of the logarithmic Ly$\alpha$ luminosity. The contours (each showing a $10\%$ number difference) contain the full sample of LAEs from \citet{Kerutt2022} and show the density of objects in the peak separation and Ly$\alpha$ luminosity plane. The gold and silver dots show the individual values for the LyC leaker candidates, with the sizes indicating the escape fraction values based on the SED models by CIGALE. The gold candidate with a green circle is the object already discovered in \citet{saxena2022} and discussed in \citet{Rivera-Thorsen2022}. Objects without double peaks are placed at peak sep. = 0 at their respective luminosities. The blue dots show data of low redshift LyC leakers from \citet{Gazagnes2020}, also featured in \citet{Izotov2016a,Izotov2016b,Izotov2018a,Izotov2018b} and \citet{Maji2022}. Right: FWHM of the Ly$\alpha$ line as a function of the logarithmic Ly$\alpha$ luminosity, again with contours showing the distribution of objects in these values.}
\label{fig:FWHM_peak_sep_logLLya}
\end{figure*}

In Table \ref{tab:goodss_Lya} we provide the rest-frame equivalent width, luminosity, flux, FWHM, and the peak separation (for objects with a double peak) of Ly$\alpha$ as well as absolute UV magnitudes for our candidates, all values taken from \citet{Kerutt2022}. It is worth mentioning that none of our candidates has a Ly$\alpha$ $\m{EW}_0 > 240\, \angstrom$ ($11\%$ of our final sample have such high $\m{EW}_0$), which is usually cited as the approximate upper limit for normal stellar populations. This is interesting, as it was expected that such unusual stellar populations (with low metallicities, potentially containing population III stars, high star formation rates, and young ages) would be more likely to produce noticeable amounts of LyC emission (see Sect.~\ref{Discussion} for a discussion). Likewise, the FWHM and peak separations are both above the average of the full sample (see \citealp{Kerutt2022}), contrary to theoretical expectations which expect small FWHM and peak separations (\citealp{Verhamme2017}). This can be seen in Fig.~\ref{fig:FWHM_peak_sep_logLLya}, where we show the distributions of FWHM and peak separation against the Ly$\alpha$ luminosity of the whole sample of LAEs, highlighting the respective values of our LyC leaker candidates. Both the FWHM and the peak separation values of the candidates are above the mean of the full sample, which is $\m{FWHM} = 218 \pm 102 \, \m{km}\,\m{s}^{-1}$ and peak sep. $= 481 \pm 244\, \m{km}\, \m{s}^{-1}$ (\citealp{Kerutt2022}), compared to $\m{FWHM} = 270 \pm 67 \, \m{km}\,\m{s}^{-1}$ and peak sep. $= 602 \pm 69\, \m{km}\, \m{s}^{-1}$ for our sample of LyC leaker candidates. 
We show the MUSE spectra of our candidates in the appendix in Fig.~\ref{fig:MUSE_spectra2}.
As can be seen from those spectra, some of the double peaks have a low S/N and are rather uncertain. However, assuming the correlation of \citet{Verhamme2017} between FWHM and velocity offset, as well as peak separation and velocity offset, there is also a correlation between FWHM and peak separation, where the peak separation is expected to be roughly twice as large as the FWHM, which is the case for our double peaks, except for ID 1521589, which has a rather high peak separation, but a convincing blue bump. Therefore we can assume that our results are not influenced by a falsely identified blue peak.

\begin{table*} 
\begin{center} 
\caption{Overview of candidates - Ly$\alpha$ properties.} 
\begin{tabular}{ l l l l l l l } 
\hline\hline 
ID$_{\m{MW}}$ & EW$_0$ & M$_{\m{UV}}$ & $\m{log_{10}(L}_{\m{Ly}\alpha})$ & $10^{-18}$ f$_{\m{Ly}\alpha}$ & FWHM & peak sep. \\ 
& [\AA] & [mag$_{\m{AB}}$] & $\m{[erg/s]}$ & [erg/s/cm$^2/\angstrom$] & [km/s] & [km/s] \\ \hline
gold candidates & & & & & & \\
\bf{\textcolor{gold}{1181371}} & $  \ph  60 \pm 13 $ & $ -18.25 \pm 0.17 $ & $ 42.0 \pm 0.4 $ & $ 12 \pm  2 $ & $ 277 \pm 43 $ & $ $ \\ 
\bf{\textcolor{gold}{3052076}} & $  \ph  18 \pm  1 $ & $ -20.89 \pm 0.04 $ & $ 42.6 \pm 0.1 $ & $ 33 \pm  1 $ & $ 304 \pm 33 $ & $ 677 \pm 32 $ \\ 
\bf{\textcolor{gold}{109004028}} & $  \ph  33 \pm  7 $ & $ -19.65 \pm 0.05 $ & $ 42.3 \pm 0.5 $ & $ 22 \pm  4 $ & $ 318 \pm 179 $ & $ $ \\ 
\bf{\textcolor{gold}{122021111}} & $  \ph  45 \pm 2 $ & $ -21.05 \pm 0.02 $ & $ 43.0 \pm 0.1 $ & $ 77 \pm  3 $ & $ 255 \pm 13 $ & $ 595 \pm 11 $ \\ 
\bf{\textcolor{gold}{126049137}} & $ 174 \pm 28 $ & $ -19.61 \pm 0.07 $ & $ 43.0 \pm 0.3 $ & $ 54 \pm  8 $ & $ 342 \pm 107 $ & $ 661 \pm 121 $ \\ \hline
silver candidates & & & & & & \\ 
\bf{\textcolor{silver}{1521589}} & $ 161 \pm 14 $ & $ -17.64 \pm 0.02 $ & $ 42.2 \pm 0.2 $ & $ 18 \pm  1 $ & $ 188 \pm 30 $ & $ 556 \pm 62 $ \\ 
\bf{\textcolor{silver}{3452147}} & $ 104 \pm 26 $ & $ -18.49 \pm 0.24 $ & $ 42.4 \pm 0.3 $ & $ 20 \pm  2 $ & $ 282 \pm 45 $ & $ 565 \pm 30 $ \\ 
\bf{\textcolor{silver}{4062373}} & $  \ph  18 \pm 2 $ & $ -20.01 \pm 0.04 $ & $ 42.2 \pm 0.3 $ & $ 13 \pm  1 $ & $ 264 \pm 78 $ & $ $ \\ 
\bf{\textcolor{silver}{4172404}} & $  \ph  94 \pm 17 $ & $ -18.56 \pm 0.18 $ & $ 42.3 \pm 0.2 $ & $ 17 \pm  1 $ & $ 249 \pm 22 $ & $ $ \\ 
\bf{\textcolor{silver}{5622786}} & $ 133 \pm 33 $ & $ -17.80 \pm 0.24 $ & $ 42.2 \pm 0.3 $ & $ 10 \pm  1 $ & $ 203 \pm 29 $ & $ 459 \pm 103 $ \\ 
\bf{\textcolor{silver}{119004004}} & $ 118 \pm 15 $ & $ -19.07 \pm 0.11 $ & $ 42.6 \pm 0.2 $ & $ 44 \pm  3 $ & $ 318 \pm 37 $ & $ 637 \pm 87 $ \\ 
\bf{\textcolor{silver}{122032127}} & $  \ph  26 \pm 8 $ & $ -19.20 \pm 0.16 $ & $ 42.0 \pm 0.7 $ & $ \ph  5 \pm  1 $ & $ 123 \pm 62 $ & $ $ \\ 
 \hline 
 mean & $ \ph 78\pm53$ & $-19.33\pm1.14$ & $42.4\pm0.3$ & $28\pm20$ & $270\pm67$ & $602\pm69$ \\ 
\end{tabular}\label{tab:goodss_Lya} 
\end{center} 
\tablefoot{ID$_{\m{MW}}$: Identifier in \citet{Urrutia2019}. EW$_0$: Ly$\alpha$ rest-frame equivalent width from MUSE-Wide in $\angstrom$. M$_{\m{UV}}$: Absolute UV continuum AB magnitude at $1500\,\angstrom$, measured from HST bands. $\m{log_{10}(L}_{\m{Ly}\alpha})$: Logarithmic Ly$\alpha$ luminosity in $\m{erg/s}$. $10^{-18}$ f$_{\m{Ly}\alpha}$: Ly$\alpha$ flux in $10^{-18}$ erg/s/cm$^2/\,\angstrom$. Ly$\alpha$ flux and luminosity were measured from MUSE-Wide data within 3 Kron radii and include the blue bump (for double-peaked lines). FWHM: Full width at half maximum of the main (red) peak of the Ly$\alpha$ line in km/s. Peak sep.: Peak separation for lines with double peaks in km/s. Both the FWHM and the peak separation are corrected for the line spread function of MUSE. The Ly$\alpha$ measurements are taken from \citet{Kerutt2022}. The values in the last line give the mean values for all candidates with the standard deviation of the measurements.}  
\end{table*}

\subsection{Connecting Lyman \texorpdfstring{$\alpha$}{Lg} Properties to Leakage} \label{Intro Connecting Properties to Leakage} 

\begin{figure*}[ht]
\centering 
\includegraphics[width=\linewidth]{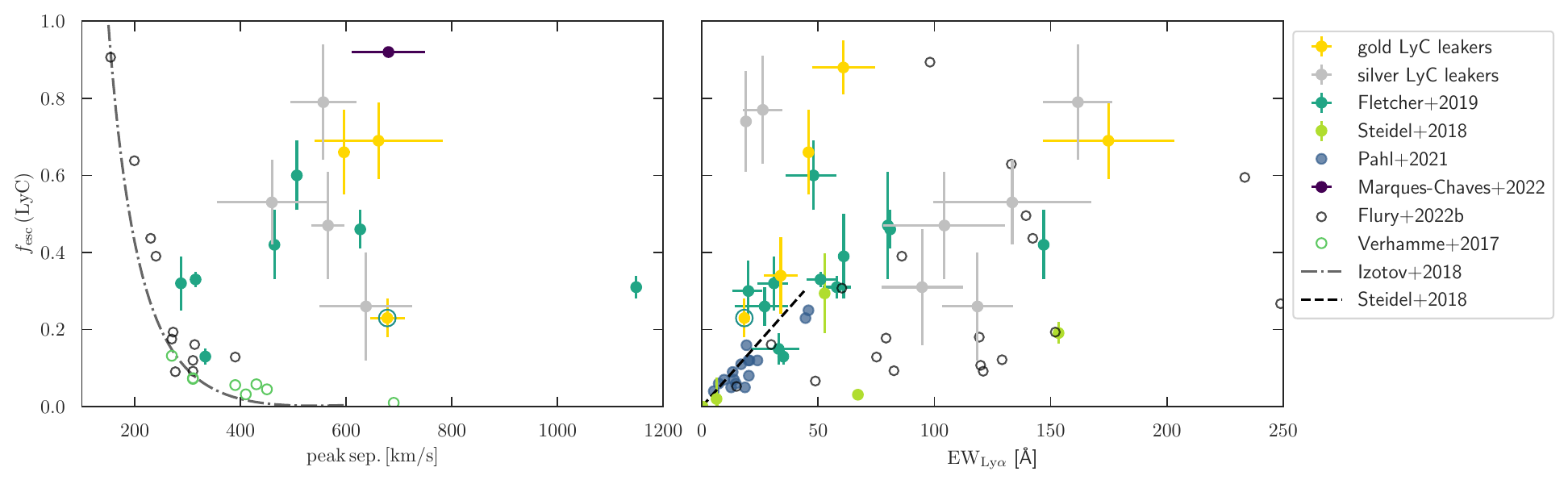}
\caption{Left: LyC escape fraction as a function of Ly$\alpha$ peak separation. The silver and gold-filled dots are from our LyC leaker candidate sample. The gold candidate with a green circle around it is again the object already discovered in \citet{saxena2022} and discussed in \citet{Rivera-Thorsen2022}. The black dash-dotted line is the relation in \citet{Izotov2018b} for low-redshift LyC leakers. 
Open circles show low redshift objects and filled dots show high redshift objects. The dark green data points are for LyC leakers at $z\approx3.1$ from \citet{Fletcher2019}, where they give velocity offsets with respect to systemic redshift, which was multiplied here by two to estimate the potential peak separation, following the correlation found between peak separation and the shift of the red peak with respect to systemic velocity from \citet{Verhamme2018}. The dark purple filled circle is from \citet{Marques-Chaves2022} for $z=3.6$. The light green open circles are taken from \citet{Verhamme2017}, also for low-redshift analogues. The black dots are taken from \citet{Flury2022b}, showing the escape fractions based on the COS UV spectra for only their strongest LyC leakers at low redshifts ($z\approx 0.3-0.4$). 
Right: LyC escape fraction as a function of the Ly$\alpha$ rest-frame equivalent width. The black dashed line is from \citet{Steidel2018}, just as the light green dots for objects at $z\approx3$. The dark green dots are from \citet{Fletcher2019} for objects at $z\approx3$. The open black dots are again taken from \citet{Flury2022b} and the open blue dots are from \citet{Pahl2021} at $z\approx3$.}
\label{fig:fesc_peak_sep_EW}
\end{figure*}

The Ly$\alpha$ properties of the LyC leakers are somewhat surprising, as it would have been expected from theoretical models and previous observations of low-redshift analogues (see e.g. \citealp{Verhamme2017,Izotov2018b}) that a small peak separation and narrow FWHM, as well as a high equivalent width, should be good indicators of LyC leakage.
In the left panel of Fig.~\ref{fig:fesc_peak_sep_EW} we show the measured escape fraction from the SED models from CIGALE over the peak separation of the Ly$\alpha$ line, with a comparison to literature values from \citet{Izotov2018b} and \citet{Verhamme2017} for low-redshift LyC leakers. 
To compare to LyC leakers at a similar redshift as ours, we use the velocity offsets given in \citet{Fletcher2019} for LyC leakers at $z\approx3.1$ and multiply them by two, assuming that there is a correlation between the velocity offset and the peak separation (as shown in \citealp{Verhamme2018}).

It has to be kept in mind that here we show the sightline-dependent escape fraction measurement from CIGALE, where we assume no dust absorption of the Lyman continuum since they escape along ionised channels with high gas temperatures. The escape fraction estimates we get are not the global LyC escape fractions of the galaxies. The same is true for Ly$\alpha$, which can vary substantially with viewing angle (see \citealp{Blaizot2023}).

The results of these comparisons are further discussed below in Sect. \ref{Discussion} below.

\section{Discussion} \label{Discussion}

Due to the need to infer LyC emission at the epoch of reionisation from other, measurable properties of high-redshift galaxies, we have looked at the Ly$\alpha$ line as a possible indicator. In this section, we discuss its potential and also suggest a way to use the number of detected LyC leakers as a possible way to predict the global escape fraction.


\subsection{The Dispersion of LyC escape fractions}

In table \ref{tab:various_fesc} we have seen that the escape fraction is sensitive to the method and assumptions applied. This makes a comparison with literature results difficult, as different studies use different values for the intrinsic ratio (L$_{1500}$/L$_{900})_{\m{int}}$ between the UV continuum and the LyC, usually in the range of 3, 5 or 7, as well as for the IGM transmission. In addition, dust extinction is also treated in different ways in the computation of the absolute escape fraction. What is more, the limiting magnitudes of different studies could result in biases, if the LyC escape is correlated with UV luminosity and/or stellar masses.
Keeping these caveats in mind, we nevertheless compile a collection of literature values for the absolute escape fraction as a function of redshift in Fig.~\ref{fig:f_esc_rel} based in part on a collection of LyC leakers from the literature in the appendix (\ref{tab:LyC_leakers_ind_high}, \ref{tab:LyC_leakers_stacks} and \ref{tab:LyC_leakers_low}). Simulations predict an increase in the escape fraction with higher redshifts between $z=4-8$ (e.g. \citealp{Trebitsch2021}). 
There are two more densely populated areas in this plot, one is the region at $z<0.5$, containing low-redshift LyC leakers as analogues of high-redshift ones. The other region is roughly between $z=2-4$, where LyC emission is observable with the HST using the filters WFC3/UVIS F275W and F336W, as done in this study. The region between these intermediate redshift objects and the low-redshift analogues is scarcely populated, mostly by stacking studies. 
In the range $z=3-4.5$, we show the escape fraction measurements of our own LyC leaker candidates in gold and silver (both for a fixed intrinsic flux ratio of 3 and based on the CIGALE SED fitting), as well as upper limits for our non-detections, based on the CIGALE SEDs. We use $2\sigma$ as the upper flux limit (since this is also our cut for a detection) and the mean IGM transmission at the redshift bin of each object (in steps of $\Delta z=0.1$), as well as the median dust extinction $\m{A}_{\m{V}}$ of the 12 LyC leaker candidates to convert to absolute escape fractions. 
Since the escape fraction measurements can vary widely based on the assumptions and parameters used for the computation, but mostly because of the different sample selections, potential biases, and completeness issues, we cannot draw conclusions about a potential trend with redshift from this plot and it should be seen as a compilation of literature values. 

\begin{figure*}
\centering 
\includegraphics[width=0.8\textwidth]{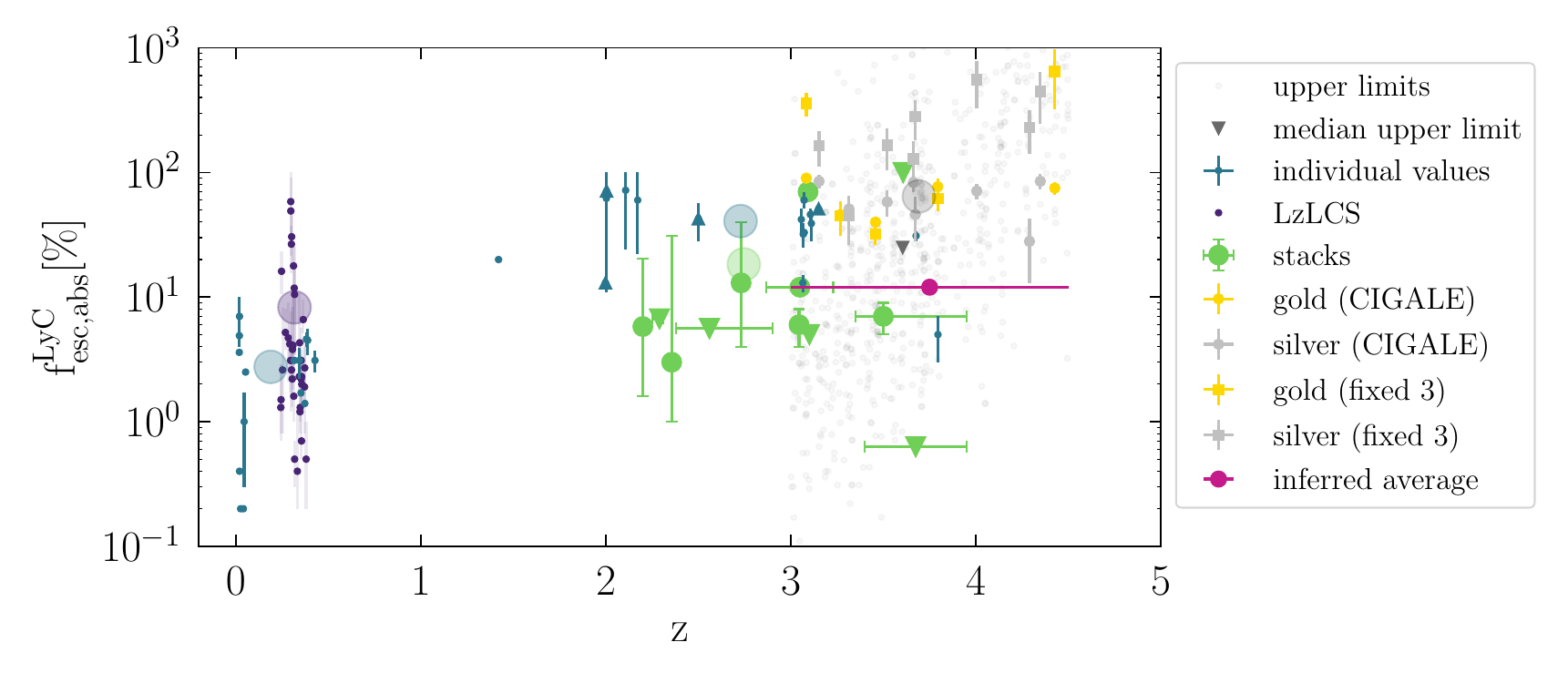}
\caption{Logarithmic absolute escape fraction of LyC over redshift for a collection of literature values and our sample of gold and silver objects, in their respective colours. Here, circles show measurements from SED fitting using CIGALE, while squares are escape fractions using an intrinsic ratio of 3 to compare to the literature. The small blue dots represent individual detections of LyC emission, for low-redshift analogues and high redshifts. The small purple dots are from the LzLCS survey (\citealp{Flury2022a,Flury2022b}). The large green symbols indicate results from stacking analyses. Stacks, where only upper or lower limits could be estimated, are shown with downwards-facing and upwards-facing triangles, respectively. The large transparent circles are the mean values of the respective sub-sample, with the grey containing both the gold and silver samples. The light grey dots between $z=3-4.5$ are the upper limits of our sample (based on an intrinsic ratio of 3), and the grey downward triangle is the median of those upper limits. The magenta dot corresponds to the inferred average escape fraction from the simulations described in Sect. \ref{Expected LyC leaker fraction}. The data shown is based on tables \ref{tab:LyC_leakers_ind_high}, \ref{tab:LyC_leakers_stacks} and \ref{tab:LyC_leakers_low} in the appendix.} 
\label{fig:f_esc_rel}
\end{figure*}


Most of the known low-redshift LyC leakers were found in the LzLCS (\citealp{Flury2022a,Flury2022b,Saldana-Lopez2022,Chisholm2022}) with a range of escape fractions between 0 and $\approx 50\%$. In their studies, \citet{Flury2022b} test among other things the assumed relation between Ly$\alpha$ properties and LyC leakage and find that the Ly$\alpha$ escape fraction, equivalent width and peak separation correlates best with LyC escape fractions. This connection is expected, as explained in the introduction and Sect. \ref{Lyman alpha properties of LyC leaker candidates}. An optical depth of hydrogen ionising photons of one corresponds to a neutral hydrogen column density of $\m{N}_{\m{HI}}=1.6\times10^{17}\,\m{cm}^{-2}$ (explained e.g. in \citealp{Verhamme2015}), however, due to the larger cross-section of Ly$\alpha$, neutral hydrogen is optically thick to Ly$\alpha$ down to $\m{N}_{\m{HI}}\sim10^{13}\,\m{cm}^{-2}$. In this density range the LyC emission can escape more easily, while Ly$\alpha$ will be scattered, which results in a double peak with increasing peak separation for increasing $\m{N}_{\m{HI}}$. 

From a theoretical point of view, it would therefore be expected that a narrow Ly$\alpha$ line ($\m{FWHM}\approx200\,\m{km/s}$) or a double peak with a small peak separation ($\m{peak\,sep.}<300\,\m{km/s}$, \citealp{Verhamme2015}), would indicate a low neutral hydrogen column density, corresponding to the limit where the optical depth of LyC is around one. 
However, if we look at Fig.~\ref{fig:FWHM_peak_sep_logLLya}, we see that our LyC leaker candidates are typically not narrow and do not have narrow peak separations (our smallest peak separation being $459\pm103\,\m{km/s}$). We also do not find a connection between the peak separation and the measured LyC escape fraction in the left panel of Fig.~\ref{fig:fesc_peak_sep_EW}, although this has been shown to be a reliable indicator for low-redshift analogues (\citealp{Verhamme2017,Izotov2018b,Flury2022b}). For higher-redshift LyC leakers as in \citet{Fletcher2019} at $z\approx3.1$, the velocity offset with respect to systemic also does not seem to be a good enough indicator for LyC escape. 
Similarly, a high Ly$\alpha$ equivalent width (shown in the right panel of Fig.~\ref{fig:FWHM_peak_sep_logLLya}) is not a requirement for a high LyC escape fraction. 
Still, \citet{Steidel2018} find a correlation for restframe Ly$\alpha$ equivalent widths of up to $\approx 50\, \angstrom$, \citet{Marchi2018} find a LyC escape fraction of $33\%$ for their stack of high Ly$\alpha$ EWs, and \citet{Begley2022} find a higher escape fraction as well for their stack of the half of their sample with higher Ly$\alpha$ equivalent widths (although it has to be noted that this half of the sample only has a median $\m{EW}_0=4.9\,\angstrom$).
In contrast, we find several of our LyC leaker candidates have high escape fractions even with smaller equivalent widths. However, the highest Ly$\alpha$ equivalent width objects among our candidates have indeed high escape fractions. 

\begin{figure}
  \includegraphics[width=\linewidth]{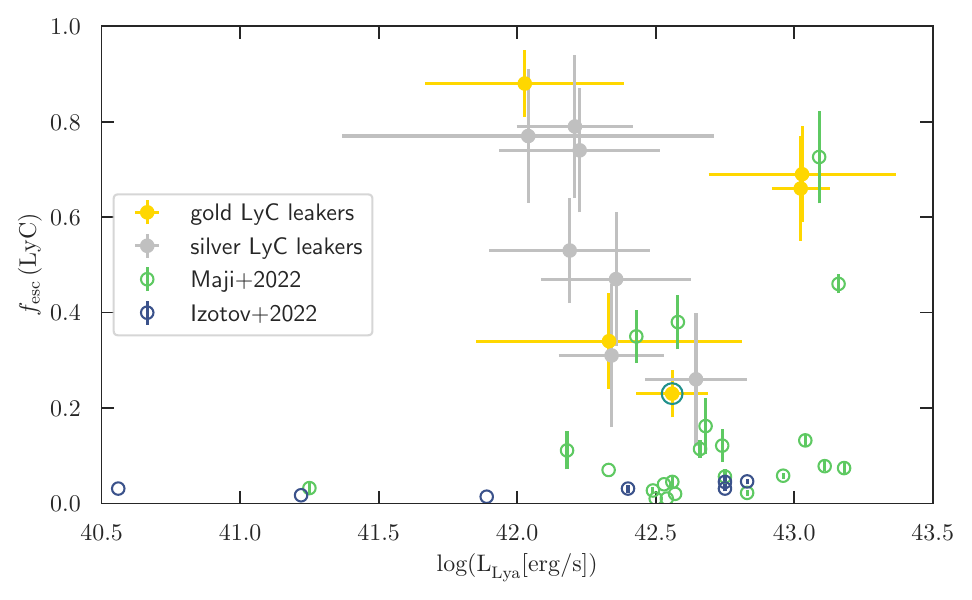}
\caption{LyC escape fraction versus the logarithmic Ly$\alpha$ luminosity for the gold and silver sample of our LyC leaker candidates. In green, we show the selection of low-redshift LyC leakers shown in \citet{Maji2022}, consisting of objects from \citet{Leitet2011,Leitet2013,Borthakur2014,Pardy2016,Izotov2016a,Izotov2016b,Verhamme2017,Puschnig2017,Izotov2018a,Izotov2018b,Micheva2020,Izotov2021,Gazagnes2021}. The blue dots show low-redshift leakers from \citet{Izotov2022}.}
\label{fig:fesc_logLLya}
\end{figure}

Another possible connection between LyC and Ly$\alpha$ has recently been proposed by \citet{Maji2022}, who analyse simulated galaxies at high redshifts and find a strong correlation between the luminosities of the two due to their production being related to the star formation rate of the galaxy, as massive stars produce LyC photons, which in turn create Ly$\alpha$ photons. Here, we look at a possible connection between the LyC escape fraction and the Ly$\alpha$ luminosity in Fig.~\ref{fig:fesc_logLLya}. We do not detect any trend, although there seems to be one for low-redshift leakers (see the figure caption and also the LzLCS, \citealp{Flury2022b}).

\subsection{Potential limitations of using Ly$\alpha$}

The question remains why these correlations do not seem to work well for LyC leakers at higher redshifts while they are sufficiently reliable at lower redshifts (e.g. \citealp{Verhamme2017,Flury2022b}). One caveat is that although there has been a lot of progress in the field in recent years, the number of confirmed and robust LyC leakers at higher redshifts is still small and we often need to rely on stacking analyses. An advantage of stacking is that we can statistically correct for the IGM absorption, which is one of the largest uncertainties for studies of individual LyC leakers.

The assumed IGM absorption is only one aspect of determining the LyC escape fraction, however. When comparing escape fractions from different papers, like in Figs. \ref{fig:fesc_peak_sep_EW}, \ref{fig:f_esc_rel} and \ref{fig:fesc_logLLya}, it has to be kept in mind that the assumptions that went into the escape fraction can vary widely, not only when it comes to the IGM absorption, but also concerning the intrinsic ratio between UV continuum and LyC, as well as the age, metallicity and dust attenuation. Furthermore, the samples might not be representative of the global population of LyC leakers, as often only the brightest ones can be detected. Another caveat is differences in resolution limits, which make it difficult to compare trends with peak separation. In the MUSE data, we can go down to a peak separation of $\sim100 \, \m{km/s}$, however, the smaller the peak separation, the harder it is to distinguish a double peak from a single peak. 
Keeping these difficulties in mind, we can now think about physical reasons why the trends with escape fraction observed at low redshifts do not seem to hold for our sample.

The galaxies at redshifts around $z\approx3-4.5$ are likely to have different properties than the ones at low redshift. The discrepancy between our study and the results from low-redshift LyC leakers might indicate that we are not comparing the same kinds of objects after all. For the current study, we use a catalogue of LAEs that was produced from a blind emission-line search without pre-selection for any galaxy properties, unlike studies like for example the LzLCs (\citealp{Wang2021,Flury2022a,Flury2022b,Saldana-Lopez2022}), which are aimed specifically at detecting LyC. One known difference, for example, is that the morphological properties of high-redshift LAEs are more irregular, which could mean that the kinematics and ISM properties are more complex (e.g. \citealp{Guaita2015}) than the often compact low-redshift leakers. A prominent example of a local LyC leaker with a complex morphology is Haro 11 (e.g. \citealp{Bergvall2006,Keenan2017,Rivera-Thorsen2017}), which has three distinct clumps, two of which show no LyC escape. However, all three show strong Ly$\alpha$ emission, with the highest Ly$\alpha$ escape coming from the clump which is likely to be the one leaking the LyC emission as well. There have been indications that this object also has a complex velocity field, indicating a possible merger (e.g. \citealp{Ostlin2015}). This shows that without detailed knowledge of the morphology and the origin of the Ly$\alpha$ and LyC emission, it is not easy to draw a connection between the two. 
At higher redshift, Ly$\alpha$ emission does not always come from one single object, but there could be clumps or two objects close by. Sometimes this can even create the impression of a double peak, although both lines come from different clumps (see Vitte et al. in prep, using data from MUSE as well). For objects that do not show a clear connection between Ly$\alpha$ and LyC, this could mean that the two are escaping from different locations in the galaxy. Indeed, when looking at the positions of Ly$\alpha$ and the LyC contours in Figs. \ref{fig:rgbs1} and \ref{fig:rgbs2}, there seems to be an offset for some of the gold candidates (roughly around $0\farcs3 - 0\farcs5$ between the respective SN peaks) and the LyC emission is typically not centred on the Ly$\alpha$ emission. Another indicator for this is, that even for the objects where we do not find a double-peaked Ly$\alpha$ line, the lines nevertheless look asymmetric, indicating scattering or expanding gas, which disfavours an ionised channel and again hints at different origins of the LyC and Ly$\alpha$. 
It has to be kept in mind, that we do not see the overall LyC escape fraction of a galaxy, but only the escape in our line of sight. It is possible that only a few sightlines have visible LyC emission, which would also explain why LyC leakers are so rare. Ly$\alpha$ emission however is scattered and there can be a combination of different sources which form the Ly$\alpha$ line, as for example \citet{Ji2020} argue that the fact they see Ly$\alpha$ in absorption in the LyC leaker Ion 1 could be caused by a combination of emission originating from the same place as the LyC photons and absorption.

If however there is not enough neutral hydrogen to produce Ly$\alpha$ in the first place, this could also explain low Ly$\alpha$ equivalent widths for high LyC escape fractions. The reverse, a high Ly$\alpha$ equivalent width but no LyC leakage, could occur in the case of a higher neutral hydrogen density in a clumpy medium, allowing Ly$\alpha$ to scatter to escape the galaxy, while LyC cannot and will be absorbed in the clumps (see \citealp{Neufeld1991}, but brought into question by \citealp{Gronke2014, Duval2014}). Another reason might be that the LyC emission could already have been transformed into Ly$\alpha$ in dense neutral hydrogen regions. Indeed, all Ly$\alpha$ emission we see is LyC emission that has been lost already, usually in the first $10\,\m{pc}$ around the stellar population where it was produced (e.g. \citealp{Paardekooper2015}).

A high LyC escape fraction in objects with small Ly$\alpha$ equivalent width can also be explained without morphological arguments. It is often assumed that extreme stellar populations are needed for a high production of LyC and a high Ly$\alpha$ equivalent width ($>240\,\angstrom$). However, it has to be kept in mind that the LyC escape depends just as much on the foreground opacity as on the stellar population properties. Indeed, \citet{Reddy2022} argue that changes in the stellar population alone cannot explain high escape fractions and only lead to moderate changes in the intrinsic LyC photon production. 
Instead, a high LyC escape fraction with small Ly$\alpha$ equivalent width could indicate a neutral hydrogen column density low enough ($\m{N}_{\m{HI}}<10^{17}\,\m{cm}^{-2}$) for LyC to escape more easily than Ly$\alpha$, which could be scattered out of the line of sight. 
A prominent example of a LyC leaker which even has Ly$\alpha$ in absorption is Ion 1 (e.g. \citealp{Vanzella2010a,Ji2020}). \citet{Ji2020} argue that since the optical thickness of Ly$\alpha$ photons is $\approx 10^4$ larger than that of LyC photons, the latter can escape more easily depending on the neutral hydrogen column density.

Therefore, there might be holes with low neutral hydrogen column densities in the ISM through which LyC could escape, even if most of the produced Ly$\alpha$ emission is scattered in the rest of the galaxy. LyC could have carved out ionised paths (as shown e.g. in \citealp{Mainali2022} for the sunburst arc) or they could be created by supernova-induced outflows and holes (e.g. \citealp{Clarke2002,Fujita2003}). \citet{Ji2020} even suggest the presence of escape channels with no neutral hydrogen, which would hinder the production of Ly$\alpha$ emission.
Another possible example of this is the low-redshift LyC leaker Tololo 1247-232 (e.g.\ \citealp{Leitet2013, Leitherer2016}), which seems to consist of multiple stellar populations with different ages. The older population of $12\, \m{Myr}$ could have cleared a path for the LyC emission to escape, while the younger component of $2\--4\, \m{Myr}$ could provide most of the ionising emission (\citealp{Micheva2018b}).

Another promising indicator of LyC escape is therefore the Ly$\alpha$ emission at the systemic redshift (measured through non-resonant lines), as shown by \citet{Naidu2022}. Unfortunately, we do not have other spectral lines for most of our sample to determine systemic redshifts. 
Another way in which emission at systemic would manifest itself, even if some or most of the Ly$\alpha$ emission was scattered in a high neutral hydrogen column density, is by showing an additional peak in the middle between the other two, thus creating a triple-peaked line. However, the resolution and depth of the spectroscopic data for our current sample of LAEs are not sufficient to find triple peaks and also result in rather large error bars on Ly$\alpha$ properties. 

Another aspect that should be mentioned, is the possibility of interlopers contaminating our sample. As discussed in Sect. \ref{Identification of LyC Counterparts}, we take several measures to avoid low-redshift interlopers and contamination from close neighbours, but there is a remaining possibility that some of our objects could be contaminated. Our candidate with the highest LyC over UV continuum ratio (ID 1181371) for example has a neighbour in the filter bands ACS F606W and F775W (see Fig.~\ref{fig:overview_goldsample}), that could be the source of the Ly$\alpha$ emission. The potential contaminant is at a distance of $\approx 0\farcs5$, though, which makes us confident that it is not connected to the Ly$\alpha$ emission.
Having otherwise excluded objects with close neighbours, objects where the rgb images indicate different colours for different parts of the objects (hinting at a chance alignment), objects that have a too strong signal in the shorter wavelength band WFC3/UVIS F275W, as well as objects where the potential LyC emission and the UV emission do not overlap significantly, we are confident that most if not all of our candidates are robust.


\subsection{Expected LyC leaker fraction} \label{Expected LyC leaker fraction}

Instead of deriving the LyC escape fraction from the individual objects, we now want to turn the question around and derive the expected number of LyC leakers that we could potentially find in our survey area, given a certain assumed global escape fraction. This way, we can compare the actual number of LyC leaker candidates to predictions for different escape fractions and thus derive a global escape fraction.

\begin{figure}
\centering
\includegraphics[width=0.5\textwidth]{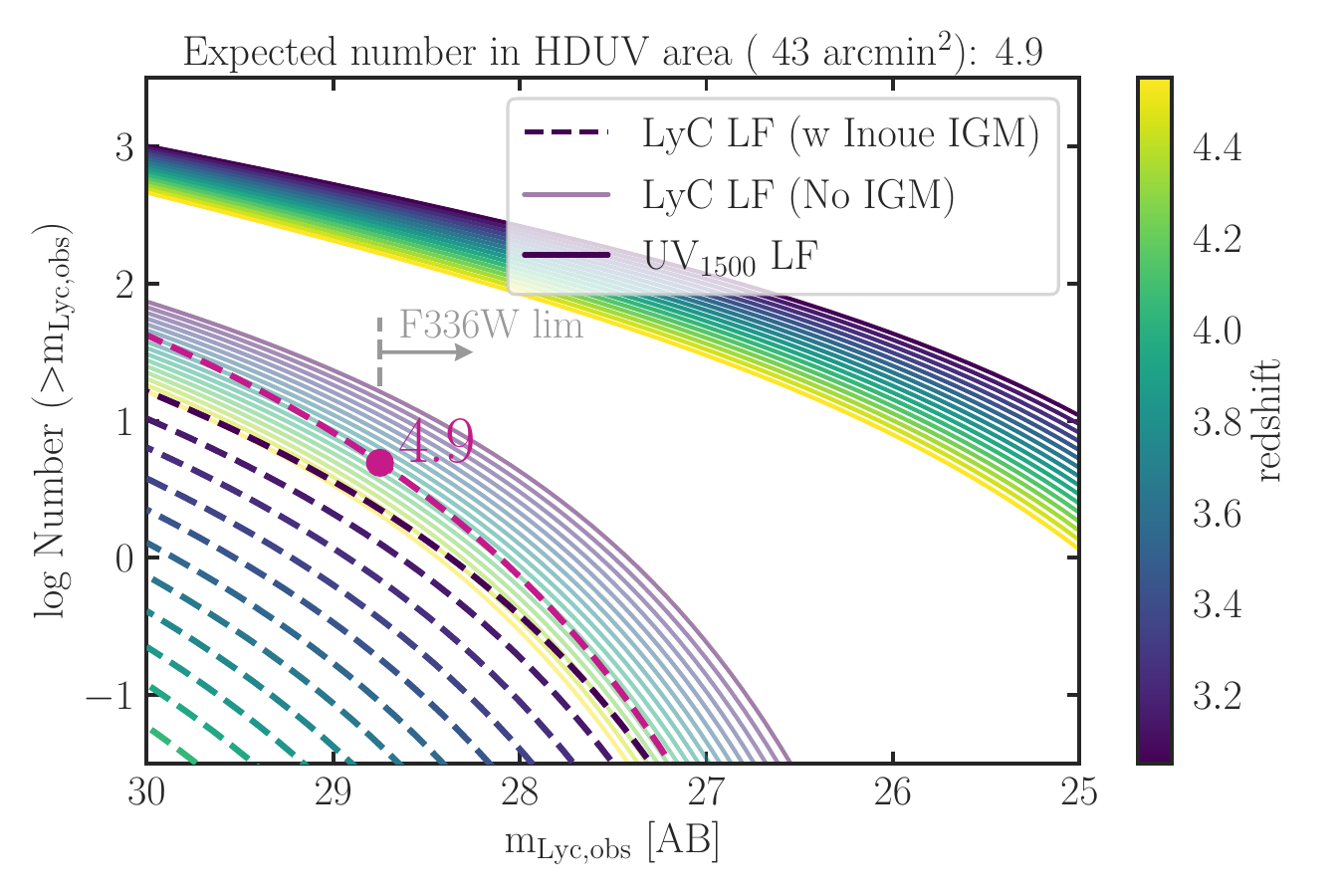}
\caption{Expected number of LyC leakers (logarithmic, y-axis) versus observed LyC AB magnitude. The thick, solid lines show the UV LF (the number of objects expected in the HDUV area at the given redshift), while the thin, solid lines show the predicted LyC LF, both colour-coded for different redshifts (see colour bar). The dashed lines indicate the expected number at different redshifts, using IGM transmission lines from \citet{Inoue2014} and assuming an escape fraction of $12\%$. The colours show redshifts from 3.0 to 4.5 in redshift bins of $\Delta z = 0.1$, which are all added up for the pink dashed line. The pink dot marks the expected number of LyC leaker candidates ($4.9$) for the HDUV area in GOODS-S (with a size of $43$ arcmin$^2$). This is given for the depth in the HST band F336W of $\approx28.75\,\m{mag}_{\m{AB}}$, which corresponds roughly to a $3\sigma$ detection.}
\label{fig:hduv_nlycexp_v0}
\end{figure}

For this approach, we first construct a LyC luminosity function. The basis for this is a UV luminosity function of the typical \citet{Schechter1976} shape:

\begin{equation}
 \phi(L) \, \m{d}L   = \phi^* \, \left( \frac{L}{L^*}\right) ^{\alpha} \, \m{e}^{-L/L^*} \, \frac{\m{d}L}{L^*}  
\end{equation}
With $\phi^*$ being the normalisation in comoving Mpc$^{-3}$, $L^*$ the characteristic luminosity in $\m{erg}\, \m{s}^{-1}$ and $\alpha$ the faint-end slope. Here we use the Schechter function in terms of absolute magnitude $ \phi(L) \, \m{d}L   = \phi(M_{\m{UV}}) dM_{\m{UV}}$, 
which can be rewritten using $\frac{L}{L^*} = 10^{-0.4(M_{\m{UV}}-M_{\m{UV}}^*)}$ as:

\begin{equation}
 \phi(M_{\m{UV}})  = \frac{\m{ln}10}{2.5} \phi^* 10^{-0.4(M_{\m{UV}}-M_{\m{UV}}^*)(\alpha+1)}  \exp \left( -10^{-0.4(M_{\m{UV}}-M_{\m{UV}}^*)} \right)
\end{equation}

where $M_{\m{UV}}^*$ is the characteristic absolute magnitude.
For the UV LF (thick lines in Fig.~\ref{fig:hduv_nlycexp_v0}) we use values from \citet{Bouwens2021} in steps of $\Delta z = 0.1$ (starting at $z=3.05$) to make a redshift grid (shown in different colours in Fig.~\ref{fig:hduv_nlycexp_v0}). 

To convert the UV LF to the LyC LF, we assume an intrinsic ratio of UV continuum to LyC luminosities of $L_{\m{UV}}/L_{\m{LyC}} = 3$ (in frequency space), see the thin lines in Fig.~\ref{fig:hduv_nlycexp_v0}.

Including the effects of IGM absorption using the transmission lines from \citet{Inoue2014} shifts the luminosity function further down (see dashed lines in Fig.~\ref{fig:hduv_nlycexp_v0}). With this function, we can predict the number of observable LyC leakers for a given survey size and escape fraction at different redshifts by integrating the luminosity function (thick pink dashed line and pink dot in Fig.~\ref{fig:hduv_nlycexp_v0}).

Taking the information that we find 5 LyC leaker candidates up to a redshift of $z=4.5$, for a survey area of 43 armin$^2$, and a detection limit for a $3\sigma$ detection in the HST band WFC3/UVIS F336W of $28.75\, \m{mag}_{\m{AB}}$, we adjust the assumed escape fraction until we match the number of detected LyC leakers. We find that for a global escape fraction of $12\%$, we can reproduce the number of $\sim 4.9$ LyC leaker candidates, which is close enough to our five candidates. From this simple model and assuming the intrinsic UV to LyC continuum flux ratio is accurate on average, we can thus conclude that the average escape fraction of our sample is at least $\approx 12\%$ (assuming a distribution of line of sight IGM transmissions based on \citealp{Inoue2014}). 
Incidentally, this matches well with the $1\sigma$ limit found in a stack of LyC non-detections of LBGs by \citet{saxena2022}.

For our silver sample, we set the detection limit to $2\sigma$ and our faintest object has an AB magnitude of $\m{m}_{\m{AB}} = 29.2$. If we use this magnitude as a limit in Fig.~\ref{fig:hduv_nlycexp_v0}, we get a number of $11.9$ expected LyC leakers, meaning seven additional ones, which again matches well with our sample of seven silver candidates. 
In a similar vein, \citet{Begley2022} predict the expected number of LyC leaker candidates among their sample from the average escape fraction they find in their stack star-forming galaxies at $z\simeq3.5$, which matches well with their individual detections for $f_{\m{esc}}=0.07\pm0.02$.
One caveat to mention is the influence of cosmic variance on our results. Since we are looking at a relatively small field of view and low number statistics of LyC leakers, the resulting escape fraction from such a calculation could be different for different fields. However, our calculation shows that it is plausible that galaxies at and after the EoR could be leaking LyC emission with an average escape fraction of $12\%$. However, most of them are not visible to us in LyC due to the absorption in the IGM. Incidentally, recently \citet{Mascia2023} find an inferred average escape fraction of $12\%$ (based on correlations with the \ion{O}{III}/\ion{O}{II} ratio and $\beta$ parameter) in their sample of gravitationally lensed galaxies at $z=4.5-8$ in the first results from the GLASS-JWST program (\citealp{Treu2022}).
The distribution of escape fractions for individual galaxies however is not clear. Some studies favour reionisation by the majority of faint galaxies (e.g. \citealp{Finkelstein2019}), while others suggest that only a small minority of bright star-forming galaxies (such as the one discovered in \citealp{Marques-Chaves2021} or \citealp{Marques-Chaves2022}) could be responsible for the entire ionising photon emission from star-forming galaxies. However, \citet{Naidu2022}, the authors claim that half of the population of LAEs at $z\approx2$ has an escape fraction of $25-50\%$, which would mean that not only the brightest ones are responsible for the ionising emission.

Coming back to Fig.~\ref{fig:f_esc_rel}, where we show the result from our exercise as the pink dot, we can see that the estimated $12\%$ escape fraction is below most of the individual measurements at $z=2-4.5$, but above or similar to values measured from stacking and for lower redshifts of $z=0-1$. It is also close to the median value of the upper limits for the non-detections. This suggests that individual measurements are biased towards higher escape fraction values.

\section{Summary and Conclusions} \label{Summary and Conclusions}

We have identified five very likely LyC leaker candidates (that we have called our "gold" sample), four of which are new, among a sample of LAEs previously detected by MUSE in the CDFS region, with an additional seven potential candidates (that we have called our "silver" sample). We measured their LyC emission in the WFC3/UVIS F336W filter from the HDUV survey (\citealp{Oesch2015,Oesch2018}) and performed our own photometry in several filters and SED fitting with the CIGALE software (\citealp{Burgarella2005,Noll2009,Boquien2019}). 

We assume that the ionising photons escape preferably through ionised channels that are low in dust due to the high gas temperature (making our escape fraction measurements from CIGALE sightline-dependent). In addition, assuming a high transmission through the IGM (using the highest $5\%$ of the distribution of \citealp{Inoue2014}), we find escape fractions ranging from $22\%$ to $88\%$ in the redshift range $z=3.08$ to $z=4.43$, one of the highest redshifts for a LyC leaker found so far. 
We used our knowledge of the Ly$\alpha$ line properties from the MUSE spectra to check several previously proposed indications of LyC leakage, such as FWHM, peak separation, equivalent width and Ly$\alpha$ luminosity. We found no reliable correlation with any of these values, even though especially the peak separation has proven to be a good indicator for low-redshift LyC leakers. We argue that this could be explained by large uncertainties on the measured escape fractions, as well as the dependence on the underlying assumptions concerning dust extinction in the ISM, absorption in the IGM and the intrinsic ratio between UV and LyC. It is also possible that LyC leakers at higher redshifts have different properties or production mechanisms of LyC photons than lower-redshift analogues. 

We show a method to infer the global escape fraction from the number of found LyC leaker candidates, by integrating the IGM corrected LyC LF down to the depth of our data, for which we use the UV LF (\citealp{Bouwens2021}) and assume an intrinsic UV to LyC ratio. Taking our five LyC leaker candidates from the gold sample, we thus derive a global escape fraction of $f_{\m{esc}}=12\%$, which would be enough to reionise the universe. From this exercise, we also see that in order to detect a significant number of reliable LyC emitters, we need to go an order of magnitude deeper in the WFC3/UVIS F336W band.

In this paper, we have focused on the detection of individual LyC leaker candidates, while also comparing to literature results from stacking (see Fig.~\ref{fig:f_esc_rel} and table \ref{tab:LyC_leakers_stacks} in the appendix). A possible next step to determine the global LyC escape fraction from our sample of LAEs is stacking analysis, either by stacking the WFC3/UVIS F336W cutouts to detect LyC from the photometry (see Oesch et al. in prep.) or by stacking the spectra of the LAEs from MUSE. This would reduce the possible redshift range for direct detection since the LyC enters the MUSE wavelength only at a redshift of $z>4$. However, a stacking analysis of LAE spectra can still give us information on UV lines that could indicate both the ionising photon production as well as the ISM properties, which are both connected to the production and escape of LyC emission (see Kramarenko et al. in prep.). For example, recent studies have suggested that the [\ion{O}{III}]/[\ion{O}{II}] ratio between [\ion{O}{III}] line(s) at $5007$ and $4959\, \angstrom$ to [\ion{O}{II}] at $3727\, \angstrom$ could be a good indicator of a strong ionisation field and low optical depths (proposed by \citealp{Jaskot2013}, see also e.g.\ \citealp{Nakajima2014, Nakajima2016, Faisst2016, Izotov2018a, Katz2020} but \citealp{Stasiska2015,Naidu2018,Bassett2019} for counter-examples). Another potential line would be \ion{Mg}{II} $\lambda\lambda 2796, 2803$ (\citealp{Henry2018,Katz2022}), which correlates strongly with Lyman~$\alpha$ emission and could be used as a substitute if Ly$\alpha$ is not observable. Similarly, $\ion{C}{III}] \lambda \lambda 1907, 1909$ (\citealp{Jaskot2016}) and $\ion{C}{IV} \lambda 1550$ (\citealp{Schaerer2022}) trace low metallicities and density-bounded regions of ionised hydrogen. Not only emission but also residual flux in saturated low-ionisation interstellar absorption lines (e.g.\ \citealp{Borthakur2014,Mauerhofer2021}) can indicate the state of the ISM and thus potential LyC emission. 

While LyC emission remains impossible to detect at the epoch of reionisation, we are gaining new insights into this era through various JWST programs. Through resolving Ly$\alpha$ emission and studying the morphology, kinematics and offset to the systemic redshift, analysing H$\alpha$ emission and obtaining Ly$\alpha$ escape fractions, and better knowledge of the spectra of galaxies at redshifts $z>6$, surveys such as the "First Reionization Epoch Spectroscopically Complete Observations" (FRESCO, \citealp{Oesch2023}), the "Cosmic Evolution Early Release Science" (CEERS, \citealp{Finkelstein2023}), or GLASS-JWST-ERS (\citealp{Treu2022}) are already well underway to shedding more light on this critical phase in the history of the universe. 


\section{Acknowledgements}
We would like to thank our colleagues for many interesting discussions and comments that helped greatly to improve this paper, among which are Rui Marques-Chaves, Ivan Kramarenko, Moupiya Maji, Laia Barrufet, Sophie van Mierlo, Vasily Kokorev, Edoardo Iani and Emil Rivera-Thorsen. 
The plots in this paper were created using \texttt{Matplotlib} (\citealp{Hunter2007}) and we also used the Python packages \texttt{Numpy} (\citealp{vanderWalt2011}) and \texttt{Astropy} (\citealp{Astropy2013}).
This work received support from the Swiss National Science Foundation through grant PP00P2\url{_}190079.  This work was supported by the Swiss State Secretariat for Education, Research and Innovation (SERI) under contract number MB22.00072.
The Cosmic Dawn Center (DAWN) is funded by the Danish National Research Foundation under grant No.\ 140. 
GI acknowledges support for this work provided by NASA through grant HST-GO-13872 awarded by the Space Telescope Science Institute, which is operated by the Association of Universities for Research in Astronomy, Inc., under NASA contract NAS 5-26555.
HA is supported by CNES, focused on the HST mission, and the Programme National Cosmology and Galaxies (PNCG) of CNRS/INSU with INP and IN2P3, co-funded by CEA and CNES. 
Support for this work was provided by NASA through the NASA Hubble Fellowship grant HST-HF2-51515.001-A awarded by the Space Telescope Science Institute, which is operated by the Association of Universities for Research in Astronomy, Incorporated, under NASA contract NAS5-26555.
MM acknowledges support from the Project PCI2021-122072-2B, financed by MICIN/AEI/10.13039/501100011033, and the European Union “NextGenerationEU”/RTRP and IAC project P/302302.
VM acknowledges support from the NWO grant 016.VIDI.189.162 (``ODIN").


\bibliographystyle{aa}
\bibliography{mybib}


\begin{appendix}

\section{Additional information on the LyC leaker candidates}

\begin{figure}[h]
\centering
\begin{minipage}{.24\textwidth}
  \centering
  \includegraphics[width=\linewidth]{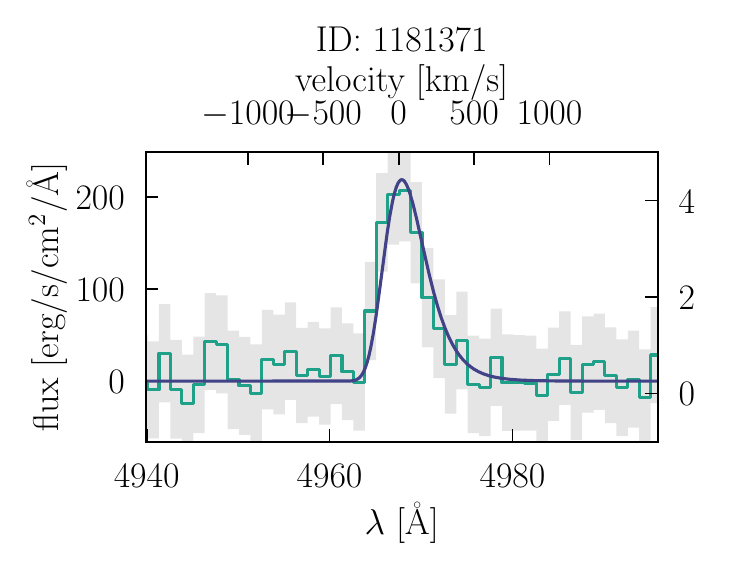}
  \includegraphics[width=\linewidth]{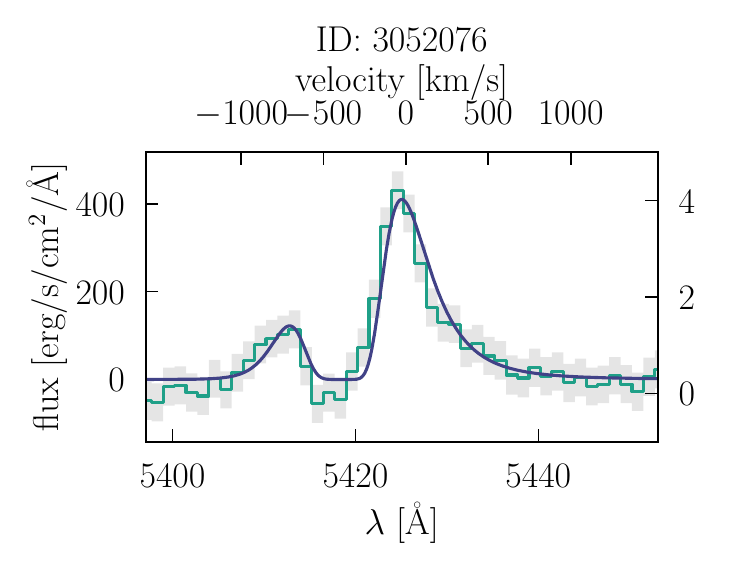} 
  \includegraphics[width=\linewidth]{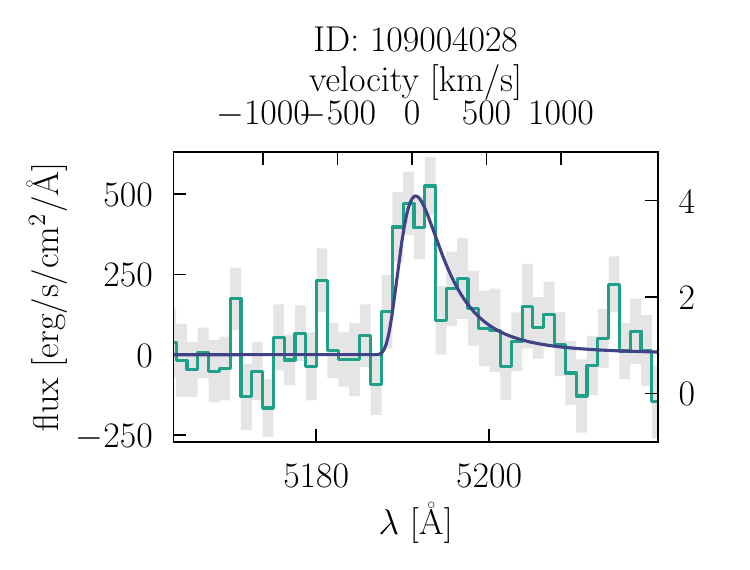}
\end{minipage}
\begin{minipage}{.24\textwidth}
  \centering
  \includegraphics[width=\linewidth]{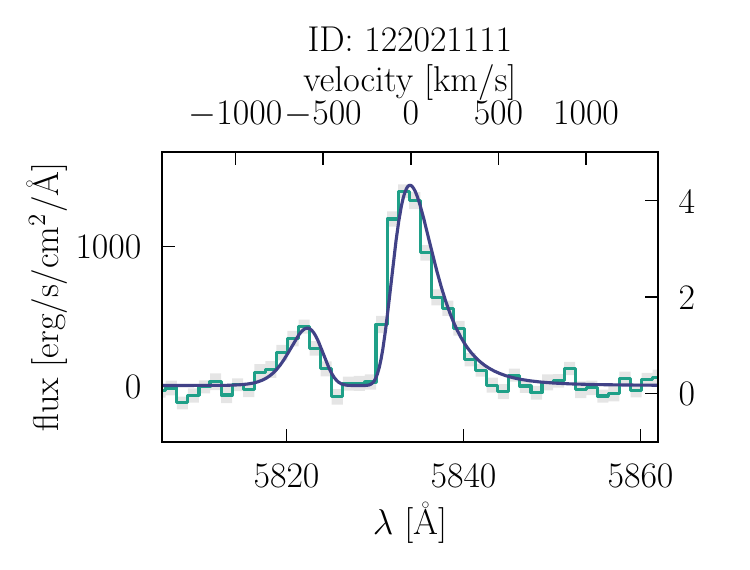}
  \includegraphics[width=\linewidth]{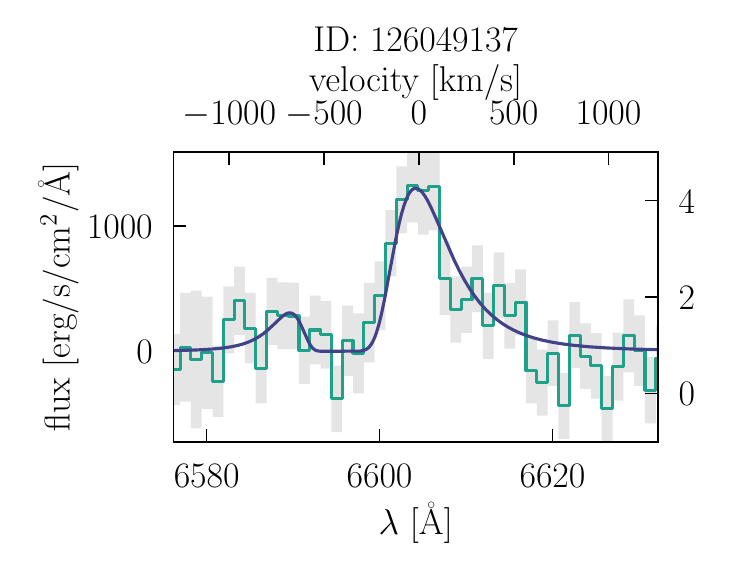}
  \includegraphics[width=\linewidth]{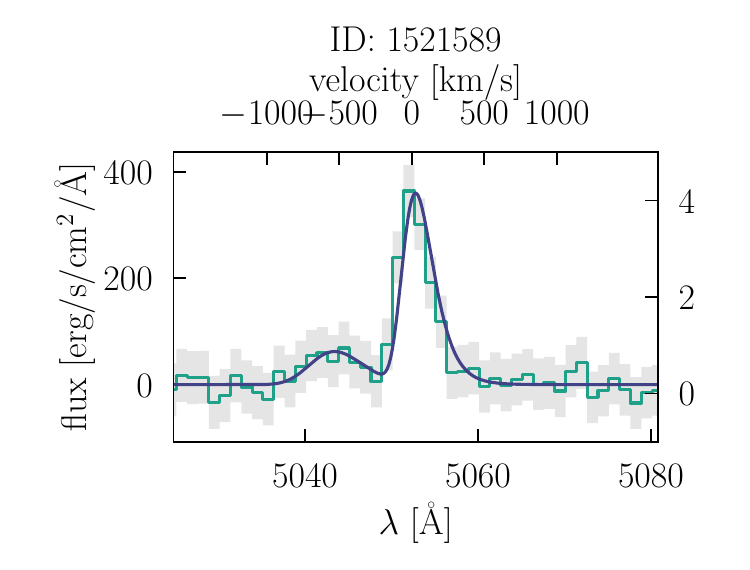}
\end{minipage}
\begin{minipage}{.24\textwidth}
  \centering
  \includegraphics[width=\linewidth]{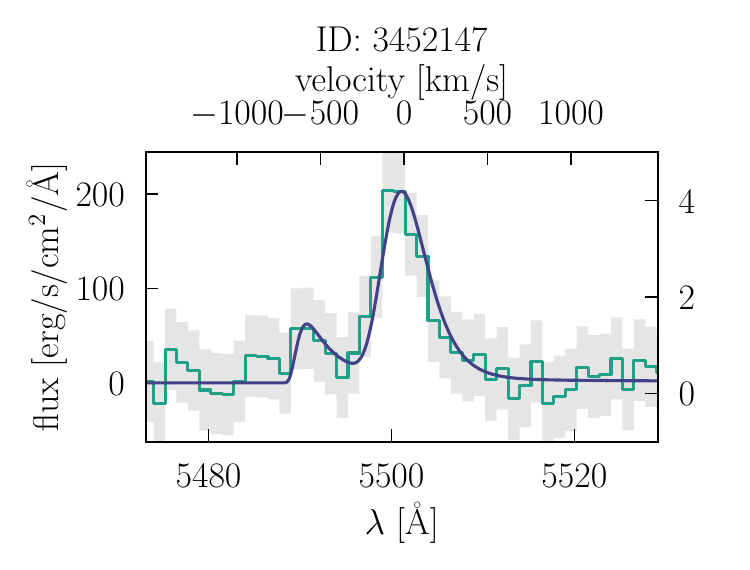}
  \includegraphics[width=\linewidth]{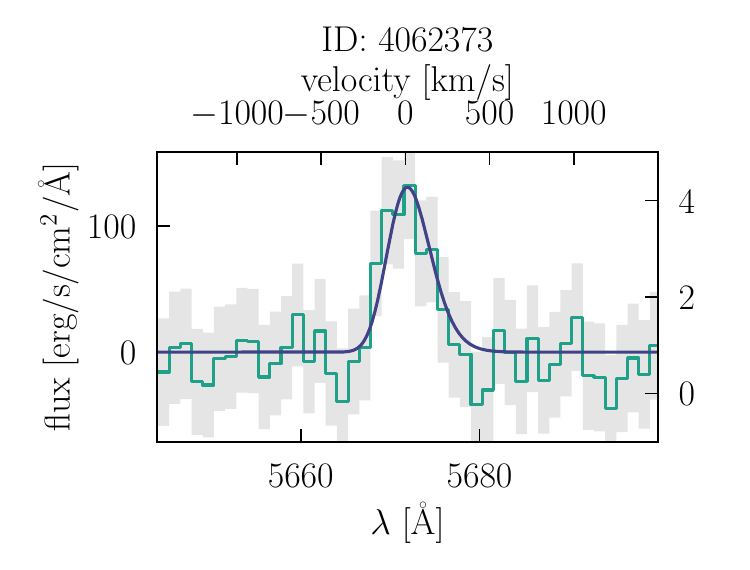}
  \includegraphics[width=\linewidth]{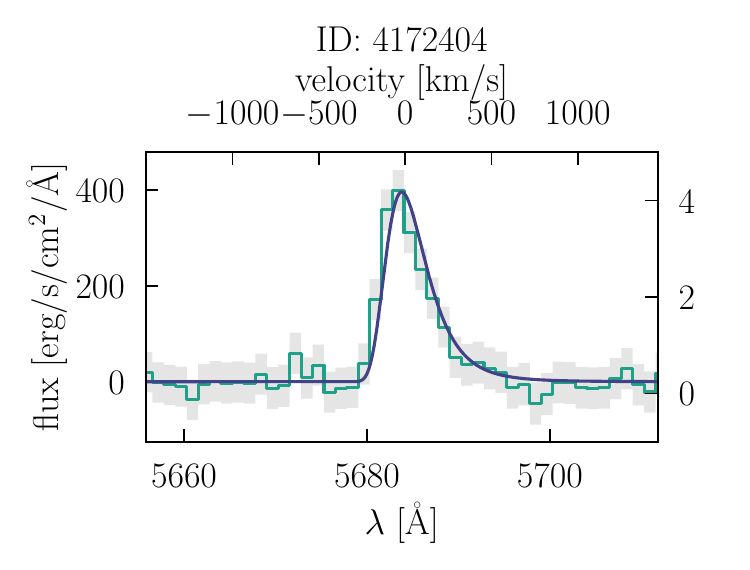}
\end{minipage}
\begin{minipage}{.24\textwidth}
  \centering
  \includegraphics[width=\linewidth]{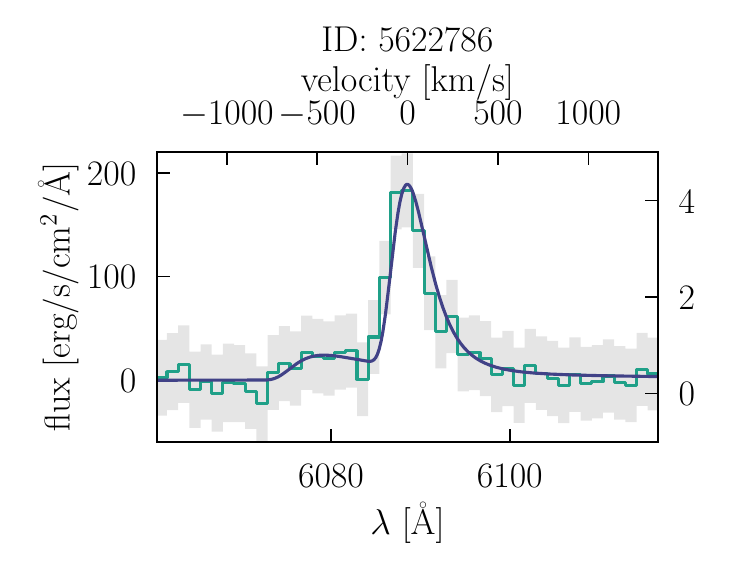}
  \includegraphics[width=\linewidth]{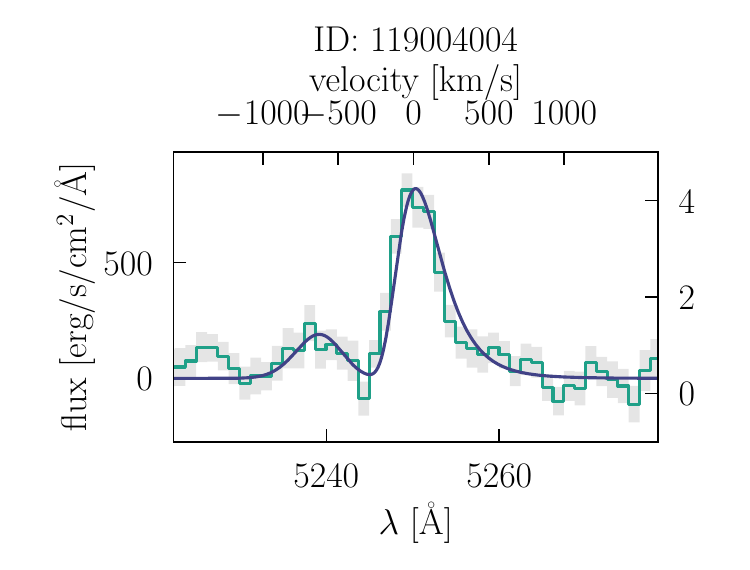}
  \includegraphics[width=\linewidth]{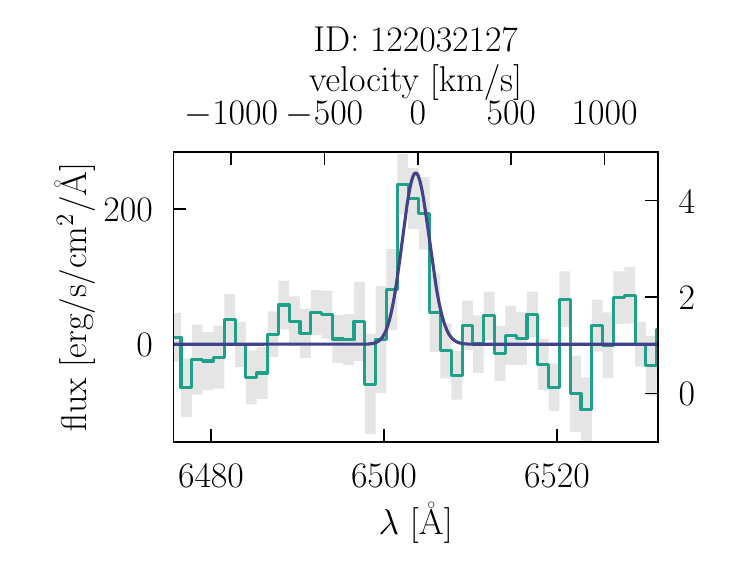}
\end{minipage}
\caption{Spectra from MUSE-Wide for the sample of LyC leaker candidates. The green line shows the data and the grey area shows the error range. The blue line shows the fit to the line consisting of one or two asymmetric Gaussians (see \citealp{Shibuya2014b} and \citealp{Kerutt2022}). Note that while the x-axis always shows a range of $50\,\angstrom$ around the Ly$\alpha$ line, the y-axis range scales with the amplitude of the line.}
\label{fig:MUSE_spectra2}
\end{figure}

\begin{figure*}
\centering
  \includegraphics[width=\linewidth]{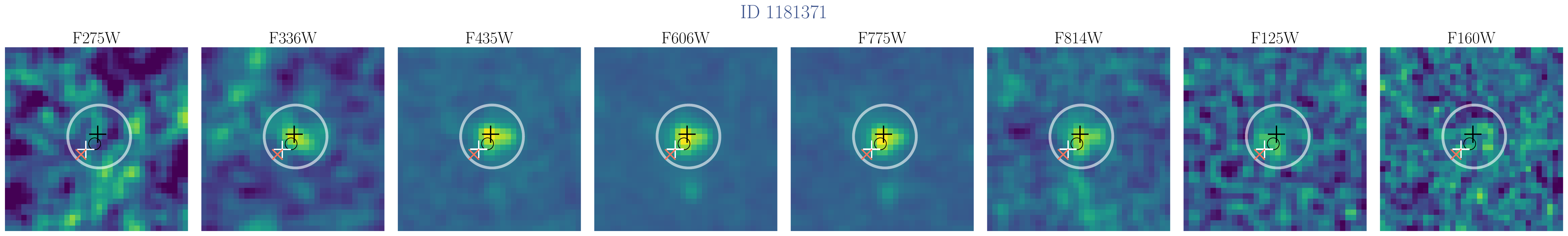}
  \includegraphics[width=\linewidth]{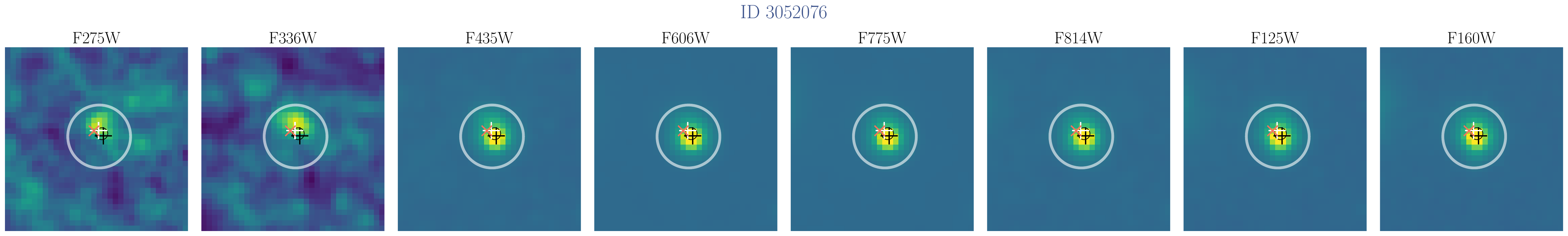}
  \includegraphics[width=\linewidth]{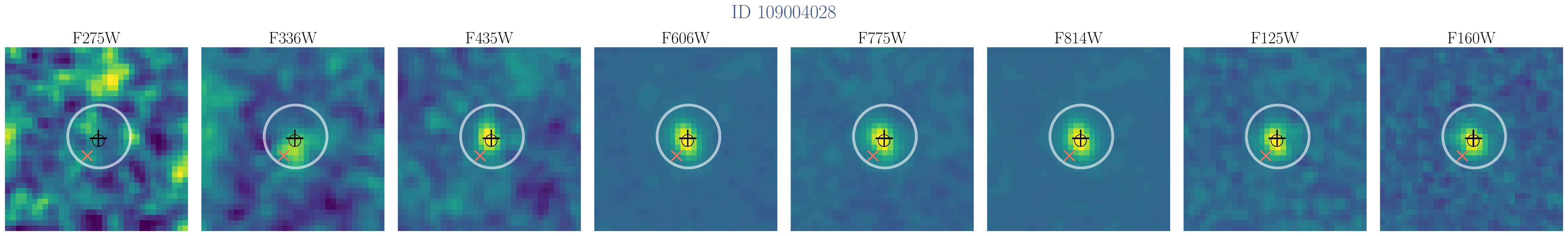}
  \includegraphics[width=\linewidth]{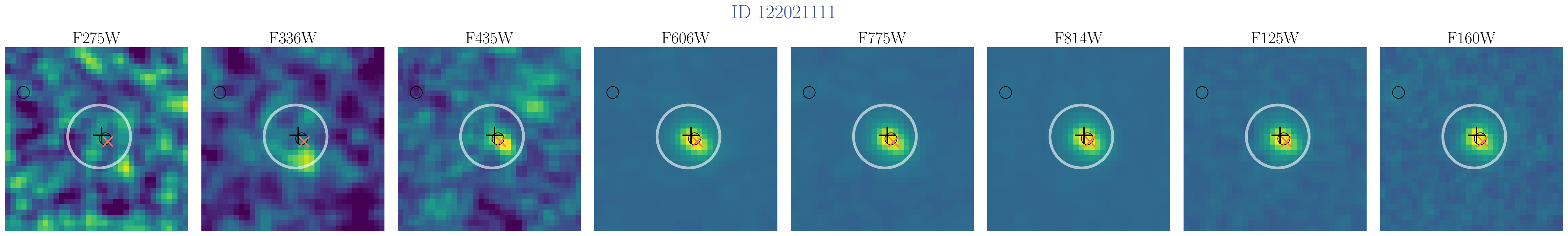}
  \includegraphics[width=\linewidth]{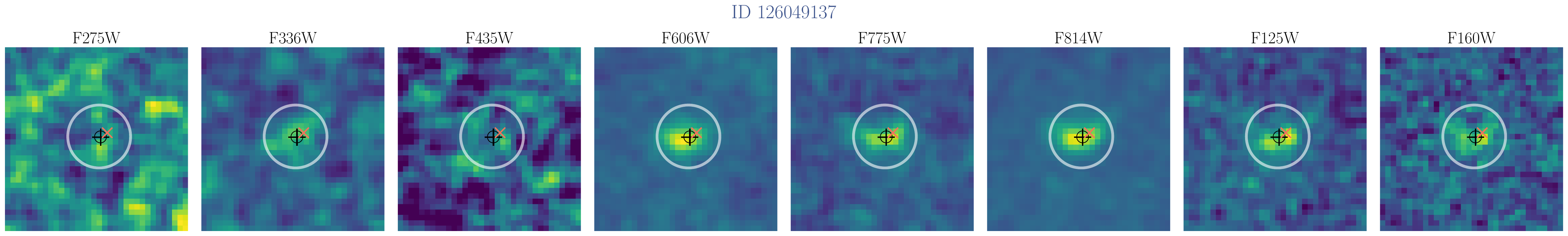}
\caption{Cutouts of different photometric bands for the gold sample objects (the IDs are written on the top of each row). From left to right, the photometric bands are the two HDUV images WFC3/UVIS F275W and F336W, followed by the HLF images ACS F435W, F606W, F775W, F814W and WFC3/IR F125W and F160W. Each cutout has a size of $2\farcs0$ on the side. The white circle has a radius of $0\farcs35$ and is centred on the MUSE-Wide position, indicated also by a black cross. The orange x marks the position of the highest SN in Ly$\alpha$. Black small circles indicate the positions of objects in the \citet{Skelton2014} catalogue, and white crosses indicate MUSE-Deep positions.}
\label{fig:overview_goldsample}
\end{figure*}

\begin{figure*}
\centering
  \includegraphics[width=\linewidth]{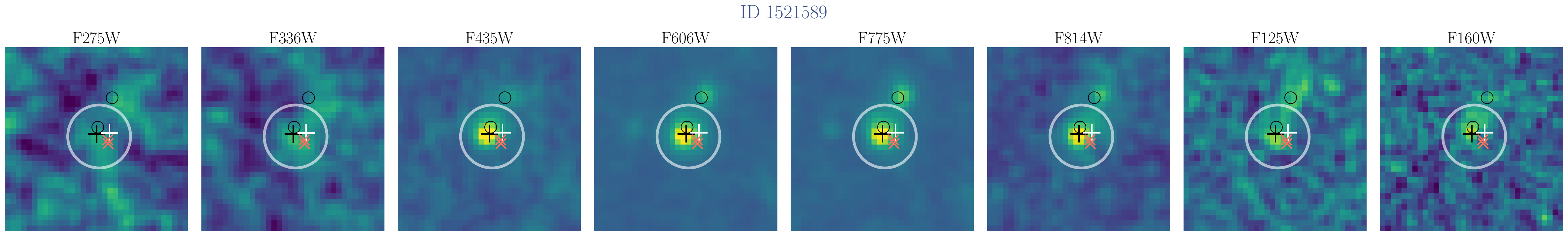}
  \includegraphics[width=\linewidth]{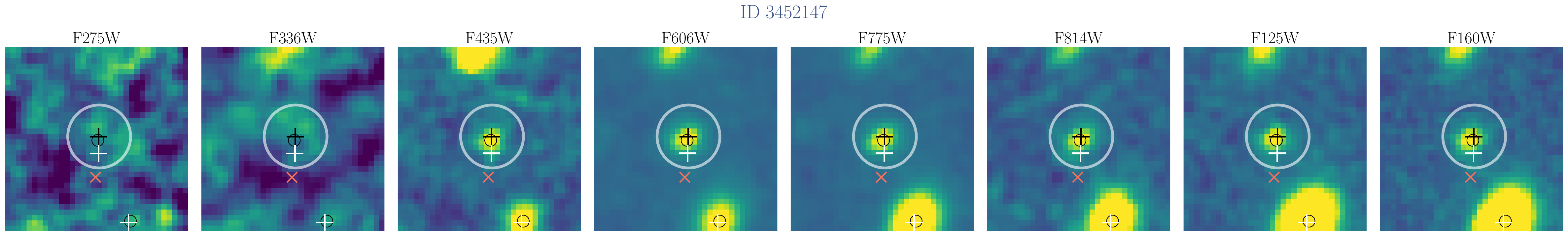}
  \includegraphics[width=\linewidth]{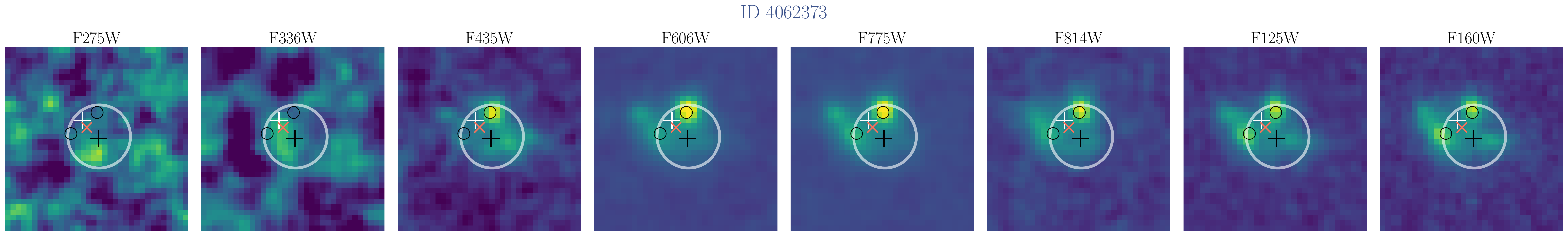}
  \includegraphics[width=\linewidth]{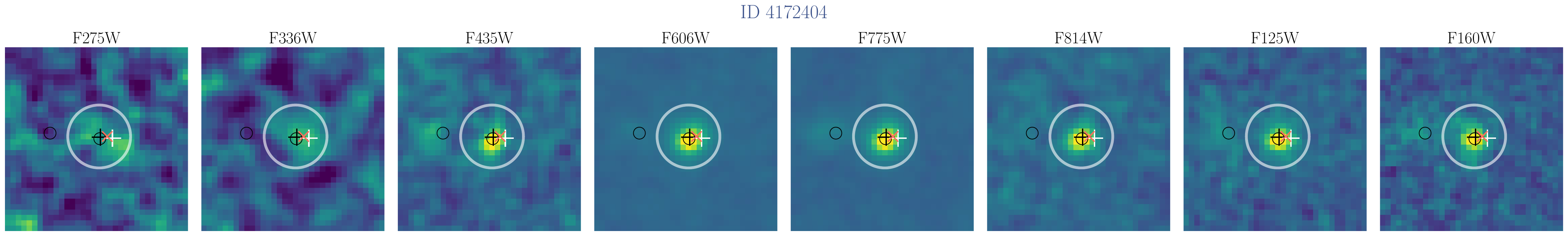}
  \includegraphics[width=\linewidth]{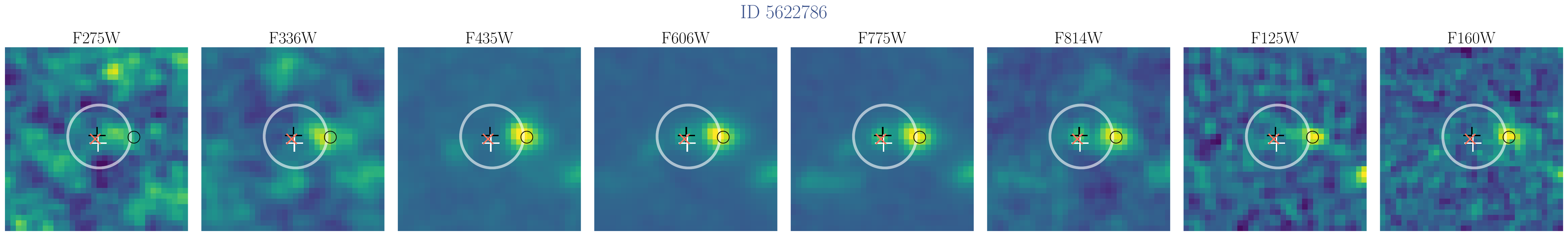}
  \includegraphics[width=\linewidth]{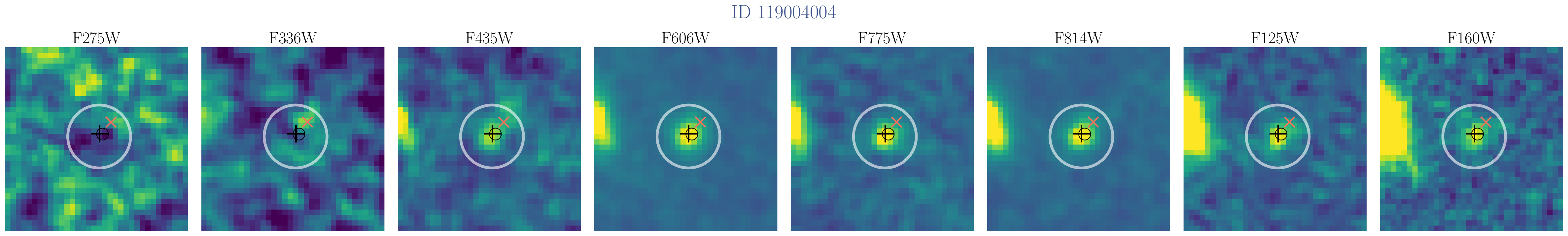}
  \includegraphics[width=\linewidth]{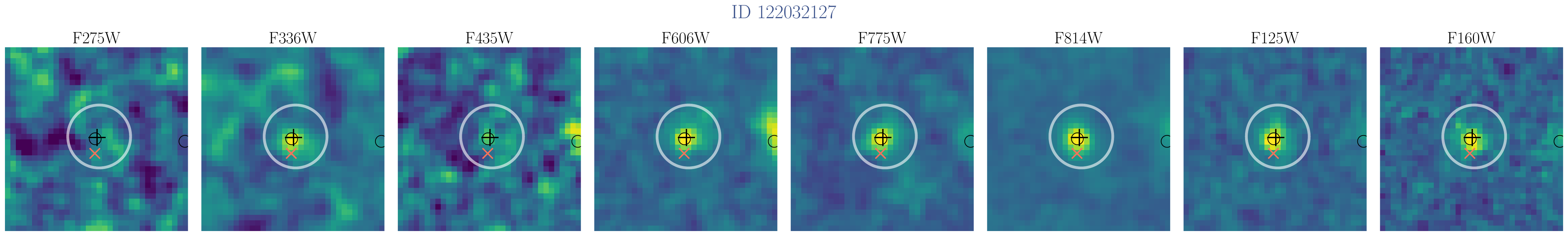}
\caption{Same as Fig.~\ref{fig:overview_goldsample}, but for the silver sample.}
\label{fig:overview_silversample}
\end{figure*}

This section contains the Ly$\alpha$ lines from MUSE for all LyC leaker candidates in Fig. \ref{fig:MUSE_spectra2} as well as a table with the results of the CIGALE SED fitting (table \ref{tab:CIGALE}) and cutouts in photometric bands from HST in Figs. \ref{fig:overview_goldsample} and \ref{fig:overview_silversample}.

\begin{table*}[h]
\begin{center} 
\caption{Overview of properties of LyC leaker candidates based on CIGALE SED fitting.} 
\begin{tabular}{  l l l l l l l } 
\hline\hline 
  ID$_{\m{MW}}$ & $\chi^2_{\mathrm{red.}}$ & E(B-V) & burst age & metallicity $Z$ & M$_{\star}$ & SFR \\ 
\bf{\textcolor{gold}{1181371}} & \ph\ph0.87 & \ph\ph0.11$\pm$ 0.02 & \ph\ph1$\pm$ 1 & \ph\ph0.08$\pm$ 0.09 & \ph\ph0.10$\pm$ 0.04 & \ph18.08$\pm$ 8.6  \\ 
\bf{\textcolor{gold}{3052076}} & \ph\ph0.80 & \ph\ph0.20$\pm$ 0.0 & \ph24$\pm$ 5 & \ph\ph0.26$\pm$ 0.09 & \ph\ph1.94$\pm$ 0.46 & \ph58.70$\pm$ 4.88  \\ 
\bf{\textcolor{gold}{109004028}} & \ph\ph0.94 & \ph\ph0.15$\pm$ 0.03 & \ph29$\pm$ 16 & \ph\ph0.44$\pm$ 0.3 & \ph\ph0.70$\pm$ 0.18 & \ph11.28$\pm$ 3.55  \\ 
\bf{\textcolor{gold}{122021111}} & \ph\ph0.64 & \ph\ph0.02$\pm$ 0.01 & \ph26$\pm$ 7 & \ph\ph0.72$\pm$ 0.3 & \ph\ph1.61$\pm$ 0.49 & \ph14.76$\pm$ 1.9  \\ 
\bf{\textcolor{gold}{126049137}} & \ph\ph0.98 & \ph\ph0.24$\pm$ 0.06 & \ph\ph2$\pm$ 1 & \ph\ph0.06$\pm$ 0.08 & \ph\ph1.54$\pm$ 0.84 & 257.36$\pm$ 176.84  \\ 
\bf{\textcolor{silver}{1521589}} & \ph\ph2.38 & \ph\ph0.04$\pm$ 0.02 & \ph\ph4$\pm$ 2 & \ph\ph0.15$\pm$ 0.15 & \ph\ph0.04$\pm$ 0.02 & \ph\ph6.60$\pm$ 5.98  \\ 
\bf{\textcolor{silver}{3452147}} & \ph\ph0.83 & \ph\ph0.18$\pm$ 0.04 & \ph21$\pm$ 17 & \ph\ph0.07$\pm$ 0.09 & \ph\ph0.36$\pm$ 0.17 & \ph\ph5.70$\pm$ 2.12  \\ 
\bf{\textcolor{silver}{4062373}} & \ph\ph1.03 & \ph\ph0.21$\pm$ 0.02 & \ph49$\pm$ 3 & \ph\ph0.28$\pm$ 0.105 & \ph\ph3.03$\pm$ 0.68 & \ph12.36$\pm$ 2.75  \\ 
\bf{\textcolor{silver}{4172404}} & \ph\ph1.63 & \ph\ph0.14$\pm$ 0.05 & \ph11$\pm$ 8 & \ph\ph0.04$\pm$ 0.07 & \ph\ph0.13$\pm$ 0.09 & \ph\ph7.42$\pm$ 2.04  \\ 
\bf{\textcolor{silver}{5622786}} & \ph\ph0.96 & \ph\ph0.30$\pm$ 0.02 & \ph\ph7$\pm$ 3 & \ph\ph0.28$\pm$ 0.13 & \ph\ph1.83$\pm$ 0.91 & \ph40.31$\pm$ 25.22  \\ 
\bf{\textcolor{silver}{119004004}} & \ph\ph0.87 & \ph\ph0.19$\pm$ 0.03 & \ph27$\pm$ 17 & \ph\ph0.10$\pm$ 0.095 & \ph\ph0.52$\pm$ 0.3 & \ph\ph9.46$\pm$ 2.63  \\ 
\bf{\textcolor{silver}{122032127}} & \ph\ph1.15 & \ph\ph0.13$\pm$ 0.05 & \ph\ph9$\pm$ 7 & \ph\ph0.24$\pm$ 0.155 & \ph\ph0.35$\pm$ 0.16 & \ph59.76$\pm$ 61.06  \\ 
\end{tabular}\label{tab:CIGALE}  
\end{center} 
\tablefoot{ID$_{\m{MW}}$: Identifier in \citet{Urrutia2019}, same as in table \ref{tab:goodss_Lya}. $\chi^2_{\mathrm{red.}}$: reduced $\chi^2$ value for SED fit. E(B-V): dust attenuation. burst age [Myr]: age of the starburst in Myr. metallicity: metallicity of the stellar population ($Z_{\odot} = 0.02$). M$_{\star}$ [$10^9 \mathrm{M}_{\odot}$]: stellar mass in $10^9 \mathrm{M}_{\odot}$. SFR: star formation rate.}  
\end{table*} 



\section{Overview of literature values}

In this section, we show tables with LyC leakers from the literature. Table \ref{tab:LyC_leakers_ind_high} gives an overview of individual LyC leakers at high redshift, table \ref{tab:LyC_leakers_stacks} shows escape fraction results from stacking and table \ref{tab:LyC_leakers_low} contains low redshift LyC leakers.

\begin{table*}
\begin{center}
\begin{tabular}{ | p{3.5cm} | p{3.2cm} | p{2.4cm} | p{2.8cm} | p{4.2cm} | }\hline

  object(s) & paper(s) & redshift & $f_{\m{esc,LyC}}$ & notes \\ \hline\hline
  
    C49 & \citet{Shapley2006,Siana2015} & $z=3.15$ & $f_{\m{esc,rel}}=65 \%$ & in the field SSA 22a, but reexamined by \citet{Siana2015} (foreground contamination) \\ \hline
    
    D3 & \citet{Shapley2006} & $z=3.07$ & $f_{\m{esc,rel}}\geq 100 \%$ & in the field SSA 22a \\ \hline
  
    Ion1 & \citet{Vanzella2010b, Vanzella2012, Vanzella2015, Vanzella2020, Ji2020} & $z = 3.794$ & $f_{\m{esc,abs}} = 5\pm2\%$, $f_{\m{esc,rel}}=32\pm11\%$ & from \citet{Ji2020}, who also stack 107 galaxies but find no LyC signal \\  \hline
    
    Ion2 & \citet{Vanzella2015, Vanzella2016, deBarros2016, Vanzella2020} & $z = 3.212$ & $f_{\m{esc,rel}}=64^{+1.1}_{-0.1} \%$ & from \citealp{deBarros2016} \\  \hline 

    Ion3 & \citet{Vanzella2018} & $z=4.0$ & $f_{\m{esc,rel}} \approx 60 \%$ & \\ \hline

    MD5b & \citet{Mostardi2015} & $z \approx 3.14$ & $f_{\m{esc,rel}}=75-100\%$ & follow-up of \citet{Mostardi2013}\\ \hline
    
  Q1549-C25 & \citet{Shapley2016} & $z = 3.15$ & $f_{\m{esc,abs}}>51 \%$ & \\  \hline
  
 Horseshoe & \citet{Vasei2016} & $z=2.38$ & $f_{\m{esc,rel}}<0.08$ & non detection in lensed galaxy \\ \hline

  A2218-Flanking & \citet{Bian2017} & $z = 2.5$ & $f_{\m{esc,abs}} >28\--57 \%$ & \\  \hline 
  
  GN-UVC-6 & \citet{Jones2018} & $z=2.439$ & & 6 candidates, one AGN, 4 contaminations \\ \hline
  
  Sunburst arc & \citet{Rivera-Thorsen2017a,Rivera-Thorsen2019,Vanzella2020,Mainali2022} & $z=2.37$ & $f_{\m{esc,rel}} = 93^{+7}_{-11}\%$ & lensed, values from \citet{Rivera-Thorsen2019} \\ \hline
  
  AUDFs01 & \citet{Saha2020} & $z=1.42$ & $f_{\m{esc,rel}} > 20 \%$ & observed with AstroSat, near the peak of star formation \\ \hline
 
  J0121+0025 & \citet{Marques-Chaves2021} & $z=3.244$ & $f_{\m{esc,abs}} \approx 40 \%$ & very luminous with a young star burst \\ \hline

  J1316-2614 & \citet{Marques-Chaves2022} & $z=3.613$ & $f_{\m{esc,abs}} \approx 90 \%$ & UV-bright starburst \\ \hline

\hline
\end{tabular}
\end{center}
\caption{Overview of some studies of individual LyC leakers at high redshift. The subscript rel. means relative escape fraction, abs. means absolute. If the information is not given in the subscript it is not clear from the paper. It has to be noted though that the definition of escape fraction also is not uniform, which explains some of the discrepancies.}
\label{tab:LyC_leakers_ind_high}
\end{table*}
 
\begin{table*}
\begin{center}
\begin{tabular}{ | p{3.5cm} | p{3.2cm} | p{2.4cm} | p{2.8cm} | p{4.2cm} | }\hline

  object(s) & paper(s) & redshift & $f_{\m{esc,LyC}}$ & notes \\ \hline\hline
  
  29 LBGs & \citet{Steidel2001} & $z = 3.4 $ & $f_{\m{esc}} \gtrsim 50$ & 5 times more LyC photons per unit comoving volume compared to quasars at z$\approx3$ \\ \hline
  
  14 SFGs & \citet{Shapley2006} & $z\approx3$ & $f_{\m{esc}} = 14$ & \\ \hline
  
  7/73 LBGs, 10/125 LAEs & \citet{Iwata2009} & $z \simeq 3.1$ & $f_{\m{esc,rel}} = 0.45$ & \\ \hline
  
    102 LBGs & \citet{Vanzella2010b} & $z=3.4\--4.5$ & $f_{\m{esc,abs}}^{\m{stack}} < 5\--20\%$ & they also find one individual candidate \\  \hline
    
    15 starburst galaxies & \citet{Siana2010} & $z\approx1.3$ & $f_{\m{esc,rel}}<0.02$ & stacked limit, $3\sigma$ \\ \hline
    
    11 LBGs & \citet{Boutsia2011} & $z\approx 3.3$ & $f_{\m{esc,rel}}<5\%$ & smallest limit at that $z$ \\ \hline
  
  6/26 LBGs, 28/130 LAEs & \citet{Nestor2011} & $z\approx 3$ & $\sim 10\%$ for LBGs & low UV/LyC ratios of LAEs explained with orientation effects, follow-up of SSA22 field \\ \hline
  
  9/41 LBGs, 20/91 LAEs & \citet{Nestor2013} & $z\approx3$ & $f_{\m{esc}}^{\m{LBG}}=5\--7\%$, $f_{\m{esc}}^{\m{LAE}}=10\--30\%$ & follow-up of SSA22 field\\ \hline
  
  4/49 LBGs, 7/91 LAEs & \citet{Mostardi2013} & $z \approx  2.85$ & $f_{\m{esc,rel}}^{\m{LBG}}=5\--8\%$, $f_{\m{esc,rel}}^{\m{LAE}}=18\--49\%$ & \\ \hline
  
  $\sim 600$ SFGs & \citet{Rutkowski2016} & $z\approx1$ & $f_{\m{esc,abs}}<2.1\%$ & $3 \sigma$ limit from individual non-detections \\ \hline
  
  37 galaxies & \citet{Grazian2016} & $z \approx 3.3$ & $f_{\m{esc,rel}}<2\%$ & two individual detections, fesc for stacks \\ \hline
  
  7 lensed galaxies & \citet{Leethochawalit2016} & $4<z<5$ & $f_{\m{esc,abs}}\simeq 19 \pm 6\%$ & \\ \hline
  
  588 H$\alpha$ galaxies, 160 LAEs& \citet{Matthee2017a} & $z\approx2$ & $f_{\m{esc}}<2.8\, (6.4)\%$ & stacking median (mean) \\ \hline
  
  69 SFGs & \citet{Grazian2017} & $z\approx3.3$ & $f_{\m{esc,rel}}^{\m{bright}}<1.7\%$, $f_{\m{esc,rel}}^{\m{faint}} \lesssim 10\%$ & \\ \hline
  
  SFGs & \citet{Rutkowski2017} & $z\approx2.5$ & $f_{\m{esc,rel}}<15\%$ & selected on [OII] and high [OIII]/[OII] \\ \hline
  
  6 galaxies & \citet{Naidu2017} & $z\approx 2$ & $f_{\m{esc}}>60\%$ & \\ \hline
  
  73 galaxies & \citet{Naidu2018} & $z\approx3.5$ & $f_{\m{esc,rel}} < 6.5^{+0.7}_{-0.7}\%$ & selected based on high [OIII]/[OII] but no LyC detection \\ \hline
  
  124 galaxies & \citet{Steidel2018} & $z\approx3$ & $f_{\m{esc, abs}}=0.09 \pm 0.01$ & \\ \hline
  
  61 galaxies & \citet{Fletcher2019} & $z\simeq 3.1$ & & $20\%$ of objects show LyC leakage \\ \hline
  
  110 galaxies & \citet{Smith2020} & $2.26<z<4.3$ & & dominated by AGN \\ \hline
  
    5 candidates & \citet{Jones2021} & $2.35<z<3.05$ & & \\ \hline
    
  11 candidates & \citet{saxena2022} & $3.1<z<3.5$ & $f_{\m{esc,abs}}=0.7 - 0.52$ & \\ \hline 
  
    6 candidates & \citet{Rivera-Thorsen2022} & $2<z<3.5$ & $f_{\m{esc,abs}}=0.36 - 1$ & \\ \hline 
        
    148 SFGs & \citet{Begley2022} & $z\simeq3.5$ & $f_{\m{esc}}=0.07 \pm 0.02 $ & \\ \hline 
 
\end{tabular}
\end{center}
\caption{Overview of some of the currently known LyC emission at high redshift, showing results from surveys or stacks, often using star-forming galaxies (SFGs), AGN, LAEs or LBGs.}
\label{tab:LyC_leakers_stacks}
\end{table*}

\begin{table*}
\begin{center}
\begin{tabular}{ | p{3.5cm} | p{3.2cm} | p{1.8cm} | p{2.8cm} | p{4.8cm} | }\hline

  object(s) & paper(s) & distance & $f_{\m{esc,LyC}}$ & notes \\ \hline\hline
  
  IRAS 08339+6517, Mrk 1267, Mrk 66, and Mrk 496 (=NGC 6090) & \citet{Leitherer1995b} & $>5000\,$km/s & $<3\%$ & \\ \hline
  
  Haro 11 & \citet{Bergvall2006} & $z=0.021$ & $f_{\m{esc,rel}} = 4--10\%$ & \\ \hline

  16 local starburst galaxies (including Haro 11) & \citet{Grimes2009} & & $<1\%$ & no detection of LyC among local starburst galaxies \\ \hline

  Haro 11 & \citet{Leitet2011} & $z=0.021$ & $f_{\m{esc}} = 16.6^{+7.4}_{-6.5}\%$ & The relative escape fraction was taken from \citet{Leitet2013}. \\ \hline
  
  Tol 1247-232 & \citet{Leitet2013, Leitherer2016}, \citet{Micheva2018b} & $207\, \m{Mpc}$ & $f_{\m{esc,rel}} = 7.4^{+7.7}_{-6.7}\%$, \,\,\,\,\, $f_{\m{esc,rel}} = 21.6 \pm 5.9 \%$ & \\  \hline
  
  J0921+4509 & \citet{Borthakur2014} & $z= 0.235$ & $f_{\m{esc}} = 21 \% \pm 5\%$ & Lyman Break Analogue. \\  \hline
  
  Mrk 54 & \citet{Leitherer2016} & $191\, \m{Mpc}$ & $f_{\m{esc,rel}} = 20.8 \pm 6.1 \%$ & \citet{Chisholm2017} did not detect any LyC emission: the candidate is likely contaminated. \\  \hline
  
  Tol 0440-381 & \citet{Leitherer2016} & $167\, \m{Mpc}$ & $f_{\m{esc,rel}} = 59.8 \pm 13 \%$ & Might be contaminated by geocoronal lines. \\  \hline
  
  J0925+1403 & \citet{Izotov2016a} & $z=0.301$ & $f_{\m{esc,rel}} = 8 \%$ & \\ \hline

  5 objects & \citet{Izotov2016b,Schaerer2016} & $z\approx 0.3$ & $f_{\m{esc,rel}} =  9 \-- 33 \%$ & high [\ion{O}{III}] /[\ion{O}{II}] ratio\\  \hline
  
  Haro 11 & \citet{Keenan2017,Rivera-Thorsen2017} & & & no LyC in knot B and C, possibly in A \\ \hline
  
  J1154+2443 & \citet{Izotov2018a} & $z = 0.369 $ & $f_{\m{esc}} = 46 \% $ &  \\ \hline
  
  6 objects & \citet{Izotov2018b} & $z= 0.3\--0.43$ & $f_{\m{esc}} = 2\--72 \%$ & \\  \hline
  
  J1503+3644 & \citet{Chisholm2020} & $z\approx 0.3$ & $f_{\m{esc}} = 6 \% $ & galaxy from \citet{Izotov2016b}, here detection of MgII\\ \hline
  
  35/66 objects in the LzLCS survey & \citet{Flury2022a} & $z=0.2-0.4$ & $f_{\m{esc}} = 0-50 \% $ & highest $f_{\m{esc}} = 58.4 \% $ \\ \hline

\end{tabular}
\end{center}
\caption{Overview of currently known LyC leakers at low redshift with their escape fractions. The same subscripts apply as for table \ref{tab:LyC_leakers_ind_high}. Note that \citet{Chisholm2017} found a lower LyC escape fraction in Tol 1247-232 and Tol 0440-381 than \citet{Leitherer2016} and no LyC emission from Mrk 54 at all, but they confirm J0921+4509 from \citet{Borthakur2014} as a leaker.}
\label{tab:LyC_leakers_low}
\end{table*}

\end{appendix}


\end{document}